\documentclass[twocolumn, aip, reprint, author-year, unsortedaddress]{revtex4-1}

\usepackage{graphics}
\usepackage{epsfig}
\usepackage{color}
\usepackage{placeins}
\graphicspath{{./fig/}}  %% PATH FOR FIGURES

\begin{document}
\title{Econophysics: Empirical facts and agent-based models}

\author{Anirban Chakraborti}
\email{anirban.chakraborti@ecp.fr}
\author{Ioane Muni Toke}
\email{ioane.muni-toke@ecp.fr}
\affiliation{Laboratoire de Math\'{e}matiques Appliqu\'{e}es aux Syst\`{e}mes, Ecole Centrale Paris, 92290 Ch\^{a}tenay-Malabry, France}

\author{Marco Patriarca}
\email{marco.patriarca@kbfi.ee}
\affiliation{National Institute of Chemical Physics and Biophysics, R\"avala 10, 15042 Tallinn, Estonia}
\affiliation{Instituto de Fisica Interdisciplinaire y Sistemas Complejos (CSIC-UIB), E-07122 Palma de Mallorca, Spain}

\author{Fr\'ed\'eric Abergel}
\email{frederic.abergel@ecp.fr}
\affiliation{Laboratoire de Math\'{e}matiques Appliqu\'{e}es aux Syst\`{e}mes, Ecole Centrale Paris, 92290 Ch\^{a}tenay-Malabry, France}

\begin{abstract}  
This article aim at reviewing recent empirical and theoretical developments usually grouped under the term \emph{Econophysics}. Since its name was coined in 1995 by merging the words ``Economics'' and ``Physics'', this new interdisciplinary field has grown in various directions: theoretical macroeconomics (wealth distributions), microstructure of financial markets (order book modelling), econometrics of financial bubbles and crashes, etc. In the first part of the review, we begin with discussions on the interactions between Physics, Mathematics, Economics and Finance that led to the emergence of Econophysics. Then we present empirical studies revealing statistical properties of financial time series. We begin the presentation with the widely acknowledged ``stylized facts'' which describe the returns of financial assets -- fat tails, volatility clustering, autocorrelation, etc. -- and recall that some of these properties are directly linked to the way ``time'' is taken into account. We continue with the statistical properties observed on order books in financial markets. For the sake of illustrating this review, (nearly) all the stated facts are reproduced using our own high-frequency financial database. Finally, contributions to the study of correlations of assets such as random matrix theory and graph theory are presented. In the second part of the review, we deal with models in Econophysics through the point of view of agent-based modelling. Amongst a large number of multi-agent-based models, we have identified three representative areas. First, using previous work originally presented in the fields of behavioural finance and market microstructure theory, econophysicists have developed agent-based models of order-driven markets that are extensively presented here. Second, kinetic theory models designed to explain some empirical facts on wealth distribution are reviewed. Third, we briefly summarize game theory models by reviewing the now classic minority game and related problems.


\end{abstract}                                                                 

\date{\today}

\maketitle
%\tableofcontents

\part{}

\section{Introduction}
\label{section:EconomicsMarketsMathematicsPhysics}

\textit{What is Econophysics}? Fifteen years after the word ``Econophysics'' was coined by H. E. Stanley by a merging of the words `Economics' and `Physics', at an international conference on Statistical Physics held in Kolkata in 1995, this is still a commonly asked question.
Many still wonder how theories aimed at explaining the physical world in terms of particles could be applied to understand complex structures, such as those found in the social and economic behaviour of human beings.
In fact, physics as a natural science is supposed to be precise or specific; its predictive powers based on the use of a few but universal properties of matter which are sufficient to explain many physical phenomena. 
But in social sciences, are there analogous precise universal properties known for human beings, who, on the contrary of fundamental particles, are certainly not identical to each other in any respect~? And what little amount of information would be sufficient to infer some of their complex behaviours~? 
There exists a positive strive in answering these questions. In the 1940's, Majorana had taken scientific interest in financial and economic systems. He wrote a pioneering paper on the essential analogy between statistical laws in physics and in social sciences (\cite{Majorana1942a,Mantegna2005a,Mantegna2006a}). 
However, during the following decades, only few physicists like \cite{Kadanoff1971a} or \cite{Montroll1974a} had an explicit interest for research in social or economic systems. It was not until the 1990's that physicists started turning to this interdisciplinary subject, and in the past years, they have made many successful attempts to approach problems in various fields of social sciences (e.g. \cite{Oliveira1999a,Stauffer2006c,Chakrabarti2006a}).
In particular, in Quantitative Economics and Finance, physics research has begun to be complementary to the most traditional approaches such as mathematical (stochastic) finance. These various investigations, based on methods imported from or also used in physics, are the subject of the present paper.

\subsection{Bridging Physics and Economics}

Economics deals with how societies efficiently use their resources to produce valuable commodities and distribute them among different people or economic agents (\cite{Samuelson1998a,Keynes1973a}). It is a discipline related to almost everything around us, starting from the marketplace through the environment to the fate of nations.
At first sight this may seem a very different situation from that of physics, whose birth as a well defined scientific theory is usually associated with the study of particular mechanical objects moving with negligible friction, such as falling bodies and planets.
However, a deeper comparison shows many more analogies than differences. 
On a general level, both economics and physics deal with ``everything around us'', despite with different perspectives. 
On a practical level, the goals of both disciplines can be either purely theoretical in nature or strongly oriented toward the improvement of the quality of life.
On a more technical side, analogies often become equivalences. Let us give here some examples.

Statistical mechanics has been defined as the
\begin{quote}
``branch of physics that combines the principles and procedures of statistics with the laws of both classical and quantum mechanics, particularly with respect to the field of thermodynamics. It aims to predict and explain the measurable properties of macroscopic systems on the basis of the properties and behaviour of the microscopic constituents of those systems.''\footnote{In Encyclop{\ae}dia Britannica. Retrieved June 11, 2010, from Encyclop{\ae}dia Britannica Online.}
\end{quote}

The tools of statistical mechanics or statistical physics (\cite{Reif1985a,Pathria1996a,Landau5}), that include extracting the average properties of a macroscopic system from the microscopic dynamics of the systems, are believed to prove useful for an economic system. Indeed, even though it is difficult or almost impossible to write down the ``microscopic equations of motion'' for an economic system with all the interacting entities, economic systems may be investigated at various size scales.  
Therefore, the understanding of the global behaviour of economic systems seems to need concepts such as stochastic dynamics, correlation effects, self-organization, self-similarity and scaling, and for their application we do not have to go into the detailed ``microscopic'' description of the economic system.

Chaos theory has had some impact in Economics modelling, e.g. in the work by \cite{Brock1998} or \cite{Chiarella2006}. 
The theory of disordered systems has also played a core role in Econophysics and study of ``complex systems''. 
The term ``complex systems'' was coined to cover the great variety of such systems which include examples from physics, chemistry, biology and also social sciences. The concepts and methods of statistical physics turned out to be extremely useful in application to these diverse complex systems including economic systems. 
Many complex systems in natural and social environments share the characteristics of
competition among interacting agents for resources and their adaptation to dynamically changing environment (\cite{Parisi1999a,Arthur1999a}). Hence, the concept of disordered systems helps for instance to go beyond the concept of representative agent, an approach prevailing in much of (macro)economics and criticized by many economists (see e.g. \cite{Kirman1992, Gallegati1999}). Minority games and their physical formulations have been exemplary.

Physics models have also helped bringing new theories explaining older observations in Economics. The Italian social economist Pareto investigated a century ago the wealth of individuals in a stable economy (\cite{Pareto1897}) by modelling them with the distribution ${P(>x)\sim x^{-\alpha}}$, where $P(>x)$ is the number of people having income greater than or equal to $x$ and $\alpha$ is an exponent (known now as the Pareto exponent) which he estimated to be $1.5$. 
To explain such empirical findings, physicists have come up with some very elegant and intriguing kinetic exchange models in recent times, and we will review these developments in the companion article.
Though the economic activities of the agents are driven by various considerations like
``utility maximization'', the eventual exchanges of money in any trade can be simply
viewed as money/wealth conserving two-body scatterings, as in the entropy
maximization based kinetic theory of gases. This qualitative analogy seems to be
quite old and both economists and natural scientists have already noted it in various
contexts (\cite{Saha1950}). Recently, an equivalence between the two maximization principles have
been quantitatively established (\cite{Chakrabarti2010}).

Let us discuss another example of the similarities of interests and tools in Physics and Economics. The frictionless systems which mark the early history of physics were soon recognized to be rare cases: not only at microscopic scale -- where they obviously represent an exception due to the unavoidable interactions with the environment -- but also at the macroscopic scale, where fluctuations of internal or external origin make a prediction of their time evolution impossible. Thus equilibrium and non-equilibrium statistical mechanics, the theory of stochastic processes, and the theory of chaos, became main tools for studying real systems as well as an important part of the theoretical framework of modern physics.
Very interestingly, the same mathematical tools have presided at the growth of classic modelling in Economics and more particularly in modern Finance. 
%The study of the nature of fluctuations has closely linked the communities of physicists and economists. 
Following the works of Mandelbrot, Fama of the 1960s, physicists from 1990 onwards have studied the fluctuation of prices and universalities in context of scaling theories, etc.
These links open the way for the use of a physics approach in Finance, complementary to the widespread mathematical one.
%Thus mathematics, and in particular the theory of stochastic processes, certainly represents the main link, common language and framework, for both Finance and Physics.
%It is hard to find other applied disciplines besides physics --- apart from probably chemistry --- where the mathematical language of stochastic processes has a central role.

\subsection{Econophysics and Finance}

Mathematical finance has benefited a lot in the past thirty years from modern probability theory -- Brownian motion, martingale theory, etc. 
Financial mathematicians are often proud to recall the most well-known source of the interactions between Mathematics and Finance: five years before Einstein's seminal work, the theory of the Brownian motion was first formulated by the French mathematician Bachelier in his doctoral thesis (\cite{Bachelier1900a,Boness1967a,Bachelier1990b}), in which he used this model to describe price fluctuations at the Paris Bourse. Bachelier had even given a course as a ``free professor'' at the Sorbonne University with the title: ``Probability calculus with applications to financial operations and analogies with certain questions from physics'' (see the historical articles in \cite{Courtault2000a,Taqqu2001a,MacTutor-Bachelier}).

Then It\=o, following the works of Bachelier, Wiener, and Kolmogorov among many, formulated the presently known It\=o calculus (\cite{Ito1996a}). The geometric Brownian motion, belonging to the class of It\=o processes, later became an important ingredient of models in Economics (\cite{Osborne1959, Samuelson1965a}), and in the well-known theory of option pricing (\cite{Black1973a,Merton1973a}).
In fact, stochastic calculus of diffusion processes combined with classical hypotheses in Economics led to the development of the \emph{arbitrage pricing theory} (\cite{Duffie1996}, \cite{Follmer2004}). The deregulation of financial markets at the end of the 1980's led to the exponential growth of the financial industry. Mathematical finance followed the trend: stochastic finance with diffusion processes and exponential growth of financial derivatives have had intertwined developments. Finally, this relationship was carved in stone when the Nobel prize was given to M.S. Scholes and R.C. Merton in 1997 (F. Black died in 1995) for their contribution to the theory of option pricing and their celebrated ``Black-Scholes'' formula.

However, this whole theory is closely linked to classical economics hypotheses and has not been grounded enough with empirical studies of financial time series. The Black-Scholes hypothesis of Gaussian log-returns of prices is in strong disagreement with empirical evidence. \cite{Mandelbrot1960a,Mandelbrot1963} was one of the firsts to observe a clear departure from Gaussian behaviour for these fluctuations. It is true that within the framework of stochastic finance and martingale modelling, more complex processes have been considered in order to take into account some empirical observations: jump processes (see e.g. \cite{ContTankov2004} for a textbook treatment) and stochastic volatility (e.g. \cite{Heston1993, Gatheral2006}) in particular. 
But recent events on financial markets and the succession of financial crashes (see e.g. \cite{Kindleberger2005} for a historical perspective) should lead scientists to re-think basic concepts of modelling. This is where Econophysics is expected to come to play. During the past decades, the financial landscape has been dramatically changing: deregulation of markets, growing complexity of products. On a technical point of view, the ever rising speed and decreasing costs of computational power and networks have lead to the emergence of huge databases that record all transactions and order book movements up to the millisecond. The availability of these data should lead to models that are better empirically founded. Statistical facts and empirical models will be reviewed in this article and its companion paper. The recent turmoil on financial markets and the 2008 crash seem to plead for new models and approaches. The Econophysics community thus has an important role to play in future financial market modelling, as suggested by contributions from \cite{Bouchaud2008}, \cite{Lux2009} or \cite{Farmer2009}.

\subsection{A growing interdisciplinary field}

The chronological development of Econophysics has been well covered in the book of \cite{Roehner2002a}.
Here it is worth mentioning a few landmarks.
The first article on analysis of finance data which appeared in a physics journal was that of \cite{Mantegna1991a}.
The first conference in Econophysics was held in Budapest in 1997 and has been since followed by numerous schools, workshops and the regular series of meetings: APFA (Application of Physics to Financial Analysis), WEHIA (Workshop on Economic Heterogeneous Interacting Agents), and Econophys-Kolkata, amongst others. 
In the recent years the number of papers has increased dramatically; the community has grown rapidly and several new directions of research have opened. 
By now renowned physics journals like the Reviews of Modern Physics, Physical Review Letters, Physical Review E, Physica A, Europhysics Letters, European Physical Journal B, International Journal of Modern Physics C, etc. publish papers in this interdisciplinary area.
Economics and mathematical finance journals, especially Quantitative Finance, receive contributions from many physicists.
The interested reader can also follow the developments quite well from the preprint server (www.arxiv.org). In fact, recently a new section called quantitative finance has been added to it. One could also visit the web sites of the \emph{Econophysics Forum} (www.unifr.ch/econophysics) and \emph{Econophysics.Org} (www.econophysics.org). The first textbook in Econophysics (\cite{Sinha2010}) is also in press.

\subsection{Organization of the review}

This article aims at reviewing recent empirical and theoretical developments that use tools from Physics in the fields of Economics and Finance.
In section~\ref{part:StatisticalProperties} of this paper, empirical studies revealing statistical properties of financial time series are reviewed. We present the widely acknowledged ``stylized facts'' describing the distribution of the returns of financial assets. In section~\ref{section:OBStats} we continue with the statistical properties observed on order books in financial markets. We reproduce most of the stated facts using our own high-frequency financial database. In the last part of this article (section~\ref{section:Correlations}), we review contributions on correlation on financial markets, among which the computation of correlations using high-frequency data, analyses based on random matrix theory and the use of correlations to build economics taxonomies.
In the companion paper to follow, Econophysics models are reviewed through the point of view of agent-based modelling. Using previous work originally presented in the fields of behavioural finance and market microstructure theory, econophysicists have developed agent-based models of order-driven markets that are extensively reviewed there. We then turn to models of wealth distribution where an agent-based approach also prevails. As mentioned above, Econophysics models help bringing a new look on some Economics observations, and advances based on kinetic theory models are presented. Finally, a detailed review of game theory models and the now classic minority games composes the final part.

\section{Statistics of financial time series: price, returns, volumes, volatility}
\label{part:StatisticalProperties}

Recording a sequence of prices of commodities or assets produce what is called time series. Analysis of financial time series has been of great interest not only to the practitioners (an empirical discipline) but also to the theoreticians for making inferences and predictions. The inherent uncertainty in the financial time series and its theory makes it specially interesting to economists, statisticians and physicists (\cite{Tsay2002a}).

Different kinds of financial time series have been recorded and studied for decades, but the scale changed twenty years ago. The computerization of stock exchanges that took place all over the world in the mid 1980's and early 1990's has lead to the explosion of the amount of data recorded. Nowadays, all transactions on a financial market are recorded \emph{tick-by-tick}, i.e. every event on a stock is recorded with a timestamp defined up to the millisecond, leading to huge amounts of data. For example, as of today (2010), the Reuters Datascope Tick History (RDTH) database records roughly 25 gigabytes of data \emph{every trading day}.

Prior to this improvement in recording market activity, statistics could be computed with daily data at best. Now scientists can compute intraday statistics in high-frequency. This allows to check known properties at new time scales (see e.g. section~\ref{section:RightTime} below), but also implies special care in the treatment (see e.g. the computation of correlation on high-frequency in section~\ref{subsection:CovarianceHighFrequencyData} below).

It is a formidable task to make an exhaustive review on this topic but we try to give a flavour of some of the aspects in this section.

\subsection{``Stylized facts'' of financial time series}
\label{subsection:StylizedFacts}

The concept of ``stylized facts'' was introduced in macroeconomics around 1960 by \cite{Kaldor1961}, who advocated that a scientist studying a phenomenon ``should be free to start off with a stylized view of the facts''. In his work, Kaldor isolated several statistical facts characterizing macroeconomic growth over long periods and in several countries, and took these robust patterns as a starting point for theoretical modelling.

This expression has thus been adopted to describe empirical facts that arose in statistical studies of financial time series and that seem to be persistent across various time periods, places, markets, assets, etc. One can find many different lists of these facts in several reviews (e.g. \cite{BollerslevEngleNelson1994, Pagan1996, Guillaume1997, Cont2001}). We choose in this article to present a minimum set of facts now widely acknowledged, at least for the prices of equities.

\subsubsection{Fat-tailed empirical distribution of returns}

\begin{figure}
\begin{center}
\includegraphics[width=\columnwidth]{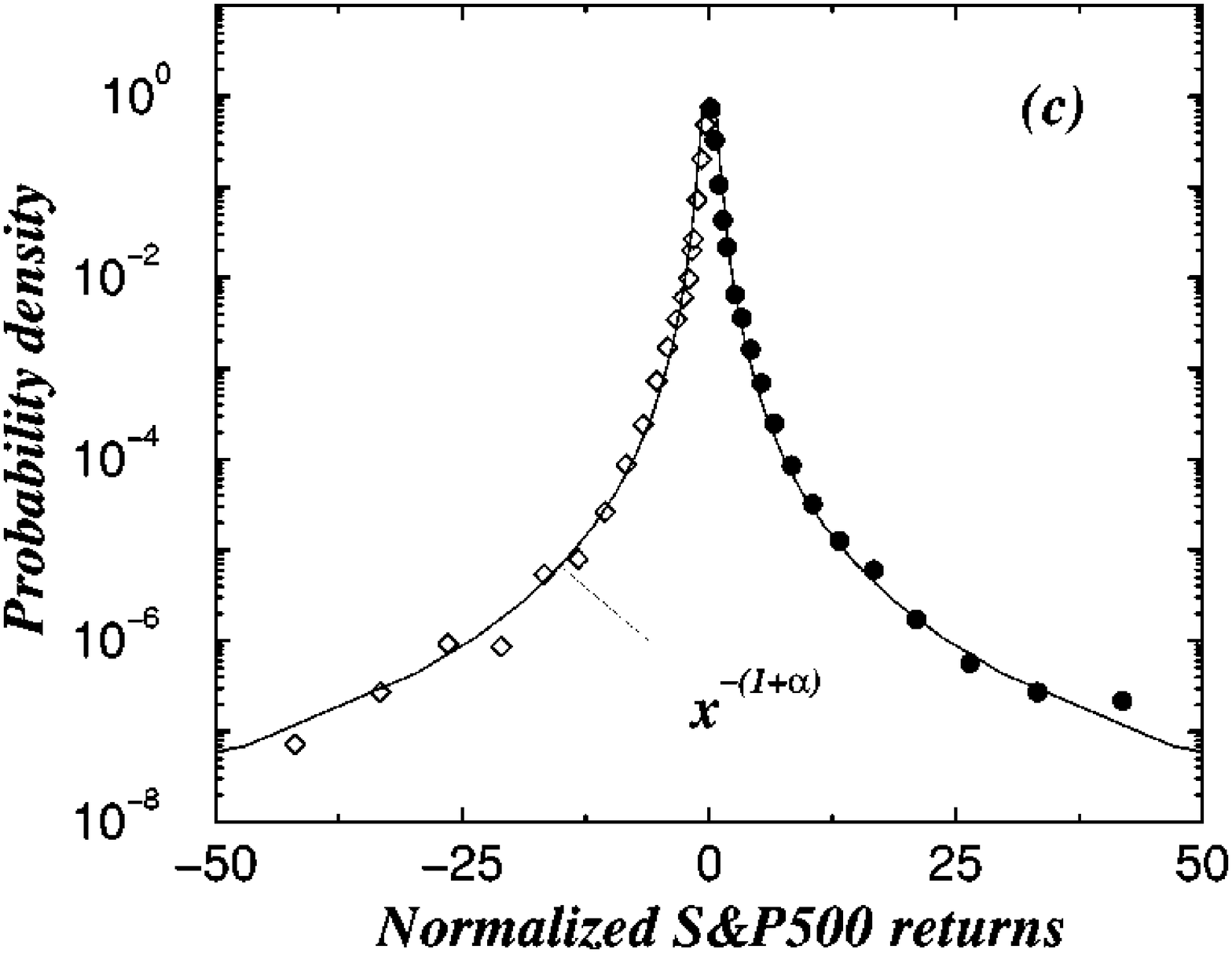}
\includegraphics[width=0.85\columnwidth]{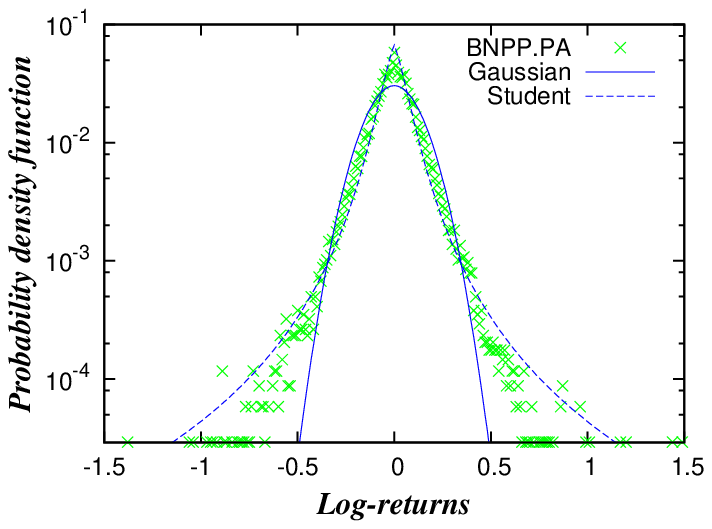}
\end{center}
\caption{(Top) Empirical probability density function of the normalized 1-minute S{\&}P500 returns between 1984 and 1996. Reproduced from \cite{Gopikrishnan1999}. (Bottom) Empirical probability density function of BNP Paribas unnormalized log-returns over a period of time $\tau=5$ minutes.}
%, computed by sampling tick-by-tick data from 9:05am til 5:20pm between January 1st, 2007 and May 30th, 2008, i.e. 15664 values.}}
\label{figure:FT_LogReturnDistribution_Log}
\end{figure}

\begin{figure}
\begin{center}
\includegraphics[width=0.85\columnwidth]{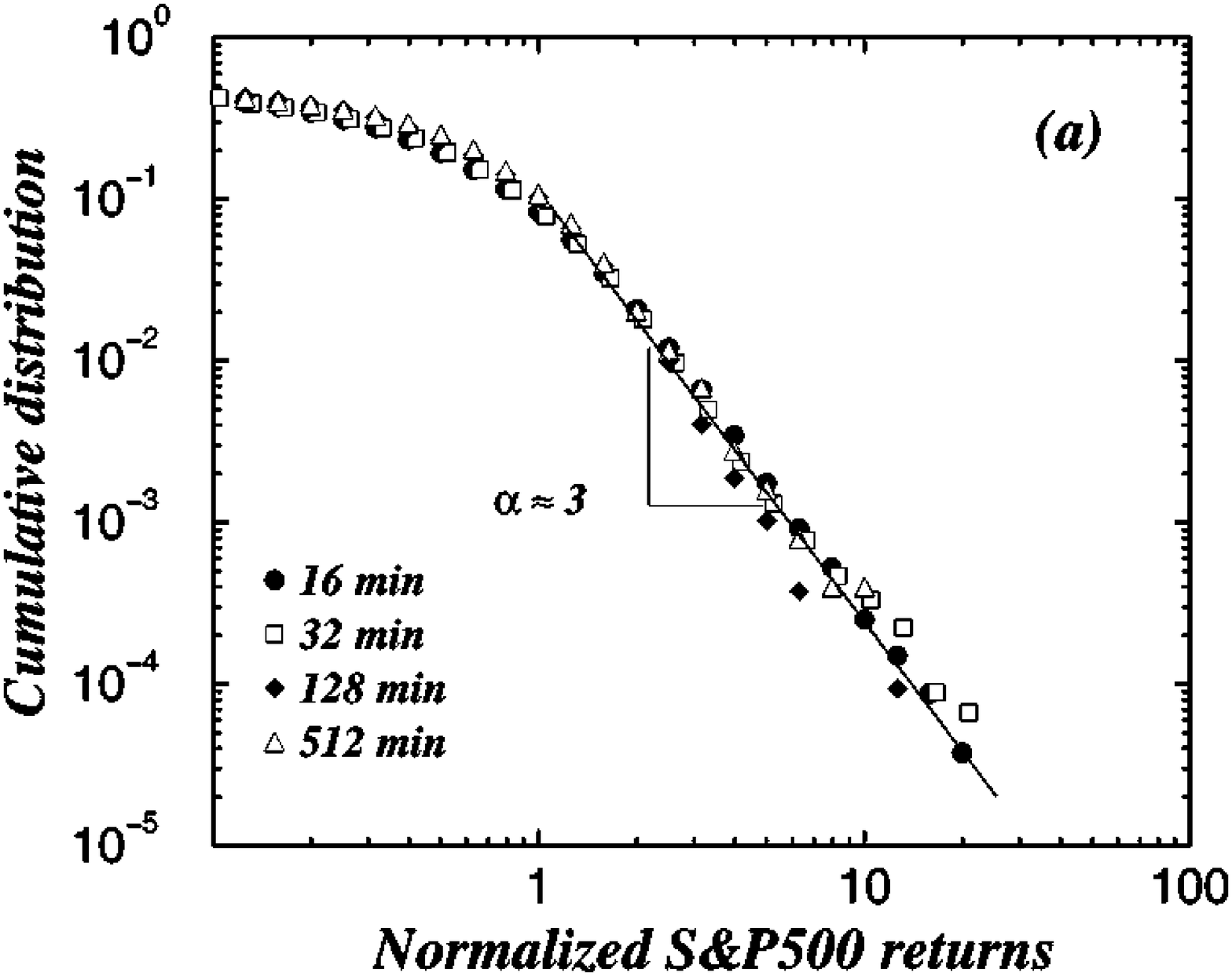}
\includegraphics[width=0.85\columnwidth]{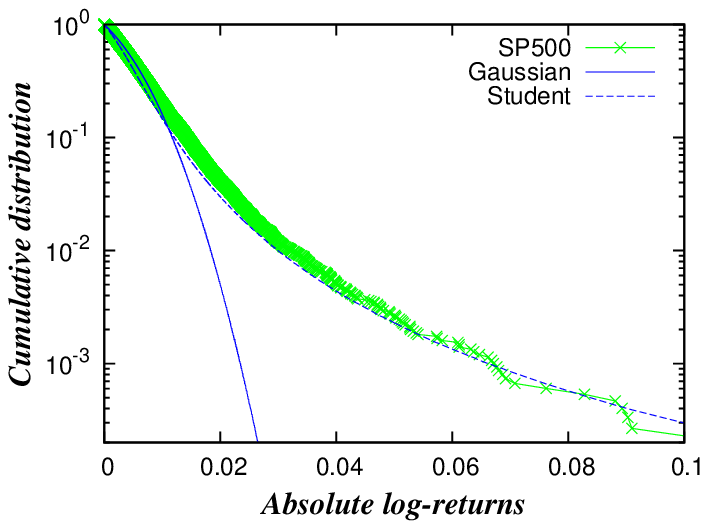}
\end{center}
\caption{Empirical cumulative distributions of S{\&}P 500 daily returns. (Top) Reproduced from \cite{Gopikrishnan1999}, in log-log scale. (Bottom) Computed using official daily close price between January 1st, 1950 and June 15th, 2009, i.e. 14956 values, in linear-log scale.}
\label{figure:SP500_LogReturnDistribution_Log}
\end{figure}

Let $p_t$ be the price of a financial asset at time $t$. We define its return over a period of time $\tau$ to be:
\begin{equation}
	r_{\tau}(t) = \frac{p(t+\tau)-p(t)}{p(t)} \approx \log(p(t+\tau)) - \log(p(t))
\end{equation}
It has been largely observed -- starting with \cite{Mandelbrot1963}, see e.g. \cite{Gopikrishnan1999} for tests on more recent data -- and it is the first stylized fact, that the empirical distributions of financial returns and log-returns are fat-tailed. On figure \ref{figure:FT_LogReturnDistribution_Log} we reproduce the empirical density function of normalized log-returns from \cite{Gopikrishnan1999} computed on the S{\&}P500 index. In addition, we plot similar distributions for unnormalized returns on a liquid French stock (BNP Paribas) with $\tau=5$ minutes. This graph is computed by sampling a set of tick-by-tick data from 9:05am till 5:20pm between January 1st, 2007 and May 30th, 2008, i.e. 356 days of trading. Except where mentioned otherwise in captions, this data set will be used for all empirical graphs in this section. On figure~\ref{figure:SP500_LogReturnDistribution_Log}, cumulative distribution in log-log scale from \cite{Gopikrishnan1999} is reproduced. We also show the same distribution in linear-log scale computed on our data for a larger time scale $\tau=1$ day, showing similar behaviour.

Many studies obtain similar observations on different sets of data. For example, using two years of data on more than a thousand US stocks, \cite{Gopikrishnan1998} finds that the cumulative distribution of returns asymptotically follow a power law $F(r_{\tau }) \sim \left|r\right|^{-\alpha }$ with $\alpha >2$ ($\alpha\approx 2.8-3$). With $\alpha > 2$, the second moment (the variance) is well-defined, excluding stable laws with infinite variance. There has been various suggestions for the form of the distribution: Student's-t, hyperbolic, normal inverse Gaussian, exponentially truncated stable, and others, but no general consensus exists on the exact form of the tails. Although being the most widely acknowledged and the most elementary one, this stylized fact is not easily met by all financial modelling. \cite{Gabaix2006} or \cite{WyartBouchaud2007} recall that efficient market theory have difficulties in explaining fat tails. \cite{LuxSornette2002} have shown that models known as ``rational expectation bubbles'', popular in economics, produced very fat-tailed distributions ($\alpha<1$) that were in disagreement with the statistical evidence.

\subsubsection{Absence of autocorrelations of returns}

\begin{figure}
\begin{center}
\includegraphics[width=\columnwidth]{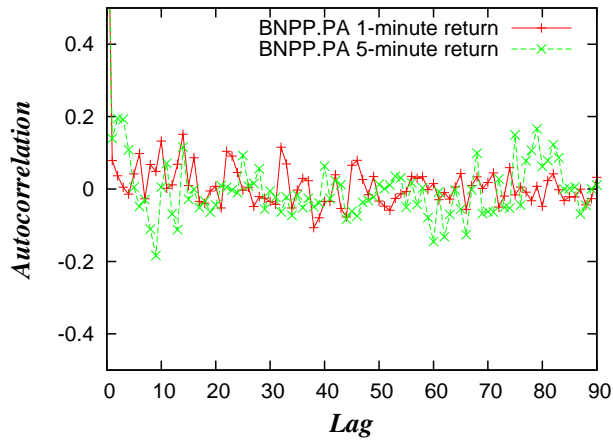}
\end{center}
\caption{Autocorrelation function of BNPP.PA returns.}
% This empirical function is computed by sampling tick-by-tick data from 9:05am til 5:20pm between January 1st, 2007 and May 30th, 2008.}}
\label{figure:AutoCorrLogReturns_BNPP.PA_20070101_20080530_Log}
\end{figure}

On figure \ref{figure:AutoCorrLogReturns_BNPP.PA_20070101_20080530_Log}, we plot the autocorrelation of log-returns defined as $\rho (T)\sim \langle r_{\tau }(t+T)r_{\tau }(t)\rangle $ with $\tau=$1 minute and 5 minutes.
We observe here, as it is widely known (see e.g. \cite{Pagan1996, Cont1997a}), that there is no evidence of correlation between successive returns, which is the second ``stylized-fact''. The autocorrelation function decays very rapidly to zero, even for a few lags of 1 minute.

\subsubsection{Volatility clustering}

The third ``stylized-fact'' that we present here is of primary importance. Absence of correlation between returns must no be mistaken for a property of independence and identical distribution: price fluctuations are not identically distributed and the properties of the distribution change with time.

In particular, absolute returns or squared returns exhibit a long-range slowly decaying auto correlation function. This phenomena is widely known as ``volatility clustering'', and was formulated by \cite{Mandelbrot1963} as ``large changes tend to be followed by large changes -- of either sign -- and small changes tend to be followed by small changes''.

\begin{figure}
\begin{center}
\includegraphics[width=\columnwidth]{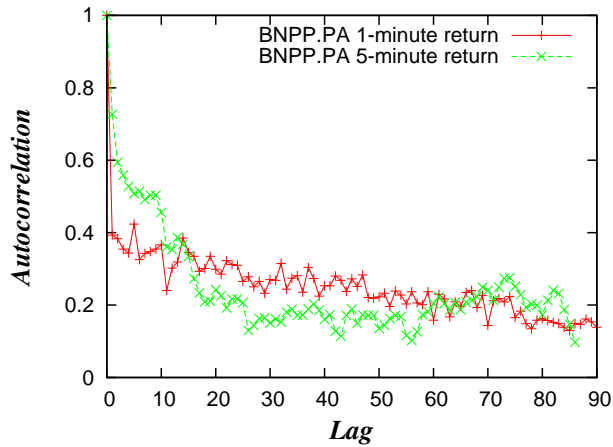}
\end{center}
\caption{Autocorrelation function of BNPP.PA absolute returns.}
% This empirical function is computed by sampling tick-by-tick data from 9:05am till 5:20pm between January 1st, 2007 and May 30th, 2008.}}
\label{figure:AutoCorrAbsoluteLogReturns_BNPP.PA_20070101_20080530_Log}
\end{figure}

On figure \ref{figure:AutoCorrAbsoluteLogReturns_BNPP.PA_20070101_20080530_Log}, the autocorrelation function of absolute returns is plotted for $\tau=$ 1 minute and 5 minutes. The levels of autocorrelations at the first lags vary wildly with the parameter $\tau$. On our data, it is found to be maximum (more than $70\%$ at the first lag) for a returns sampled every five minutes. However, whatever  the sampling frequency, autocorrelation is still above 10\% after several hours of trading. On this data, we can grossly fit a power law decay with exponent $0.4$. Other empirical tests report exponents between $0.1$ and $0.3$ (\cite{Cont1997a,Liu1997a,Cizeau1997a}).

\subsubsection{Aggregational normality}

\begin{figure}
\begin{center}
\includegraphics[width=\columnwidth]{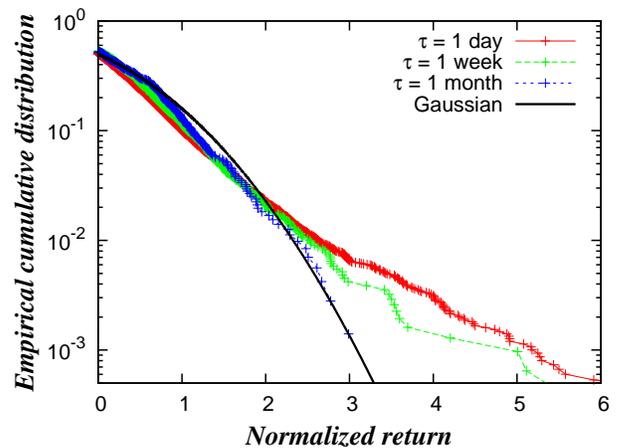}
\end{center}
\caption{Distribution of log-returns of S\&P 500 daily, weekly and monthly returns. Same data set as figure~\ref{figure:SP500_LogReturnDistribution_Log} bottom.}
\label{figure:AggregationalNormality_CalendarTime_SP500_19500101_20090615}
\end{figure}

It has been observed that as one increases the time scale over which the returns are calculated, the fat-tail property becomes less pronounced, and their distribution approaches the Gaussian form, which is the fourth ``stylized-fact''. This cross-over phenomenon is documented in \cite{Kullmann1999} where the evolution of the Pareto exponent of the distribution with the time scale is studied.
On figure \ref{figure:AggregationalNormality_CalendarTime_SP500_19500101_20090615}, we plot these standardized distributions for S\&P 500 index between January 1st, 1950 and June 15th, 2009.
It is clear that the larger the time scale increases, the more Gaussian the distribution is. The fact that the shape of the distribution changes with $\tau$ makes it clear that the random process underlying prices must have non-trivial temporal structure.

\subsection{Getting the right ``time''}
\label{section:RightTime}
\subsubsection{Four ways to measure ``time''}

In the previous section, all ``stylized facts'' have been presented in \emph{physical time}, or \emph{calendar time}, i.e. time series were indexed, as we expect them to be,  in hours, minutes, seconds, milliseconds. Let us recall here that tick-by-tick data available on financial markets all over the world is time-stamped up to the millisecond, but the order of magnitude of the guaranteed precision is much larger, usually one second or a few hundreds of milliseconds.

Calendar time is the time usually used to compute statistical properties of financial time series. This means that computing these statistics involves sampling, which might be a delicate thing to do when dealing for example with several stocks with different liquidity. Therefore, three other ways to keep track of time may be used.

Let us first introduce \emph{event time}. Using this count, time is increased by one unit each time one order is submitted to the observed market. This framework is natural when dealing with the simulation of financial markets, as it will be showed in the companion paper. The main outcome of event time is its ``smoothing'' of data. In event time, intraday seasonality (lunch break) or outburst of activity consequent to some news are smoothed in the time series, since we always have one event per time unit.

Now, when dealing with time series of prices, another count of time might be relevant, and we call it \emph{trade time} or \emph{transaction time}. Using this count, time is increased by one unit each time a transaction happens. The advantage of this count is that limit orders submitted far away in the order book, and may thus be of lesser importance with respect to the price series, do not increase the clock by one unit.

Finally, going on with focusing on important events to increase the clock, we can use \emph{tick time}. Using this count, time is increased by one unit each time the price changes. Thus consecutive market orders that progressively ``eat'' liquidity until the first best limit is removed in an order book are counted as one unit time.

Let us finish by noting that with these definitions, when dealing with mid prices, or bid and ask prices, a time series in event time can easily be extracted from a time series in calendar time. Furthermore, one can always extract a time series in trade time or in price time from a time series in event time. However, one cannot extract a series in price time from a series in trade time, as the latter ignores limit orders that are submitted inside the spread, and thus change mid, bid or ask prices without any transaction taking place.

\subsubsection{Revisiting ``stylized facts'' with a new clock}
\label{subsubsection:StylizedFactsNewClock}

Now, using the right clock might be of primary importance when dealing with statistical properties and estimators. For example, \cite{Griffin2008} investigates the standard realized variance estimator (see section \ref{subsection:CovarianceHighFrequencyData}) in trade time and tick time. \cite{MuniToke2010} also recalls that the differences observed on a spread distribution in trade time and physical time are meaningful. In this section we compute some statistics complementary to the ones we have presented in the previous section \ref{subsection:StylizedFacts} and show the role of the clock in the studied properties.

\paragraph{Aggregational normality in trade time}

\begin{figure}
\begin{center}
\includegraphics[width=\columnwidth]{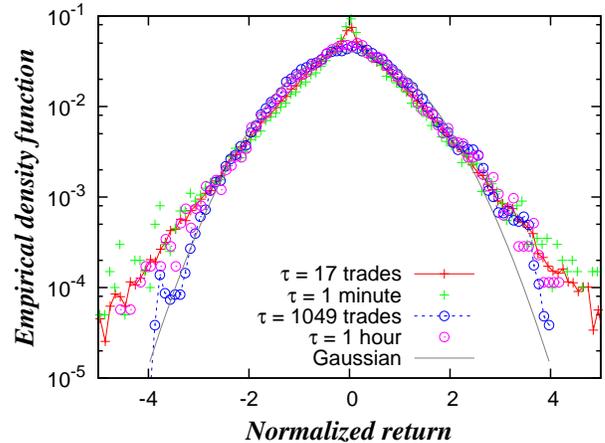}
\end{center}
\caption{Distribution of log-returns of stock BNPP.PA. This empirical distribution is computed using data from 2007, April 1st until 2008, May 31st.}
\label{figure:Normality_TradeVSCalendarTime_BNPP.PA_20070101_20080530}
\end{figure}

We have seen above that when the sampling size increases, the distribution of the log-returns tends to be more Gaussian. This property is much better seen using trade time. On figure \ref{figure:Normality_TradeVSCalendarTime_BNPP.PA_20070101_20080530}, we plot the distributions of the log-returns for BNP Paribas stock using 2-month-long data in calendar time and trade time. Over this period, the average number of trade per day is 8562, so that 17 trades (resp. 1049 trades) corresponds to an average calendar time step of 1 minute (resp. 1 hour). We observe that the distribution of returns sampled every 1049 trades is much more Gaussian than the one sampled every 17 trades (aggregational normality), and that it is also more Gaussian that the one sampled every 1 hour (quicker convergence in trade time).

Note that this property appears to be valid in a multidimensional setting, see \cite{Huth2009}.

\paragraph{Autocorrelation of trade signs in tick time}

It is well-known that the series of the signs of the trades on a given stock (usual convention: $+1$ for a transaction at the ask price, $-1$ for a transaction at the bid price) exhibit large autocorrelation. It has been observed in \cite{LilloFarmer2004} for example that the autocorrelation function of the signs of trades $(\epsilon_n)$ was a slowly decaying function in $n^{-\alpha}$, with $\alpha \approx 0.5$.
\begin{figure}
\begin{center}
\includegraphics[width=\columnwidth]{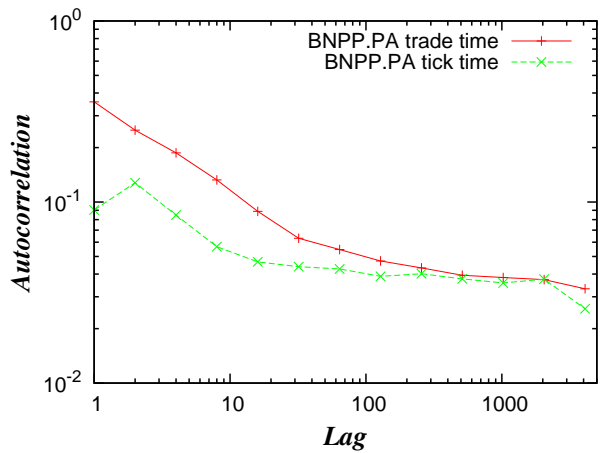}
\end{center}
\caption{Auto-correlation of trade signs for stock BNPP.PA.}
\label{figure:Autocorrelation_OrderSigns_BNPP.PA_20070101_20080530_LogLog}
\end{figure}
We compute this statistics for the trades on BNP Paribas stock from 2007, January 1st until 2008, May 31st. We plot the result in figure \ref{figure:Autocorrelation_OrderSigns_BNPP.PA_20070101_20080530_LogLog}. We find that the first values for short lags are about $0.3$, and that the log-log plot clearly shows some power-law decay with roughly $\alpha \approx 0.7$. 

A very plausible explanation of this phenomenon relies on the execution strategies of some major brokers on a given markets. These brokers have large transaction to execute on the account of some clients. In order to avoid market making move because of an inconsiderably large order (see below section \ref{subsection:MarketImpact} on market impact), they tend to split large orders into small ones. We think that these strategies explain, at least partly, the large autocorrelation observed. Using data on markets where orders are publicly identified and linked to a given broker, it can be shown that the autocorrelation function of the  order signs \emph{of a given broker}, is even higher. See \cite{Bouchaud2009} for a review of these facts and some associated theories.

We present here another evidence supporting this explanation. We compute the autocorrelation function of order signs \emph{in tick time}, i.e. taking only into account transactions that make the price change. Results are plotted on figure \ref{figure:Autocorrelation_OrderSigns_BNPP.PA_20070101_20080530_LogLog}. We find that the first values for short lags are about $0.10$, which is much smaller than the values observed with the previous time series. This supports the idea that many small transactions progressively ``eat'' the available liquidity at the best quotes. Note however that even in tick time, the correlation remains positive for large lags also.

\subsubsection{Correlation between volume and volatility}

\begin{figure}
\begin{center}
\includegraphics[width=\columnwidth]{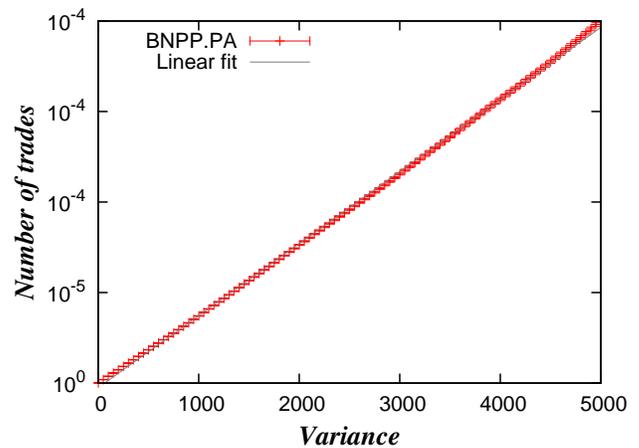}
\end{center}
\caption{Second moment of the distribution of returns over $N$ trades for the stock BNPP.PA.}
\label{figure:LogReturnVarianceVSNtrades_20070101_20080530_BNPP.PA}
\end{figure}

\begin{figure}
\begin{center}
\includegraphics[width=\columnwidth]{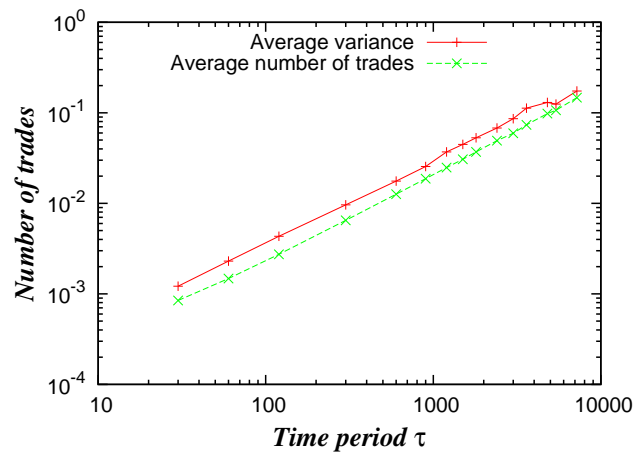}
\end{center}
\caption{Average number of trades and average volatility on a time period $\tau$ for the stock BNPP.PA.}
\label{figure:LogReturnVarianceOverNtradesDeltaT_average_BNPP.PA_20070101_20080530}
\end{figure}

Investigating time series of cotton prices, \cite{Clark1973} noted that ``trading volume and price change variance seem to have a curvilinear relationship''. \emph{Trade time} allows us to have a better view on this property: \cite{Plerou2000}  and \cite{SilvaYakovenko2007} among others, show that the variance of log-returns after $N$ trades, i.e. over a time period of $N$ in trade time, is proprtional to $N$. 
We confirm this observation by plotting the second moment of the distribution of log-returns after $N$ trades as a function of $N$ for our data, as well as the average number of trades and the average volatility on a given time interval. The results are shown on figure \ref{figure:LogReturnVarianceVSNtrades_20070101_20080530_BNPP.PA} and \ref{figure:LogReturnVarianceOverNtradesDeltaT_average_BNPP.PA_20070101_20080530}.

\begin{figure}
\begin{center}
\includegraphics[angle=270,origin=c,width=\columnwidth]{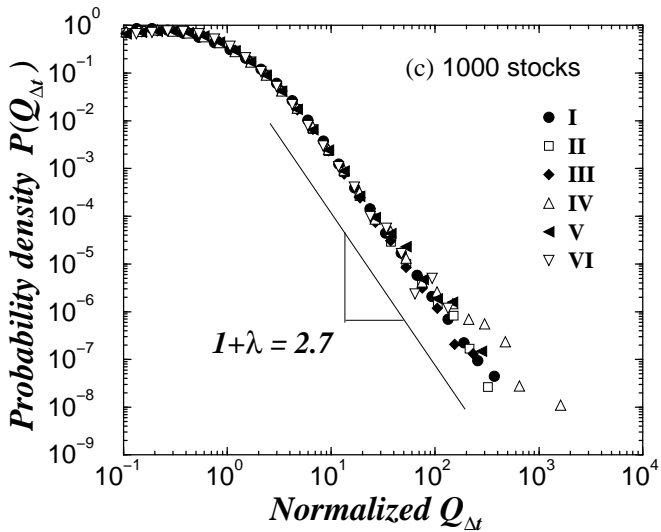}
\end{center}
\caption{Distribution of the number of shares traded $Q_{\Delta t}$. Adapted from \cite{Gopikrishnan2000a}.}
\label{fig:volume-density}
\end{figure}

This results are to be put in relation to the one presented in \cite{Gopikrishnan2000a}, where the statistical properties of the number of shares traded $Q_{\Delta t}$ for a given stock in a fixed time interval $\Delta t$ is studied. They analyzed transaction data for the largest 1000 stocks for the two-year period 1994-95, using a database that recorded every transaction for all securities in three major US stock markets. They found that the distribution $P(Q_{\Delta t})$ displayed a power-law decay as shown in Fig. \ref{fig:volume-density}, and that the time correlations in $Q_{\Delta t}$ displayed long-range persistence. Further, they investigated the relation between $Q_{\Delta t}$ and the number of transactions $N_{\Delta t}$ in a time interval $\Delta t$, and found that the long-range correlations in $Q_{\Delta t}$ were largely due to those of $N_{\Delta t}$. 
Their results are consistent with the interpretation that the large equal-time correlation previously found between $Q_{\Delta t}$ and the absolute value of price change $|G_{\Delta t}|$ (related to volatility) were largely due to $N_{\Delta t}$.

%%%%%%%%%%%%%%%%%%%
%%%%%%%%%%%%%%%%%%%%

Therefore, studying variance of price changer in \emph{trade time} suggests that the number of trade is a good proxy for the unobserved volatility.

\subsubsection{A link with stochastic processes: subordination}

These empirical facts (aggregational normality in trade time, relationship between volume and volatility) reinforce the interest for models based on the subordination of stochastic processes, which had been introduced in financial modeling by \cite{Clark1973}.

Let us introduce it here. Assuming the proportionality between the variance $\langle x \rangle_{\tau}^2$ of the centred returns $x$ and the number of trades $N_\tau$ over a time period $\tau$, we can write:
\begin{equation}
	\langle x \rangle_{\tau}^2 = \alpha N_{\tau}.
\end{equation}
Therefore, assuming the normality in trade time, we can write the density function of log-returns after $ N $ trades as 
\begin{equation}
	f_N(x) = \frac{ e^{\frac{-x^2}{2 \alpha N}} }{\sqrt{2\pi\alpha N}},
\end{equation}
Finally, denoting $K_{\tau}(N)$ the probability density function of having $N$ trades in a time period $\tau$, the distribution of log returns in calendar time can be written
\begin{equation}
	P_{\tau}(x) = \int_0^\infty \frac{ e^{\frac{-x^2}{2 \alpha N}} }{\sqrt{2\pi\alpha N}} K_{\tau}(N) dN.
\end{equation}

This is the subordination of the Gaussian process $x_N$ using the number of trades $N_\tau$ as the \emph{directing process}, i.e. as the new clock. With this kind of modelization, it is expected, since $P_N$ is gaussian, the observed non-gaussian behavior will come from $K_{\tau}(N)$. For example, some specific choice of directing processes may lead to a symmetric stable distribution (see \cite{Feller1968}). \cite{Clark1973} tests empirically a log-normal subordination with time series of prices of cotton. In a similar way, \cite{SilvaYakovenko2007} find that an exponential subordination with a kernel:
\begin{equation}
	K_\tau(N) = \frac{1}{\eta\tau} e^{-\frac{N}{\eta\tau}}.
	\label{equation:KSilva}
\end{equation}
is in good agreement with empirical data.
If the orders were submitted to the market in a independent way and at a constant rate $\eta$, then the distribution of the number of trade per time period $\tau$ should be a Poisson process with intensity $\eta\tau$. Therefore, the empirical fit of equation~(\ref{equation:KSilva}) is inconsistent with such a simplistic hypothesis of distribution of time of arrivals of orders. We will suggest in the next section some possible distributions that fit our empirical data.

%%%%%%%%%%%%%%%%%%%%%%%%%%%%%%%%%%%%%%%%%%%%%%%%%%%%%%%%%%%%%%%%%%%%%%%%
%%%%%%%%%%%%%%%%%%%%%%%%%%%%%%%%%%%%%%%%%%%%%%%%%%%%%%%%%%%%%%%%%%%%%%%%
%%%%%%%%%%%%%%%%%%%%%%%%%%%%%%%%%%%%%%%%%%%%%%%%%%%%%%%%%%%%%%%%%%%%%%%%
%%%%%%%%%%%%%%%%%%%%%%%%%%%%%%%%%%%%%%%%%%%%%%%%%%%%%%%%%%%%%%%%%%%%%%%%
%%%%%%%%%%%%%%%%%%%%%%%%%%%%%%%%%%%%%%%%%%%%%%%%%%%%%%%%%%%%%%%%%%%%%%%%
%%%%%%%%%%%%%%%%%%%%%%%%%%%%%%%%%%%%%%%%%%%%%%%%%%%%%%%%%%%%%%%%%%%%%%%%
\FloatBarrier
\section{Statistics of order books}
\label{section:OBStats}

The computerization of financial markets in the second half of the 1980's provided the empirical scientists with easier access to extensive data on order books. \cite{Biais1995} is an early study of the new data flows on the newly (at that time) computerized Paris Bourse. Variables crucial to a fine modeling of order flows and dynamics of order books are studied: time of arrival of orders, placement of orders, size of orders, shape of order book, etc. Many subsequent papers offer complementary empirical findings and modeling, e.g. \cite{Gopikrishnan2000}, \cite{ChalletStinchcombe2001}, \cite{MaslovMills2001}, \cite{BouchaudPotters2002}, \cite{BouchaudPotters2003}. Before going further in our review of available models, we try to summarize some of these empirical facts.

For each of the enumerated properties, we present new empirical plots. We use Reuters tick-by-tick data on the Paris Bourse. We select four stocks: France Telecom (FTE.PA) , BNP Paribas (BNPP.PA), Societe G\'en\'erale (SOGN.PA) and Renault (RENA.PA). For any given stocks, the data displays time-stamps, traded quantities, traded prices, the first five best-bid limits and the first five best-ask limits. From now on, we will denote $a_i(t)$ (resp. ($b_j(t)$) the price of the $i$-th limit at ask (resp. $j$-th limit at bid). Except when mentioned otherwise, all statistics are computed using all trading days from Oct, 1st 2007 to May, 30th 2008, i.e. 168 trading days. On a given day, orders submitted between 9:05am and 5:20pm are taken into account, i.e. first and last minutes of each trading days are removed.

Note that we do not deal in this section with the correlations of the signs of trades, since statistical results on this fact have already been treated in section \ref{subsubsection:StylizedFactsNewClock}.
Note also that although most of these facts are widely acknowledged, we will not describe them as new ``stylized facts for order books'' since their ranges of validity are still to be checked among various products/stocks, markets and epochs, and strong properties need to be properly extracted and formalized from these observations. However, we will keep them in mind as we go through the new trend of ``empirical modeling'' of order books.

Finally, let us recall that the markets we are dealing with are electronic order books with no official market maker, in which orders are submitted in a double auction and executions follow price/time priority. This type of exchange is now adopted nearly all over the world, but this was not obvious as long as computerization was not complete. Different market mechanisms have been widely studied in the microstructure literature, see e.g. \cite{Garman1976,Kyle1985,Glosten1994,OHara1997,Biais1997,Hasbrouck2007}. We will not review this literature here (except \cite{Garman1976} in our companion paper), as this would be too large a digression. However, such a literature is linked in many aspects to the problems reviewed in this paper.

\subsection{Time of arrivals of orders}

As explained in the previous section, the choice of the time count might be of prime importance when dealing with ``stylized facts'' of empirical financial time series. When reviewing the subordination of stochastic processes (\cite{Clark1973,SilvaYakovenko2007}), we have seen that the Poisson hypothesis for the arrival times of orders is not empirically verified.

\begin{figure}
\begin{center}
\includegraphics[width=\columnwidth]{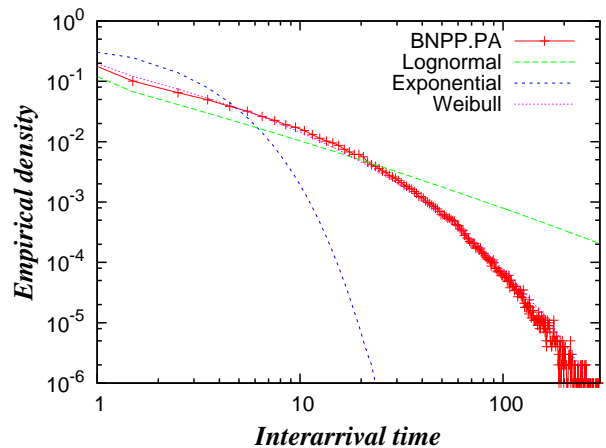}
\end{center}
\caption{Distribution of interarrival times for stock BNPP.PA in log-scale.}
\label{figure:MarketOrdersInterarrivalTimes_BNPP.PA_20071001_20080530_LogLog}
\end{figure}

\begin{figure}
\begin{center}
\includegraphics[width=\columnwidth]{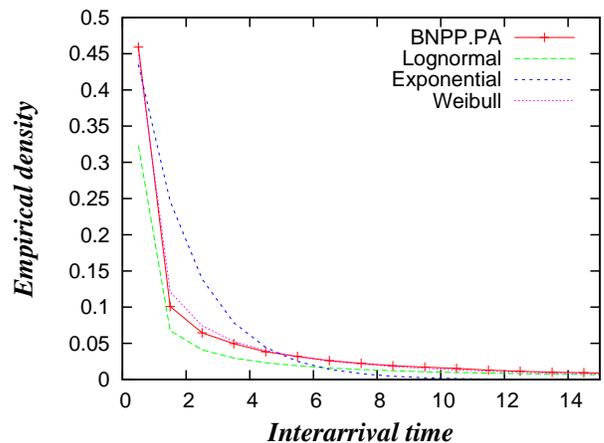}
\end{center}
\caption{Distribution of interarrival times for stock BNPP.PA (Main body, linear scale).}
\label{figure:MarketOrdersInterarrivalTimes_BNPP.PA_20071001_20080530}
\end{figure}

We compute the empirical distribution for interarrival times -- or durations -- of market orders on the stock BNP Paribas using our data set described in the previous section. The results are plotted in figures \ref{figure:MarketOrdersInterarrivalTimes_BNPP.PA_20071001_20080530_LogLog} and \ref{figure:MarketOrdersInterarrivalTimes_BNPP.PA_20071001_20080530}, both in linear and log scale. It is clearly observed that the exponential fit is not a good one. We check however that the Weibull distribution fit is potentially a very good one. Weibull distributions have been suggested for example in \cite{Ivanov2004}. \cite{Politi2008} also obtain good fits with $q$-exponential distributions. 

In the Econometrics literature, these observations of non-Poissonian arrival times have given rise to a large trend of modelling of irregular financial data. \cite{EngleRussell1997} and \cite{Engle2000} have introduced autoregressive condition duration or intensity models that may help modelling these processes of orders' submission. See \cite{Hautsch2004} for a textbook treatment.

\begin{figure}
\begin{center}
\includegraphics[width=\columnwidth]{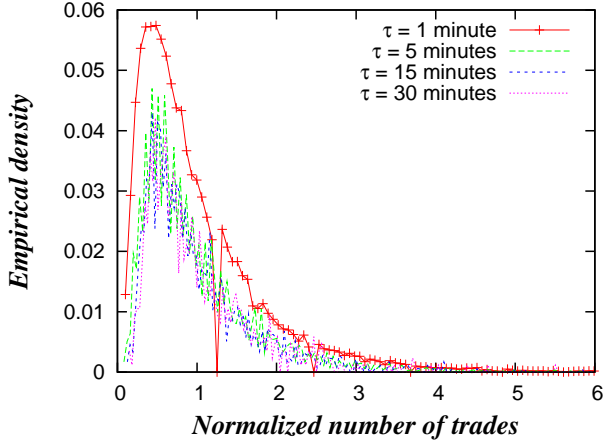}
\end{center}
\caption{Distribution of the number of trades in a given time period $\tau$ for stock BNPP.PA. This empirical distribution is computed using data from 2007, October 1st until 2008, May 31st.}
\label{figure:DistributionNtradesPerDeltaT_BNPP.PA_20071001_20080530}
\end{figure}

Using the same data, we compute the empirical distribution of the number of transactions in a given time period $\tau$. Results are plotted in figure \ref{figure:DistributionNtradesPerDeltaT_BNPP.PA_20071001_20080530}.
It seems that the log-normal and the gamma distributions are both good candidates, however none of them really describes the empirical result, suggesting a complex structure of arrival of orders. A similar result on Russian stocks was presented in \cite{Dremin2005}.

\subsection{Volume of orders}

Empirical studies show that the unconditional distribution of order size is very complex to characterize. \cite{Gopikrishnan2000} and \cite{MaslovMills2001} observe a power law decay with an exponent ${1+\mu\approx 2.3-2.7}$ for market orders and ${1+\mu\approx 2.0}$ for limit orders. \cite{ChalletStinchcombe2001} emphasize on a clustering property: orders tend to have a ``round'' size in packages of shares, and clusters are observed around 100's and 1000's. As of today, no consensus emerges in proposed models, and it is plausible that such a distribution varies very wildly with products and markets.

\begin{figure}[h]
\begin{center}
\includegraphics[width=\columnwidth]{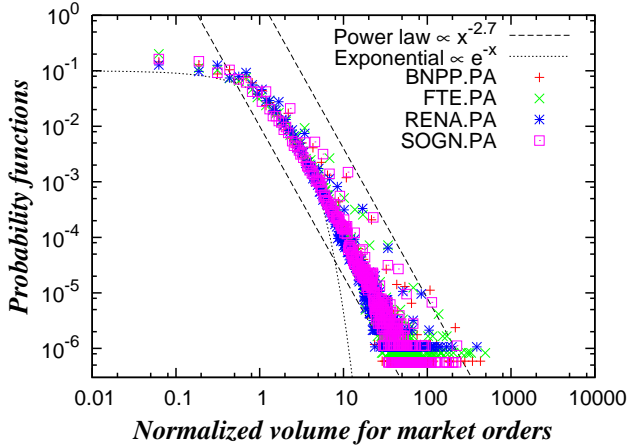}
\end{center}
\caption{Distribution of volumes of market orders. Quantities are normalized by their mean.}
\label{figure:MarketOrderVolume_Log}
\end{figure}

\begin{figure}[h]
\begin{center}
\includegraphics[width=\columnwidth]{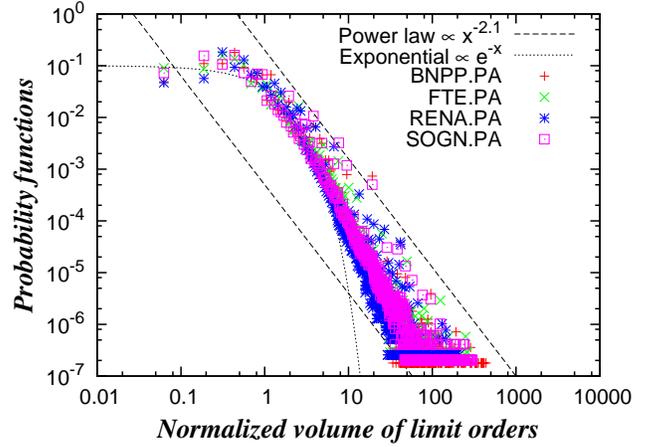}
\end{center}
\caption{Distribution of normalized volumes of limit orders. Quantities are normalized by their mean.}
\label{figure:LimitOrderVolume_Log}
\end{figure}

In figure \ref{figure:MarketOrderVolume_Log}, we plot the distribution of volume of market orders for the four stocks composing our benchmark. Quantities are normalized by their mean. Power-law coefficient is estimated by a Hill estimator (see e.g. \cite{Hill1975,Haan2000}). We find a power law with exponent ${1+\mu\approx 2.7}$ which confirms studies previously cited. Figure \ref{figure:LimitOrderVolume_Log} displays the same distribution for limit orders (of all available limits). We find an average value of ${1+\mu\approx 2.1}$, consistent with previous studies. However, we note that the power law is a poorer fit in the case of limit orders: data normalized by their mean collapse badly on a single curve, and computed coefficients vary with stocks.

\subsection{Placement of orders}

\paragraph{Placement of arriving limit orders} \cite{BouchaudPotters2002} observe a broad power-law placement around the best quotes on French stocks, confirmed in \cite{BouchaudPotters2003} on US stocks. Observed exponents are quite stable across stocks, but exchange dependent: ${1+\mu\approx 1.6}$ on the Paris Bourse, ${1+\mu\approx 2.0}$ on the New York Stock Exchange, ${1+\mu\approx 2.5}$ on the London Stock Exchange. \cite{MikeFarmer2008} propose to fit the empirical distribution with a Student distribution with $1.3$ degree of freedom.

\begin{figure}[h]
\begin{center}
\includegraphics[width=\columnwidth]{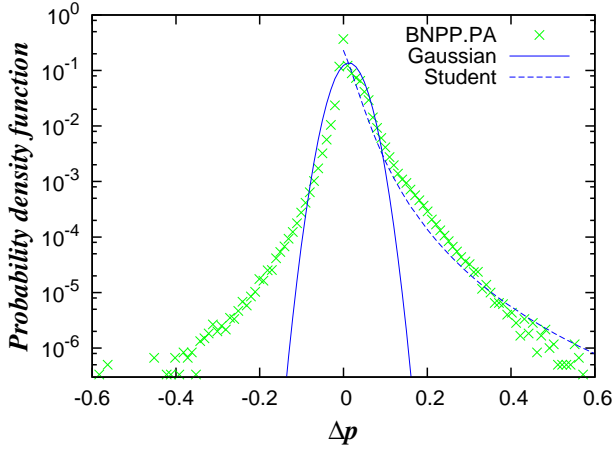}
\end{center}
\caption{Placement of limit orders using the same best quote reference in semilog scale. Data used for this computation is BNP Paribas order book from September 1st, 2007,  until May 31st, 2008.}
\label{figure:LimitOrderDeltaPrice_Log}
\end{figure}

\begin{figure}[h]
\begin{center}
\includegraphics[width=\columnwidth]{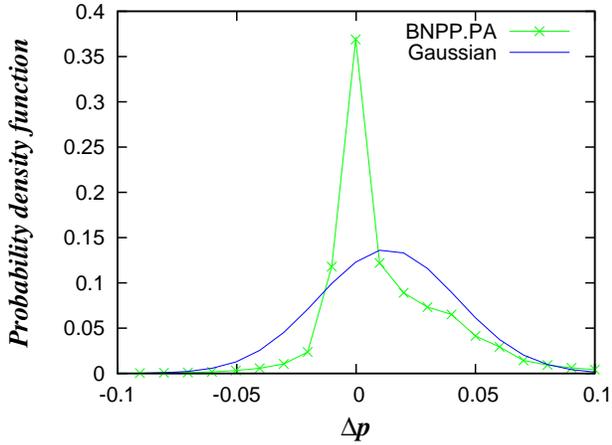}
\end{center}
\caption{Placement of limit orders using the same best quote reference in linear scale. Data used for this computation is BNP Paribas order book from September 1st, 2007,  until May 31st, 2008.}
\label{figure:LimitOrderDeltaPrice}
\end{figure}

We plot the distribution of the following quantity computed on our data set, i.e. using only the first five limits of the order book:	
${\Delta p = b_0(t-) - b(t)}$ (resp. ${a(t) - a_0(t-)}$) if an bid (resp. ask) order arrives at price $b(t)$ (resp. $a(t)$), where $b_0(t-)$ (resp.$a_0(t-)$) is the best bid (resp. ask) before the arrival of this order.
Results are plotted on figures \ref{figure:LimitOrderDeltaPrice_Log} (in semilog scale) and \ref{figure:LimitOrderDeltaPrice} (in linear scale).
These graphs being computed with incomplete data (five best limits), we do not observe a placement as broad as in \cite{BouchaudPotters2002}. However, our data makes it clear that fat tails are observed. We also observe an asymmetry in the empirical distribution: the left side is less broad than the right side. Since the left side represent limit orders submitted \emph{inside} the spread, this is expected. Thus, the empirical distribution of the placement of arriving limit orders is maximum at zero (same best quote). We then ask the question: How is it translated in terms of shape of the order book~?

\begin{figure}[h]
\begin{center}
\includegraphics[width=\columnwidth]{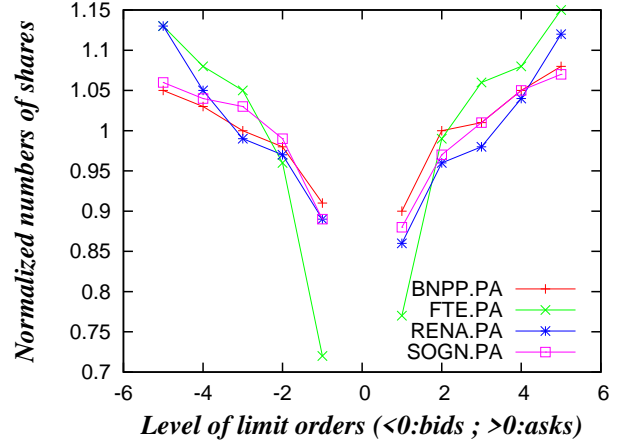}
\end{center}
\caption{Average quantity offered in the limit order book.}
\label{figure:MeanOrderBookPerLevel}
\end{figure}

\begin{figure}[h]
\begin{center}
\includegraphics[width=\columnwidth]{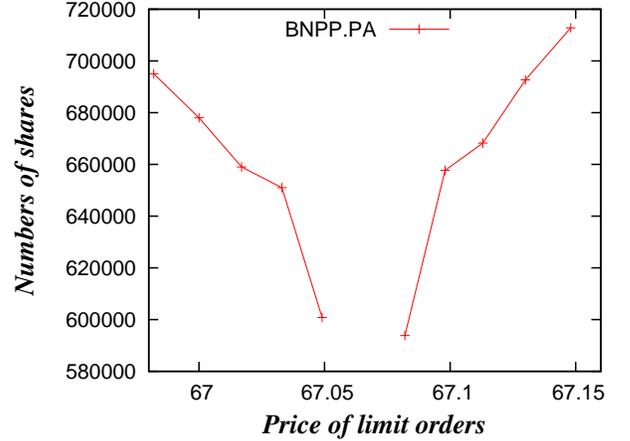}
\end{center}
\caption{Average limit order book: price and depth.}
\label{figure:MeanOrderBookPerLevel_PriceDepth}
\end{figure}

\paragraph{Average shape of the order book}  Contrary to what one might expect, it seems that the maximum of the average offered volume in an order book is located away from the best quotes (see e.g. \cite{BouchaudPotters2002}). Our data confirms this observation: the average quantity offered on the five best quotes grows with the level. This result is presented in figure \ref{figure:MeanOrderBookPerLevel}. We also compute the average price of these levels in order to plot a cross-sectional graph similar to the ones presented in \cite{Biais1995}. Our result is presented for stock BNP.PA in figure \ref{figure:MeanOrderBookPerLevel_PriceDepth} and displays the expected shape. Results for other stocks are similar. We find that the average gap between two levels is constant among the five best bids and asks (less than one tick for FTE.PA, $1.5$ tick for BNPP.PA, $2.0$ ticks for SOGN.PA, $2.5$ ticks for RENA.PA). We also find that the average spread is roughly twice as large the average gap (factor 1.5 for FTE.PA, 2 for BNPP.PA, 2.2 for SOGN.PA, 2.4 for RENA.PA).


%Finally, we compute the average shape of the orderbook as a function of its distance to the opposite best quote, without taking the level into account:  at each time step in event time, we take a snapshot of the pairs $(\Delta p^X_i, q^X_i), X=A,B, i=1,\ldots,5$, and compute the empirical average. Results are presented in figure \ref{figure:MeanOrderBookPerDistanceSameBest}.
%\begin{figure}[ht]
%\begin{center}
%\includegraphics[width=\columnwidth]{include/meanLimitOrderBookPerDistance_BNPP.PA_20070901_20080530_Log.eps}
%\end{center}
%\caption{Average limit order book: price and depth.}
%\label{figure:MeanOrderBookPerDistanceSameBest}
%\end{figure}

\subsection{Cancelation of orders} \cite{ChalletStinchcombe2001} show that the distribution of the average lifetime of limit orders fits a power law with exponent ${1+\mu\approx2.1}$ for cancelled limit orders, and ${1+\mu\approx1.5}$ for executed limit orders. \cite{MikeFarmer2008} find that in either case the exponential hypothesis (Poisson process) is not satisfied on the market.

\begin{figure}[h]
\begin{center}
\includegraphics[width=\columnwidth]{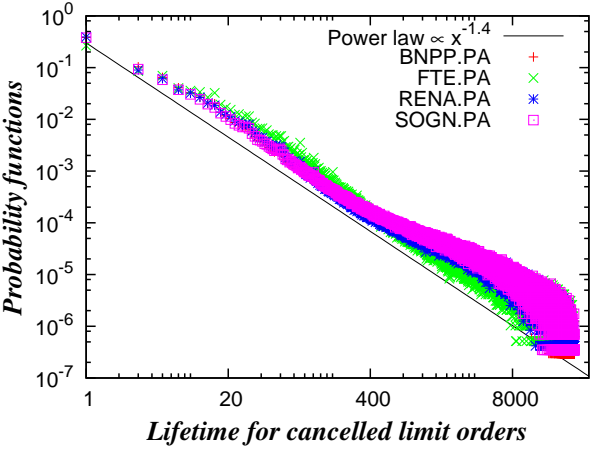}
\end{center}
\caption{Distribution of estimated lifetime of cancelled limit orders.}
\label{figure:CancelledOrdersLifeTime_Log}
\end{figure}

\begin{figure}[h]
\begin{center}
\includegraphics[width=\columnwidth]{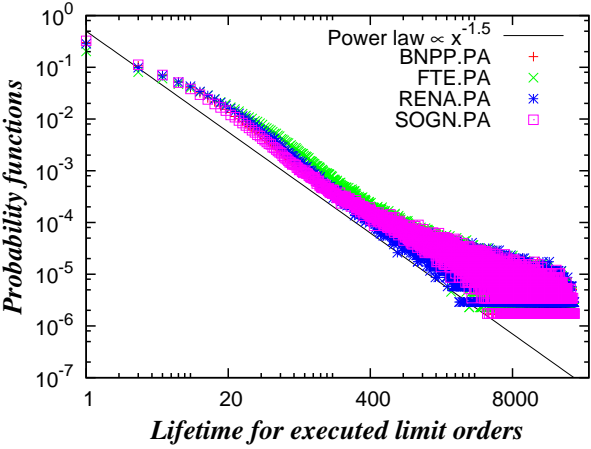}
\end{center}
\caption{Distribution of estimated lifetime of executed limit orders.}
\label{figure:ExecutedOrdersLifeTime_Log}
\end{figure}

We compute the average lifetime of cancelled and executed orders on our dataset. Since our data does not include a unique identifier of a given order, we reconstruct life time orders as follows: each time a cancellation is detected, we go back through the history of limit order submission and look for a matching order with same price and same quantity. If an order is not matched, we discard the cancellation from our lifetime data. Results are presented in figure \ref{figure:CancelledOrdersLifeTime_Log} and \ref{figure:ExecutedOrdersLifeTime_Log}. We observe a power law decay with coefficients ${1+\mu\approx1.3-1.6}$ for both cancelled and executed limit orders, with little variations among stocks. These results are a bit different than the ones presented in previous studies: similar for executed limit orders, but our data exhibits a lower decay as for cancelled orders. Note that the observed cut-off in the distribution for lifetimes above 20000 seconds is due to the fact that we do not take into account execution or cancellation of orders submitted on a previous day.


\subsection{Intraday seasonality}

\begin{figure}[h]
\begin{center}
\includegraphics[width=\columnwidth]{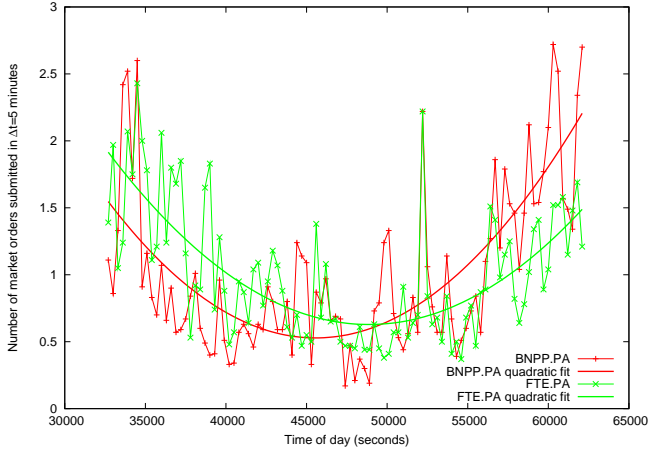}
\end{center}
\caption{Normalized average number of market orders in a 5-minute interval.}
\label{figure:nMarketOrderSeasonality}
\end{figure}

\begin{figure}[h]
\begin{center}
\includegraphics[width=\columnwidth]{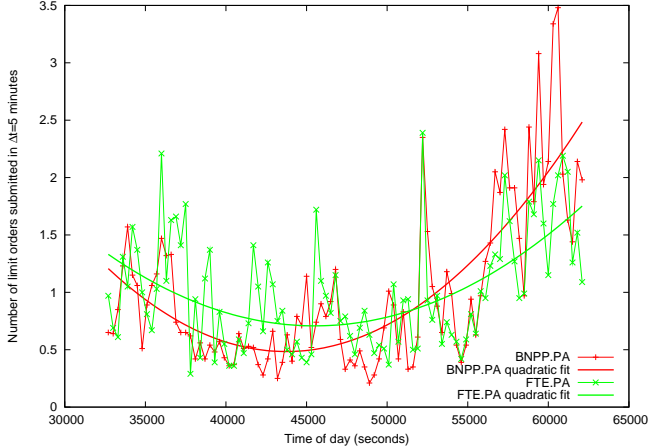}
\end{center}
\caption{Normalized average number of limit orders in a 5-minute interval.}
\label{figure:nLimitOrderSeasonality}
\end{figure}

Activity on financial markets is of course not constant throughout the day. Figure \ref{figure:nMarketOrderSeasonality} (resp. \ref{figure:nLimitOrderSeasonality}) plots the (normalized) number of market (resp. limit) orders arriving in a 5-minute interval. It is clear that a U-shape is observed (an ordinary least-square quadratic fit is plotted): the observed market activity is larger at the beginning and the end of the day, and more quiet around mid-day. Such a U-shaped curve is well-known, see \cite{Biais1995}, for example. On our data, we observe that the number of orders on a 5-minute interval can vary with a factor 10 throughout the day.


\cite{ChalletStinchcombe2001} note that the average number of orders submitted to the market in a period $\Delta T$ vary wildly during the day. The authors also observe that these quantities for market orders and limit orders are highly correlated. Such a type of intraday variation of the global market activity is a well-known fact, already observed in \cite{Biais1995}, for example.

\subsection{Market impact}
\label{subsection:MarketImpact}

The statistics we have presented may help to understand a phenomenon of primary importance for any financial market practitioner: the market impact, i.e. the relationship between the volume traded and the expected price shift once the order has been executed. On a first approximation, one understands that it is closely linked with many items described above: the volume of market orders submitted, the shape of the order book (how much pending limit orders are hit by one large market orders), the correlation of trade signs (one may assume that large orders are splitted in order to avoid a large market impact), etc.

Many empirical studies are available. An empirical study on the price impact of individual transactions on 1000 stocks on the NYSE is conducted in \cite{Lillo2003}. It is found that proper rescaling make all the curve collapse onto a single concave master curve. This function increases as a power that is the order of $1/2$ for small volumes, but then increases more slowly for large volumes. They obtain similar results in each year for the period 1995 to 1998. 

We will not review any further the large literature of market impact, but rather refer the reader to the recent exhaustive synthesis proposed in \cite{Bouchaud2009}, where different types of impacts, as well as some theoretical models are discussed.

\FloatBarrier
\section{Correlations of assets}
\label{section:Correlations}

The word ``correlation'' is defined as ``a relation existing between phenomena or things or between mathematical or statistical variables which tend to vary, be associated, or occur together in a way not expected on the basis of chance alone''\footnote{In Merriam-Webster Online Dictionary. Retrieved June 14, 2010, from http://www.merriam-webster.com/dictionary/correlations}.
When we talk about correlations in stock prices, what we are really interested in are relations between variables such as stock prices, order signs, transaction volumes, etc. and more importantly how these relations affect the nature of the statistical distributions and laws which govern the price time series.
This section deals with several topics concerning linear correlation observed in financial data. The first part deals with the important issue of computing correlations in high-frequency. As mentioned earlier, the computerization of financial exchanges has lead to the availability of huge amount of tick-by-tick data, and computing correlation using these intraday data raises lots of issues concerning usual estimators. The second and third parts deals with the use of correlation in order to cluster assets with potential applications in risk management problems.

\subsection{Estimating covariance on high-frequency data}
\label{subsection:CovarianceHighFrequencyData}

Let us assume that we observe $d$ time series of prices or log-prices $p_i, {i=1,\ldots,d}$, observed at times $t_m, {m=0,\ldots,M}$. The usual estimator of the covariance of prices $i$ and $j$ is the \emph{realized covariance estimator}, which is computed as:
\begin{equation}
	\hat\Sigma^{RV}_{ij}(t)=\sum_{m=1}^M (p_{i}(t_m)-p_{i}(t_{m-1}))(p_{j}(t_m)-p_{j}(t_{m-1})).
\end{equation}

The problem is that high-frequency tick-by-tick data record changes of prices when they happen, i.e. at random times. Tick-by-tick data is thus asynchronous, contrary to daily close prices for example, that are recorded at the same time for all the assets on a given exchange. Using standard estimators without caution, could be one cause for the ``Epps effect'', first observed in \cite{Epps1979}, which stated that ``[c]orrelations among price changes in common stocks of companies in one industry are found to decrease with the length of the interval for which the price changes are measured.'' This has largely been verified since, e.g. in \cite{Bonanno2001} or \cite{Reno2003}. \cite{HayashiYoshida2005} shows that non-synchronicity of tick-by-tick data and necessary sampling of time series in order to compute the usual realized covariance estimator partially explain this phenomenon. We very briefly review here two covariance estimators that do not need any synchronicity (hence, sampling) in order to be computed. 

\subsubsection{The Fourier estimator} The Fourier estimator has been introduced by \cite{Malliavin2002}. Let us assume that we have $d$ time series of log-prices that are observations of Brownian semi-martingales $p_i$:
\begin{equation}
	dp_i=\sum_{j=1}^K \sigma_{ij} dW_j + \mu_i dt, i=1,\ldots,d.
\end{equation}
The coefficient of the covariance matrix are then written $\Sigma_{ij}(t)=\sum_{k=1}^K \sigma_{ik}(t)\sigma_{jk}(t)$. \cite{Malliavin2002} show that the Fourier coefficient of $\Sigma_{ij}(t)$ are, with $n_0$ a given integer:
\begin{eqnarray}
	a_k(\Sigma_{ij}) & = & \lim_{N\rightarrow\infty} \frac{\pi}{N+1-n_0} \sum_{s=n_0}^N \frac{1}{2} \left[a_s(dp_i) a_{s+k}(dp_j)\right. \nonumber
	\\ & & \left. + b_{s+k}(dp_i) b_{s}(dp_j) \right] ,
\end{eqnarray}
\begin{eqnarray}
	b_k(\Sigma_{ij}) & = & \lim_{N\rightarrow\infty} \frac{\pi}{N+1-n_0} \sum_{s=n_0}^N \frac{1}{2} \left[a_s(dp_i) b_{s+k}(dp_j)\right. \nonumber
	\\ & & \left. - b_{s}(dp_i) a_{s+k}(dp_j) \right] ,
\end{eqnarray}
where the Fourier coefficients $a_k(dp_i)$ and $b_k(dp_i)$ of $dp_i$ can be directly computed on the time series. Indeed, rescaling the time window on $[0,2\pi]$ and using integration by parts, we have:
\begin{equation}
	a_k(dp_i) = \frac{p(2\pi)-p(0)}{\pi} - \frac{k}{\pi}\int_0^{2\pi} \sin(kt)p_i(t) dt.
\end{equation}
This last integral can be discretized and approximately computed using the times $t^i_m$ of observations of the process $p_i$. Therefore, fixing a sufficiently large $N$, one can compute an estimator $\Sigma^F_{ij}$ of the covariance of the processes $i$ and $j$. See \cite{Reno2003} or \cite{IoriPrecup2007}, for examples of empirical studies using this estimator.

\subsubsection{The Hayashi-Yoshida estimator} \cite{HayashiYoshida2005} have proposed a simple estimator in order to compute covariance/correlation without any need for synchronicity of time series. As in the Fourier estimator, it is assumed that the observed process is a Brownian semi-martingale. The time window of observation is easily partitioned into $d$ family of intervals $\Pi^i=(U^i_m), i=1,\ldots,d$, where $t^i_m=\inf\{U^i_{m+1}\}$ is the time of the $m$-th observation  of the process $i$. Let us denote $\Delta p_i(U^i_m)= p_i(t^i_m)-p_i(t^i_{m-1})$. The \emph{cumulative covariance estimator} as the authors named it, or the \emph{Hayashi-Yoshida estimator} as it has been largely refered to, is then built as follows:
\begin{equation}
	\displaystyle \hat\Sigma^{HY}_{ij}(t)=\sum_{m,n} \Delta p_i(U^i_m) \Delta p_j(U^j_n) \mbox{\boldmath$1$}_{\{U^i_m \cap U^j_n \neq \emptyset\}}.
\end{equation}

There is a large literature in Econometrics that tackles the new challenges posed by high-frequency data. We refer the reader, wishing to go beyond this brief presentation, to the econometrics reviews by \cite{BarndorffNielsen2007} or \cite{McAleer2008}, for example.

\subsection{Correlation matrix and Random Matrix Theory}

The stock market data being essentially a \emph{multivariate} time series data, we construct correlation matrix to study its spectra and contrast it with the random multivariate data from coupled map lattice. 
It is known from previous studies that the empirical spectra of correlation matrices drawn from time series data, for most part, follow random matrix theory (RMT, see e.g.~\cite{Gopikrishnan2001a}).

%%%%%%%%%%%%%%
\subsubsection{Correlation matrix and Eigenvalue density}
%The correlations and anti-correlations in the prices of stocks belonging
%to a given portfolio play an important role in the selection of the
%efficient portfolio, and in many other studies.

\paragraph{Correlation matrix}
%\subparagraph{Financial Correlation matrix}
If there are $N$ assets with price $P_{i}(t)$ for asset $i$ at time $t$, then the logarithmic return of stock $i$ is $r_{i}(t)=\ln P_{i}(t)-\ln P_{i}(t-1)$, which for a certain consecutive sequence of trading days forms the return vector $\boldmath{r}_{i}$. In order to characterize the synchronous time evolution of stocks, the equal time correlation coefficients between stocks $i$ and $j$ is defined as
\begin{equation}
\rho _{ij}
= \frac{\langle \boldmath r_{i}\boldmath r_{j}\rangle -\langle \boldmath r_{i}\rangle \langle \boldmath r_{j}\rangle }
{\sqrt{[\langle \boldmath r_{i}^{2}\rangle -\langle \boldmath r_{i}\rangle ^{2}][\langle \boldmath r_{j}^{2}\rangle -\langle \boldmath r_{j}\rangle ^{2}]}} \, ,
\label{corrcoeff}
\end{equation}
where $\left\langle ...\right\rangle $ indicates a time average over the trading days included in the return vectors. 
These correlation coefficients form an $N\times N$ matrix with $-1\leq \rho _{ij}\leq 1$.
If $\rho _{ij}=1$, the stock price changes are completely correlated; if $\rho _{ij}=0$, the stock price changes are uncorrelated, and if $\rho _{ij}=-1$, then the stock price changes are completely anti-correlated.

%The probability density function $P(\rho _{ij})$ is plotted in Fig.
%\ref{fig:cordis}.%
%\begin{figure}
%\begin{center}
%\resizebox{0.6\textwidth}{!}
%{\includegraphics{cordis} }
%\end{center}

%\caption{Plot of the probability density function of correlation coefficients
%of returns against time.}

%\label{fig:cordis}
%\end{figure}

%\subparagraph{Correlation matrix from spatio-temporal series from coupled map lattices}
\paragraph{Correlation matrix of spatio-temporal series from coupled map lattices}
Consider a time series of the form $z'(x,t)$, where $x=1,2,...n$ and $t=1,2....p$ denote the discrete space and time, respectively.
In this, the time series at every spatial point is treated as a different variable. 
We define the normalised variable as 
\begin{equation}
z(x,t) = \frac{z'(x,t)-\langle z'(x)\rangle}{\sigma(x)} \, ,
\end{equation}
where the brackets $\langle . \rangle$ represent temporal averages and $\sigma(x)$ the standard deviation of $z'$ at position $x$.
Then, the equal-time cross-correlation matrix that represents the spatial correlations can be written as
\begin{equation}
S_{x,x'} = \langle z(x,t) \, z(x',t) \rangle\, , 
~~~~
x, x' = 1, 2,\dots, n \, .
\end{equation}
The correlation matrix is symmetric by construction. 
In addition, a large class of processes are translation invariant and the correlation matrix can contain that additional symmetry too. 
We will use this property for our correlation models in the context of coupled map lattice.
In time series analysis, the averages $\langle . \rangle$ have to be replaced by estimates obtained from finite samples. 
As usual, we will use the maximum likelihood estimates, $\langle a(t) \rangle \approx
\frac{1}{p}\sum_{t=1}^p a(t)$. 
These estimates contain statistical uncertainties, which disappears for $p\to\infty$.
Ideally, one requires $p \gg n$ to have reasonably correct correlation estimates. See \cite{Chakraborti2007} for details of parameters.

\paragraph{Eigenvalue Density}

The interpretation of the spectra of empirical correlation matrices should be done carefully if one wants to be able to distinguish between system specific signatures and universal
features. 
The former express themselves in the smoothed level density, whereas the latter usually are represented by the fluctuations on top of this smooth curve. 
In time series analysis, the matrix elements are not only prone to uncertainty such as measurement noise on the time series data, but also statistical fluctuations due to finite sample effects. 
When characterizing time series data in terms of random matrix theory, one is not interested in these trivial sources of fluctuations which are present on every data set, but one would like to identify the significant features which would be shared, in principle, by an ``infinite'' amount of data without measurement noise.
The eigenfunctions of the correlation matrices constructed from such empirical time series carry the information contained in the original time series data in a ``graded'' manner and they also provide a compact representation for it. 
Thus, by applying an approach based on random matrix theory, one tries to identify non-random components of the correlation matrix spectra as deviations from random matrix theory predictions (\cite{Gopikrishnan2001a}).

\begin{figure}[ht]
%\centering
\includegraphics[width=1.8in,angle=-90]{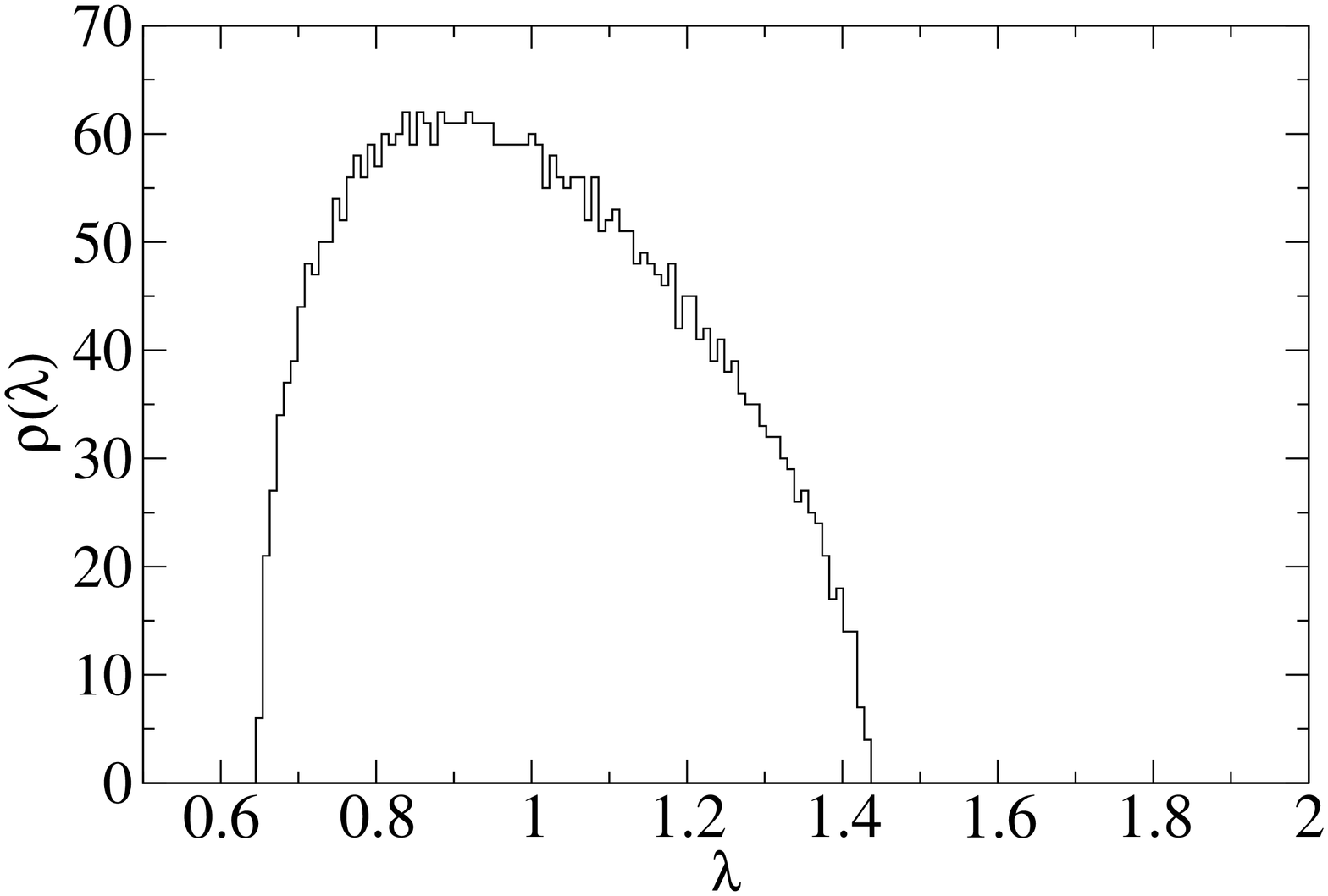}
\includegraphics[width=1.8in,angle=-90]{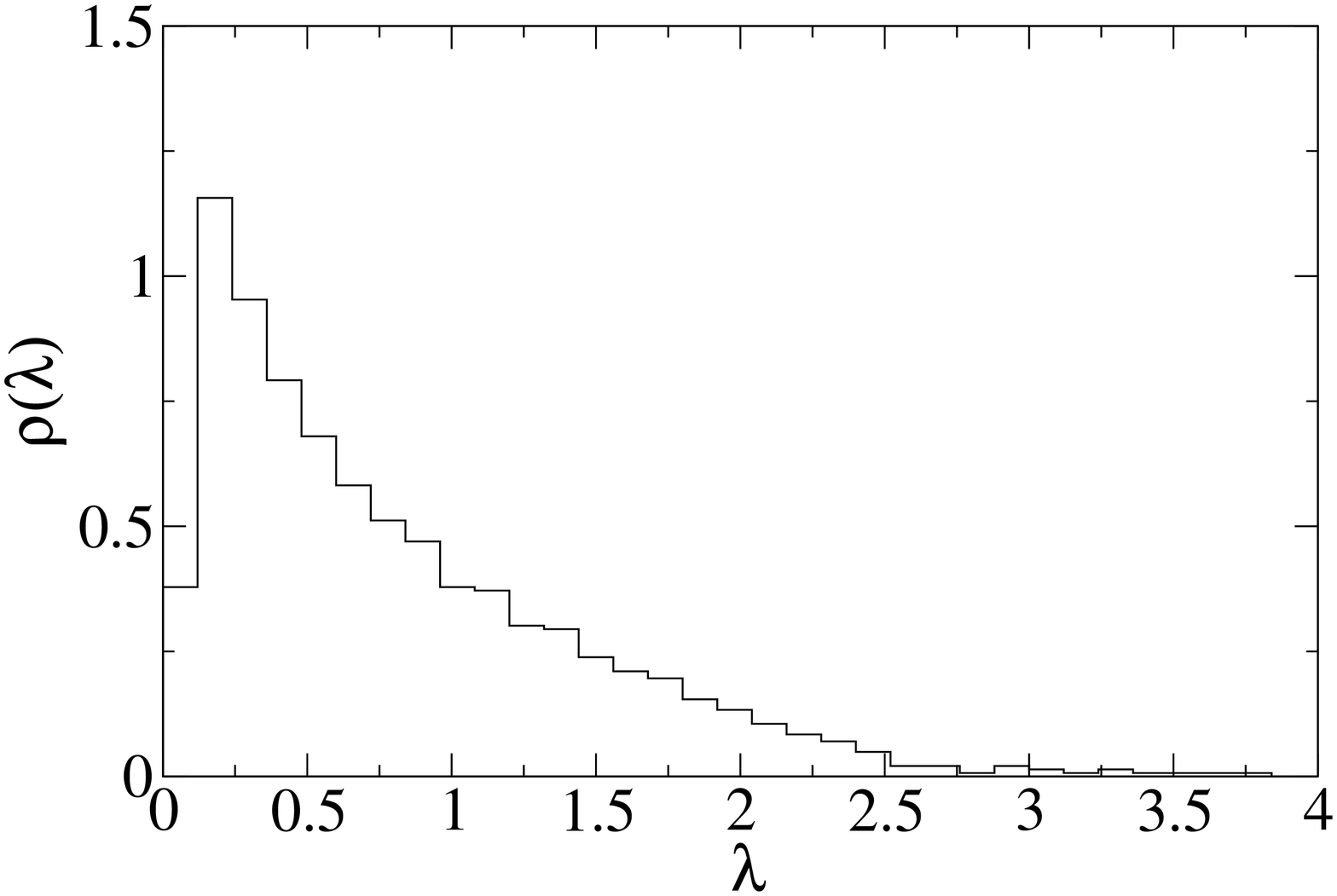}
\caption{ The upper panel shows spectral density for multivariate spatio-temporal time series drawn from coupled map lattices. The lower panel shows the eigenvalue density for the return time series of the S\&P500 stock market data (8938 time steps).
}
\label{fig3}
\end{figure}

We will look at the eigenvalue density that has been studied in the context of applying random matrix theory methods to time series correlations.
Let ${\mathcal N}(\lambda)$ be the integrated eigenvalue density which gives the number of eigenvalues less than a given value $\lambda$.
Then, the eigenvalue or level density is given by $\rho(\lambda) = \frac{d {\mathcal N}(\lambda)}{d\lambda}$.
This can be obtained assuming random correlation matrix and is found to be in good agreement with the empirical time series data from stock market fluctuations. 
From Random Matrix Theory considerations, the eigenvalue density for random correlations is given by
\begin{equation}
\rho_{rmt}(\lambda) 
= \frac{Q}{2 \pi \lambda} \sqrt{{(\lambda_{max}-\lambda})({\lambda-\lambda_{min}})} \, ,
\end{equation}
where $Q=N/T$ is the ratio of the number of variables to the length of each time series.
Here, $\lambda_{max}$ and $\lambda_{min}$, representing the maximum and minimum eigenvalues of the random correlation matrix respectively, are given by $\lambda_{max,min} = 1 + 1/Q \pm 2 \sqrt{1/Q}$.
However, due to presence of correlations in the empirical correlation matrix, this eigenvalue density is often violated for a certain number of dominant eigenvalues. 
They often correspond to system specific information in the data.
In Fig.~\ref{fig3} we show the eigenvalue density for S\&P500 data and also for the chaotic data from coupled map lattice. 
Clearly, both curves are qualitatively different. 
Thus, presence or absence of correlations in data is manifest in the spectrum of the corresponding correlation matrices.
%%%%%%%%%%%%%%%%%%%%%%%%%%%%%%%%%%%%%%
%%%%%%%%%%%%%%%%%%%%%%%%%%%%%%%%%%%%%%

\subsubsection{Earlier estimates and studies using Random Matrix Theory}
\begin{figure}
\begin{center}
\includegraphics[width=\columnwidth]{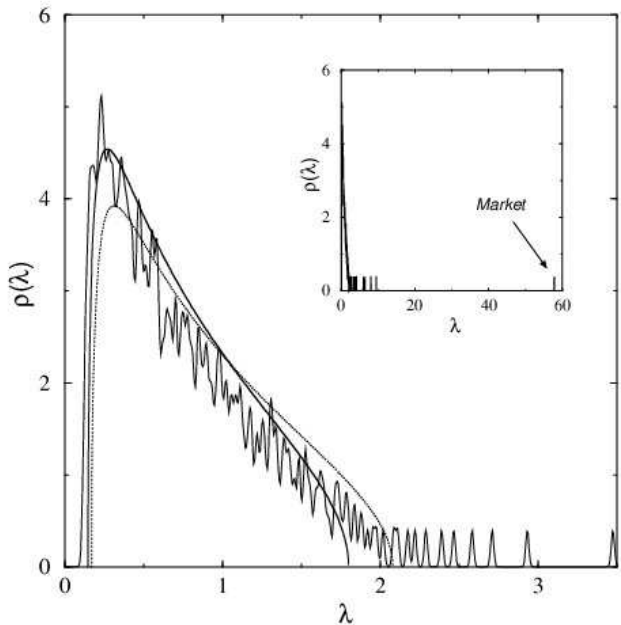}
\end{center}
\caption{Eigenvalue spectrum of the correlation matrices. Adapted from \cite{Laloux1999a}.}

\label{fig:eigensp}
\end{figure}

\cite{Laloux1999a} showed that results from the random matrix theory were useful to understand the statistical structure of the empirical correlation matrices appearing in the study of price fluctuations. 
The empirical determination of a correlation matrix is a difficult task. 
If one considers $N$ assets, the correlation matrix contains $N(N-1)/2$ mathematically independent elements, which must be determined from $N$ time series of length $T$. 
If $T$ is not very large compared to $N$, then generally the determination of the covariances is noisy, and therefore the empirical correlation matrix is to a large extent random. 
The smallest eigenvalues of the matrix are the most sensitive to this `noise'. 
But the eigenvectors corresponding to these smallest eigenvalues determine the minimum
risk portfolios in Markowitz theory. 
It is thus important to distinguish ``signal'' from ``noise'' or, in other words, to extract the eigenvectors and eigenvalues of the correlation matrix containing real information (those important for risk control), from those which do not contain any useful information and are unstable in time. 
It is useful to compare the properties of an empirical correlation matrix to a  ``null hypothesis''--- a random matrix which arises for example from a finite time series of strictly uncorrelated assets. 
Deviations from the random matrix case might then suggest the presence of true information. The main result of their study was the remarkable agreement between the theoretical prediction (based on the assumption that the correlation matrix is random) and empirical data concerning the density of eigenvalues (shown in Fig. \ref{fig:eigensp}) associated to the time series of the different stocks of the S\&P 500 (or other stock markets). 
%%%%%%%%%%%%%%%%%%%
%%%%%%%%%%%%%%%%%%%%
Cross-correlations in financial data were also studied by \cite{Plerou1999a,Plerou2002a}. They analysed cross-correlations between price fluctuations of different stocks using methods of RMT. 
Using two large databases, they calculated cross-correlation matrices of returns constructed from (i) 30-min returns of 1000 US stocks for the 2-yr period 1994--95, (ii) 30-min returns of 881 US stocks for the 2-yr period 1996--97, and (iii) 1-day returns of 422 US stocks for the 35-yr period 1962--96. 
They also tested the statistics of the eigenvalues $\lambda _{i}$ of cross-correlation matrices against a  ``null hypothesis''. 
They found that a majority of the eigenvalues of the cross-correlation matrices were within the RMT bounds $[\lambda _{min},\lambda _{max}]$, as defined above, for the eigenvalues of random correlation matrices. 
They also tested the eigenvalues of the cross-correlation matrices within the RMT bounds for universal properties of random matrices and found good agreement with the results for the Gaussian orthogonal ensemble (GOE) of random matrices --- implying a large degree of randomness in the measured cross-correlation coefficients. 
Furthermore, they found that the distribution of eigenvector components for the eigenvectors corresponding to the eigenvalues outside the RMT bounds displayed systematic deviations from the RMT prediction and that these  ``deviating eigenvectors'' were stable in time. 
They analysed the components of the deviating eigenvectors and found that the largest eigenvalue corresponded to an influence common to all stocks. 
Their analysis of the remaining deviating eigenvectors showed distinct groups, whose identities corresponded to conventionally-identified business sectors.

%%%%%%%%%%%%%%%%%%%%%%%%%%%%%%%%%%%%%%%%%%%%%%%%%%%%
%%%%%%%%%%%%%%%%%%%%%%%%%%%%%%%%%%%%%%%%%%%%%%%%%%%%
%%%%%%%%%%%%%%%%%%%%%%%%%%%%%%%%%%%%%%%%%%%%%%%%%%%%

\subsection{Analyses of correlations and economic taxonomy}

\subsubsection{Models and theoretical studies of financial correlations}

\cite{Podobnik2000a} studied how the presence of correlations in physical variables contributes to the form of probability distributions.
They investigated a process with correlations in the variance generated by a Gaussian or a truncated Levy distribution. 
For both Gaussian and truncated Levy distributions, they found that due to the correlations
in the variance, the process  ``dynamically'' generated power-law tails in the distributions, whose exponents could be controlled through the way the correlations in the variance were introduced. 
For a truncated Levy distribution, the process could extend a truncated distribution beyond the \emph{truncation cutoff}, leading to a crossover between a Levy stable power law and their  ``dynamically-generated'' power law. 
It was also shown that the process could explain the crossover behavior observed in the S\&P 500 stock index.

\cite{Noh2000a} proposed a model for correlations in stock markets in which the markets were composed of several groups, within which the stock price fluctuations were correlated. 
The spectral properties of empirical correlation matrices (\cite{Plerou1999a,Laloux1999a}) were studied in relation to this model and the connection between the spectral properties of the empirical correlation matrix and the structure of correlations in stock markets was established.

The correlation structure of extreme stock returns were studied by \cite{Cizeau2000a}. 
It has been commonly believed that the correlations between stock returns increased in high volatility periods. 
They investigated how much of these correlations could be explained within a simple non-Gaussian one-factor description with time independent correlations.
Using surrogate data with the true market return as the dominant factor, it was shown that most of these correlations, measured by a variety of different indicators, could be accounted for. 
In particular, their one-factor model could explain the level and asymmetry of empirical exceeding correlations. 
However, more subtle effects required an extension of the one factor model, where the variance and skewness of the residuals also depended on the market return.

\cite{Burda2001a} provided a statistical analysis of three S\&P 500 covariances with evidence for raw tail distributions. 
They studied the stability of these tails against reshuffling for the S\&P 500 data and showed that the covariance with the strongest tails was robust, with a spectral density in remarkable agreement with random Levy matrix theory. 
They also studied the inverse participation ratio for the three covariances. 
The strong localization observed at both ends of the spectral density was analogous to the localization exhibited in the random Levy matrix ensemble. 
They showed that the stocks with the largest scattering were the least susceptible to correlations and were the likely candidates for the localized states.

%An attempt to find scale-free network in financial correlations were made by Hyun-Joo Kim et al.~\cite{Kim2001a}. 
%They studied the cross-correlations in stock price changes between the S\&P 500 companies
%by introducing a weighted random graph, where all vertices (companies) were fully connected, and each edge was weighted, given by the normalized covariance of the two modified returns connected which ranged from $-1$ to $1$. 
%The modified return meant the deviation of a return from its average over all companies. 
%They defined the influence-strength at each vertex as the sum of the weights on the edges incident upon that vertex and found that the influence-strength distribution in its absolute magnitude $|s|$ followed a power-law, $P(|s|)\sim |s|^{-\delta }$, with exponent $\delta \approx 1.8(1)$.

%%%%%%%%%%%%%%
\subsubsection{Analyses using graph theory and economic taxonomy}

\cite{Mantegna1999a} introduced a method for finding a hierarchical arrangement of stocks traded in financial market, through studying the clustering of companies by using correlations
of asset returns. 
With an appropriate metric -- based on the earlier explained correlation matrix coefficients $\rho _{ij}$'s between all pairs of stocks $i$ and $j$ of the portfolio, computed in Eq.~\ref{corrcoeff} by considering the synchronous time evolution of the difference of the logarithm of daily stock price -- a fully connected graph was defined in which the nodes are companies, or stocks, and the ``distances'' between them were obtained from the corresponding correlation coefficients. 
The minimum spanning tree (MST) was generated from the graph by selecting the most important
correlations and it was used to identify clusters of companies. 
The hierarchical tree of the sub-dominant ultrametric space associated with the graph provided information useful to investigate the number and nature of the common economic factors affecting the time evolution of logarithm of price of well defined groups of stocks.
Several other attempts have been made to obtain clustering from the huge correlation matrix.

\cite{Bonanno2001} studied the high-frequency cross-correlation existing between pairs of stocks traded in a financial market in a set of 100 stocks traded in US equity markets. 
A hierarchical organization of the investigated stocks was obtained by determining a metric distance between stocks and by investigating the properties of the sub-dominant ultrametric associated with it. A clear modification of the hierarchical organization of the set of stocks investigated was detected when the time horizon used to determine stock returns was changed. 
The hierarchical location of stocks of the energy sector was investigated as a function of the time horizon. The hierarchical structure explored by the minimum spanning tree also seemed to give information about the influential power of the companies.

\begin{figure}
\begin{center}
\includegraphics[width=\columnwidth]{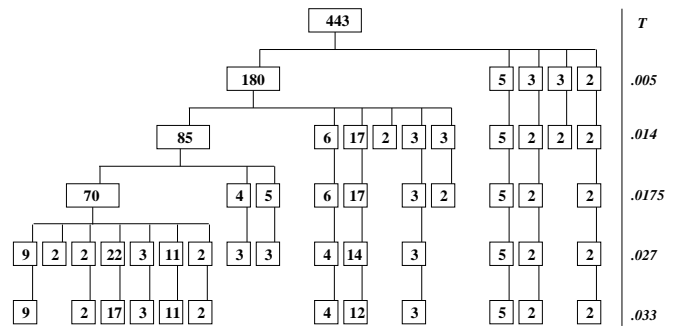}
\caption{The hierarchical structure of clusters of the S\&P 500 companies in the ferromagnetic case. In the boxes the number of elements of the cluster are indicated. The clusters consisting of single companies are not indicated. Adapted from \cite{Kullmann2000}.}
\label{fig:potts}
\end{center}
\end{figure}

It also turned out that the hierarchical structure of the financial market could be identified in accordance with the results obtained by an independent clustering method, based on Potts super-paramagnetic transitions as studied by \cite{Kullmann2000}, where the spins correspond to companies and the interactions are functions of the correlation coefficients determined from the time dependence of the companies' individual stock
prices. 
The method is a generalization of the clustering algorithm by \cite{Blatt1996} to the case of anti-ferromagnetic interactions corresponding to anti-correlations. 
For the Dow Jones Industrial Average, no anti-correlations were observed in the investigated time period and the previous results obtained by different tools were well reproduced. 
For the S\&P 500, where anti-correlations occur, repulsion between stocks modified the cluster structure of the $N=443$ companies studied, as shown in Fig.~\ref{fig:potts}. 
The efficiency of the method is represented by the fact that the figure matches well with the corresponding result obtained by the minimal spanning tree method, including the specific composition of the clusters. 
For example, at the lowest level of the hierarchy (highest temperature in the super-paramagnetic phase) the different industrial branches can be clearly identified: Oil, electricity, gold mining, etc. companies build separate clusters.
%%%%%%%%%%%%%%%%%%%%%%%%%%%%%%%%%%%%%%
%%%%%%%%%%%%%%%%%%%%%%%%%%%%%%%%%%%%%%
%
The network of influence was investigated by means of a time-dependent correlation method by \cite{Kullmann2000}. 
They studied the correlations as the function of the time shift between pairs of stock return 
time series of tick-by-tick data of the NYSE. 
They investigated whether any ``pulling effect'' between stocks existed or not, i.e. whether at any given time the return value of one stock influenced that of another stock at a different time or not. 
They found that, in general, two types of mechanisms generated significant correlation between any two given stocks. 
One was some kind of external effect (say, economic or political news) that influenced both stock prices simultaneously, and the change for both prices appeared at the same time, such that the maximum of the correlation was at zero time shift. 
The second effect was that, one of the companies had an influence on the other company indicating that one company's operation depended on the other, so that the price change of the influenced stock appeared latter because it required some time to react on the price change of the first stock displaying a ``pulling effect''.
A weak but significant effect with the real data set was found, showing that in many cases the maximum correlation was at non-zero time shift indicating directions of influence between the companies, and the characteristic time was of the order of a few minutes, which was compatible with efficient market hypothesis. 
In the pulling effect, they found that in general, more important companies (which were traded more) pulled the relatively smaller companies.

The time dependent properties of the minimum spanning tree (introduced by Mantegna), called a `dynamic asset tree', were studied by \cite{Onnela2003a}. 
The nodes of the tree were identified with stocks and the distance between them was a unique function of the corresponding element of the correlation matrix. 
By using the concept of a central vertex, chosen as the most strongly connected node of the tree, the mean occupation layer was defined, which was an important characteristic of the tree. 
During crashes the strong global correlation in the market manifested itself by a low value of the mean occupation layer. 
The tree seemed to have a scale free structure where the scaling exponent of the degree distribution was different for `business as usual' and `crash' periods. 
The basic structure of the tree topology was very robust with respect to time.
%The clustering of companies (into different industrial sectors as shown in 
%Fig. \ref{fig:asstree}) was also 
%studied in details. They also found that classic Markowitz portfolio with 
%minimum risk were 
%always located on the outer layers of the tree. 
Let us discuss in more details how the dynamic asset tree was applied to studies of economic taxonomy.
%
%\begin{figure}
%\begin{center}\resizebox{0.46\textwidth}{!}{\includegraphics{asstree} }\end{center}
%\caption{Snapshot of a dynamic asset tree connecting the examined 116 stocks
%of the S\&P 500 index. The tree was produced using four-year window
%width and it is centered on January 1, 1998. Business sectors are
%indicated according to Forbes. In this tree, General Electric (GE)
%was used as the central vertex and eight layers can be identified.
%Adapted from cond-mat/0302546.}
%\label{fig:asstree}
%\end{figure}

\paragraph{Financial Correlation matrix and constructing Asset Trees}

% Data used in the study
Two different sets of financial data were used.
The first set from the Standard \& Poor's 500 index (S\&P500) of the New York Stock Exchange (NYSE) from July 2, 1962 to December 31, 1997 contained 8939 daily closing values. The second set recorded the split-adjusted daily closure prices for a total of $N=477$ stocks traded at the New York Stock Exchange (NYSE) over the period of 20 years, from 02-Jan-1980 to 31-Dec-1999.
This amounted a total of 5056 prices per stock, indexed by time variable $\tau = 1, 2, \ldots, 5056$. 
For analysis and smoothing purposes, the data was divided time-wise into $M$ \emph{windows} $t=1,\, 2,...,\, M$ of width $T$, where $T$ corresponded to the number of daily returns included in the window. 
Note that several consecutive windows overlap with each other, the extent of which is dictated by the window step length parameter $\delta T$, which describes the displacement of the window and is also measured in trading days. 
The choice of window width is a trade-off between too noisy and too smoothed data for small and large window widths, respectively. 
The results presented here were calculated from monthly stepped four-year windows, i.e. $\delta T = 250/12 \approx 21$ days and $T=1000$ days. 
A large scale of different values for both parameters were explored, and the cited values were found optimal(\cite{Onnela2000a}). 
With these choices, the overall number of windows is $M=195$.

% Correlations
The earlier definition of correlation matrix, given by Eq. \ref{corrcoeff} is used.
%In order to investigate correlations between stocks we first denote 
%the closure price of stock $i$ at time $\tau$ by $P_{i}(\tau)$ 
%(Note that $\tau$ refers to a date, not a time window). We focus 
%our attention to the logarithmic return of stock $i$, given by 
%$r_{i}(\tau)=\ln P_{i}(\tau)-\ln P_{i}(\tau-1)$ which for a sequence 
%of consecutive trading days, i.e. those encompassing the given window 
%$t$, form the return vector $\boldmath r_{i}^t$. In order to 
%characterize the synchronous time evolution of assets, we use the equal 
%time correlation coefficients between assets $i$ and $j$ defined as
%\begin{equation}
%\rho _{ij}^t=\frac{\langle \boldmath r_{i}^t \boldmath r_{j}^t \rangle -\langle \boldmath r_{i}^t \rangle \langle \boldmath r_{j}^t \rangle }{\sqrt{[\langle {\boldmath r_{i}^t}^{2} \rangle -\langle \boldmath r_{i}^t\rangle ^{2}][\langle {\boldmath r_{j}^t}^{2} \rangle -\langle \boldmath r_{j}^t \rangle ^{2}]}},
%\end{equation}
%\noindent where $\left\langle ...\right\rangle $ indicates a time 
%average over the consecutive trading days included in the return 
%vectors. These correlation 
%coefficients fulfill the condition $-1\leq \rho _{ij}\leq 1$. 
%If $\rho _{ij}=1$, the stock price changes are completely correlated;
%if $\rho _{ij}=0$, the stock price changes are uncorrelated and if
%$\rho _{ij}=-1$, then the stock price changes are completely 
%anti-correlated \cite{jp}.  
These correlation coefficients form an $N\times N$ correlation matrix $\mathbf{C}^t$, which serves as the basis for trees discussed below.
% tree construction
An asset tree is then constructed according to the methodology by \cite{Mantegna1999a}. 
For the purpose of constructing asset trees, a distance is defined between a pair of stocks.
This distance is associated with the edge connecting the stocks and it is expected to reflect the level at which the stocks are correlated.
A simple non-linear transformation $d^t_{ij}=\sqrt{2(1-\rho _{ij}^t)}$ is used to obtain distances with the property $2\geq d_{ij}\geq 0$, forming an $N\times N$ symmetric distance matrix $\mathbf{D}^t$. 
So, if $d_{ij}=0$, the stock price changes are completely correlated; if $d_{ij}=2$, the stock price changes are completely anti-uncorrelated.
The trees for different time windows are not independent of each other, but form a series through time. 
Consequently, this multitude of trees is interpreted as a sequence of evolutionary steps of a single \emph{dynamic asset tree}.
An additional hypothesis is required about the topology of the metric space: the ultrametricity hypothesis. 
In practice, it leads to determining the minimum spanning tree (MST) of the distances, denoted $\mathbf{T}^t$. 
The spanning tree is a simply connected acyclic (no cycles) graph that connects all $N$ nodes (stocks) with $N-1$ edges such that the sum of all edge weights, $\sum _{d_{ij}^t \in \mathbf{T}^t}d_{ij}^t$, is minimum. 
We refer to the minimum spanning tree at time $t$ by the notation $\mathbf{T}^t=(V,E^t)$, where $V$ is a set of vertices and $E^t$ is a corresponding set of unordered pairs of vertices, or edges.
Since the spanning tree criterion requires all $N$ nodes to be always present, the set of vertices $V$ is time independent, which is why the time superscript has been dropped from notation. 
The set of edges $E^t$, however, does depend on time, as it is expected that edge lengths in the matrix  $\mathbf{D}^t$ evolve over time, and thus different edges get selected in the tree at different times.

% normalized lengths 
\paragraph{Market characterization}

\begin{figure}[ht]
\begin{center}
\includegraphics[width=\columnwidth]{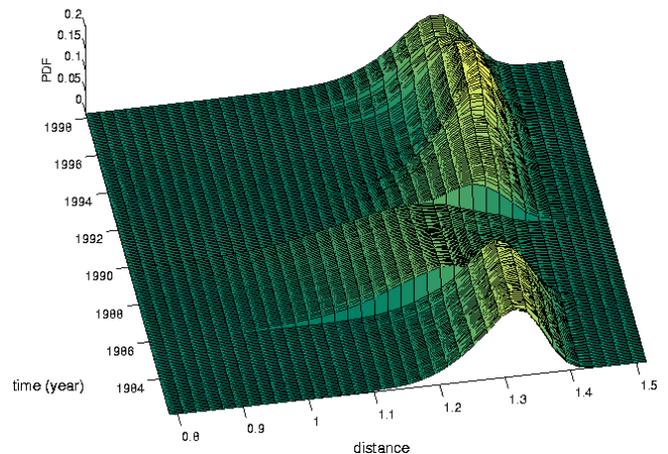}
\end{center}
\caption{Distribution of all $N(N-1)/2$ distance elements $d_{ij}$ contained in the distance matrix $\mathbf{D}^t$ as a function of time.}
\label{all_distances}
\end{figure}

\begin{figure}[hb]
\begin{center}
\includegraphics[width=\columnwidth]{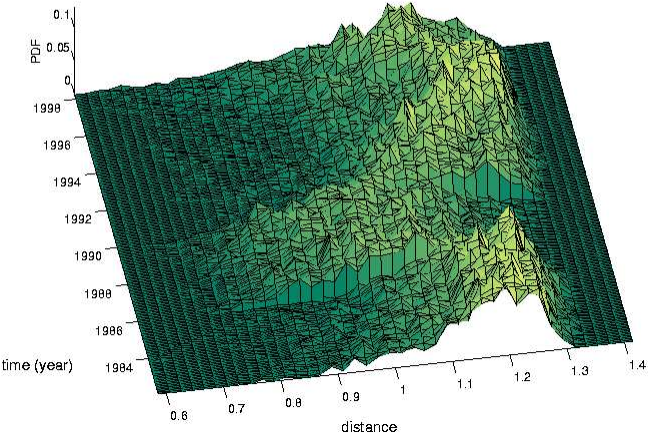}
\end{center}
\caption{Distribution of the $(N-1)$ distance elements $d_{ij}$ contained in the asset (minimum spanning) tree $\mathbf{T}^t$ as a function of time.}
\label{mst_distances}
\end{figure}

We plot the distribution of (i) distance elements $d^t_{ij}$ contained in the distance matrix $\mathbf{D}^t$ (Fig.~\ref{all_distances}), (ii) distance elements $d_{ij}$ contained in the asset (minimum spanning) tree $\mathbf{T}^t$ (Fig.~\ref{mst_distances}). 
In both plots, but most prominently in Fig.~\ref{all_distances}, there appears to be a discontinuity in the distribution between roughly 1986 and 1990. 
The part that has been cut out, pushed to the left and made flatter, is a manifestation of 
Black Monday (October 19, 1987), and its length along the time axis is related to the choice of window width $T$~\cite{Onnela2002a,Onnela2003a}. 
%
%\begin{figure}
%\begin{center}\resizebox{0.46\textwidth}{!}{\includegraphics{asstree} }\end{center}
%\caption{Snapshot of a dynamic asset tree connecting the examined 116 stocks
%of the S\&P 500 index. The tree was produced using four-year window
%width and it is centered on January 1, 1998. Business sectors are
%indicated according to Forbes. In this tree, General Electric (GE)
%was used as the central vertex and eight layers can be identified.
%Adapted from cond-mat/0302546.}
%\label{fig:asstree}
%\end{figure}
Also, note that in the distribution of tree edges in Fig.~\ref{mst_distances} most edges included in the tree seem to come from the area to the right of the value 1.1 in Fig.~\ref{all_distances}, and the largest distance element is $d_{max}=1.3549$.

\subparagraph{Tree occupation and central vertex}

% mean occupation layer
Let us focus on characterizing the spread of nodes on the tree, by introducing 
the quantity of \emph{mean occupation layer}  
\begin{equation}
 l(t,v_c) 
 = \frac{1}{N}\sum _{i=1}^{N}\mathop {\mathrm{lev}}(v_{i}^{t}) \, ,
\end{equation}
where $\mathop {\mathrm{lev}}(v_{i})$ denotes the level of vertex $v_{i}$. 
The levels, not to be confused with the distances $d_{ij}$ between nodes, are measured in natural numbers in relation to the \emph{central vertex} $v_c$, whose level is taken to be zero. 
Here the mean occupation layer indicates the layer on which the mass of the tree, on average, is conceived to be located.  
The central vertex is considered to be the parent of all other nodes in the tree, and is 
also known as the root of the tree. 
It is used as the {\it reference} point in the tree, against which the locations of all other nodes are relative. 
Thus all other nodes in the tree are children of the central vertex. 
Although there is an {\it arbitrariness} in the choice of the central vertex, it is proposed that the vertex is central, in the sense that any change in its price strongly affects the course of events in the market on the whole. 
Three alternative definitions for the central vertex were proposed in the studies, all yielding similar and, in most cases, identical outcomes. 
The idea is to find the node that is most strongly connected to its nearest neighbors. 
For example, according to one definition, the central node is the one with the highest \emph{vertex degree}, i.e. the number of edges which are incident with (neighbor of) the vertex.
Also, one may have either (i) static (fixed at all times) or (ii) dynamic (updated at each time step) central vertex, but again the results do not seem to vary significantly. 
The study of the variation of the topological properties and nature of the trees, with time were done. 
%This type of visualization tool sometimes provide deeper insight of the dynamical system.

\subparagraph{Economic taxonomy}

Mantegna's idea of linking stocks in an ultrametric space was motivated \emph{a posteriori} by the property of such a space to provide a meaningful economic taxonomy (\cite{Onnela2002b}).
%%%%%%%%%%%%%%%%%%%
\begin{figure}
\begin{center}
\includegraphics[width=\columnwidth]{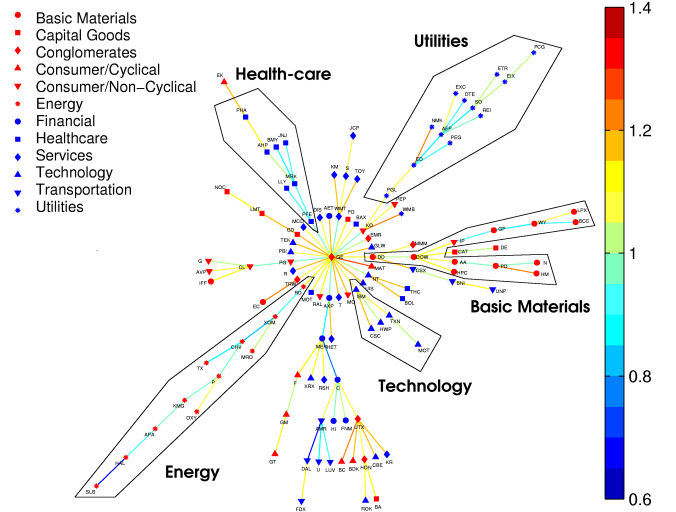}
\end{center}
\caption{Snapshot of a dynamic asset tree connecting the examined 116 stocks of the S\&P 500 index. The tree was produced using four-year window width and it is centered on January 1, 1998. 
Business sectors are indicated according to Forbes (www.forbes.com). 
In this tree, General Electric (GE) was used as the central vertex and eight layers can be identified.
}
\label{samplegraph}
\end{figure}
%%%%%%%%%%%%%%%%%%%%
Mantegna examined the meaningfulness of the taxonomy, by comparing the grouping of stocks in the tree with a third party reference grouping of stocks e.g. by their industry classifications (\cite{Mantegna1999a}). 
In this case, the reference was provided by Forbes (www.forbes.com), which uses its own classification system, assigning each stock with a sector (higher level) and industry (lower level) category. 
In order to visualize the grouping of stocks, a sample asset tree is constructed for a smaller dataset (shown in Fig.~\ref{samplegraph}), which consists of 116 S\&P 500 stocks, extending from the beginning of 1982 to the end of 2000, resulting in a total of 4787 price 
quotes per stock (\cite{Onnela2003a}). 
The window width was set at $T=1000$, and the shown sample tree is located time-wise at $t=t^*$, corresponding to 1.1.1998. 
The stocks in this dataset fall into 12 \emph{sectors}, which are Basic Materials, Capital Goods, Conglomerates, Consumer/Cyclical, Consumer/Non-Cyclical, Energy, Financial, Healthcare, Services, Technology, Transportation and Utilities. 
The sectors are indicated in the tree (see Fig.~\ref{samplegraph}) with different markers, 
while the industry classifications are omitted for reasons of clarity.
The term sector is used exclusively to refer to the given third party classification system of stocks. 
The term \emph{branch} refers to a subset of the tree, to all the nodes that share the specified common parent. 
In addition to the parent, it is needed to have a reference point to indicate the generational direction (i.e. who is who's parent) in order for a branch to be well defined. 
Without this reference there is absolutely no way to determine where one branch ends and the other begins. 
In this case, the reference is the central node. 
There are some branches in the tree, in which most of the stocks belong to just one sector, indicating that the branch is fairly homogeneous with respect to business sectors. 
This finding is in accordance with those of \cite{Mantegna1999a}
, although there are branches that are fairly heterogeneous, such as the one extending directly downwards from the central vertex (see Fig.~\ref{samplegraph}).

\section{Partial conclusion}
\label{part:Conclusion}
This first part of our review has shown statistical properties of financial data (time series of prices, order book structure, assets correlations). Some of these properties, such as fat tails of returns or volatility clustering, are widely known and acknowledged as ``financial stylized facts''. They are now largely cited in order to compare financial models, and reveal the lacks of many classical stochastic models of financial assets. Some other properties are newer findings that are obtained by studying high-frequency data of the whole order book structure. Volume of orders, interval time between orders, intraday seasonality, etc. are essential phenomenons to be understood when working in financial modelling.
The important role of studies of correlations has been emphasized. Beside the technical challenges raised by high-frequency, many studies based for example on random matrix theory or clustering algorithms help getting a better grasp on some Economics problems.
It is our belief that future modelling in finance will have to be partly based on Econophysics work on agent-based models in order to incorporate these ``stylized facts'' in a comprehensive way. Agent-based reasoning for order book models, wealth exchange models and game theoretic models will be reviewed in the following part of the review, to appear in a following companion paper.

%\section*{Acknowledgements}
%The authors would like to thank their collaborators and two anonymous reviewers whose comments greatly helped improving the review. AC is grateful to B.K. Chakrabarti, K. Kaski, J. Kertesz, T. Lux, M. Marsili, D. Stauffer and V. Yakovenko for invaluable suggestions and criticisms.

%%%%%%%%%%%%%%%%%%%%%%%%%%%%%%%%%%%%%%%%%%%%%%%%%%%%%%%%%%%%%
%%%%%%%%%%%%%%%%%%%%%%%%%%%%%%%%%%%%%%%%%%%%%%%%%%%%%%%%%%%%%

\part{}

\section{Introduction}
In the first part of the review, empirical developments in Econophysics have been studied. We have pointed out that some of these widely known ``stylized facts'' are already at the heart of financial models. But many facts, especially the newer statistical properties of order books, are not yet taken into account. As advocated by many during the financial crisis in 2007-2008 (see e.g. \cite{Bouchaud2008,Lux2009,Farmer2009}), agent-based models should have a great role to play in future financial modelling. In economic models, there is usually the representative agent, who is ``perfectly rational'' and uses the ``utility maximization'' principle while taking actions. Instead the multi-agent models that have originated from statistical physics considerations have allowed to go beyond the prototype theories with the ``representative'' agent in traditional economics. In this second part of our review, we present recent developments of agent-based models in Econophysics.

There are, of course, many reviews and books already published in this areas (see e.g. \cite{Bouchaud2009}, \cite{Lux2009}, \cite{Samanidou2007}, \cite{YakovenkoRosser2009}, \cite{Chatterjee2007b}, \cite{ChalletBook2004}, \cite{Coolen2005}, etc.). We will present here our perspectives in three representative areas.

\section{Agent-based modelling of order books}
\label{section:AgentBasedModelingOrderBooks}
\subsection{Introduction}

Although known, at least partly, for a long time -- \cite{Mandelbrot1963} gives a reference for a paper dealing with non-normality of price time series in 1915, followed by several others in the 1920's -- ``stylized facts'' have often been left aside when modelling financial markets. They were even often referred to as ``anomalous'' characteristics, as if observations failed to comply with theory.
Much has been done these past fifteen years in order to address this challenge and provide new models that can reproduce these facts. These recent developments have been built on top of early attempts at modelling mechanisms of financial markets with agents. For example, \cite{Stigler1964}, investigating some rules of the SEC\footnote{Security Exchange Commission}, or \cite{Garman1976}, investigating double-auction microstructure, belong to those historical works.
It seems that the first modern attempts at that type of models were made in the field of behavioural finance. This field aims at improving financial modelling based on the psychology and sociology of the investors. Models are built with agents who can exchange shares of stocks according to exogenously defined utility functions reflecting their preferences and risk aversions. \cite{LeBaron2006-1} shows that this type of modelling offers good flexibility for reproducing some of the stylized facts and \cite{LeBaron2006} provides a review of that type of model. However, although achieving some of their goals, these models suffer from many drawbacks: first, they are very complex, and it may be a very difficult task to identify the role of their numerous parameters and the types of dependence to these parameters; second, the chosen utility functions do not necessarily reflect what is observed on the mechanisms of a financial market. 

A sensible change in modelling appears with much simpler models implementing only well-identified and presumably realistic ``behaviour'': \cite{ContBouchaud2000} uses noise traders that are subject to ``herding'', i.e. form random clusters of traders sharing the same view on the market. The idea is used in \cite{Raberto2001} as well. A complementary approach is to characterize traders as fundamentalists, chartists or noise traders. \cite{LuxMarchesi2000} propose an agent-based model in which these types of traders interact. In all these models, the price variation directly results from the excess demand: at each time step, all agents submit orders and the resulting price is computed. Therefore, everything is cleared at each time step and there is no structure of order book to keep track of orders.

One big step is made with models really taking into account limit orders and keeping them in an order book once submitted and not executed. \cite{ChiarellaIori2002} build an agent-based model where all traders submit orders depending on the three elements identified in \cite{LuxMarchesi2000}: chartists, fundamentalists, noise. Orders submitted are then stored in a persistent order book. In fact, one of the first simple models with this feature was proposed in \cite{BPS1997}. In this model, orders are particles moving along a price line, and each collision is a transaction. Due to numerous caveats in this model, the authors propose in the same paper an extension with fundamentalist and noise traders in the spirit of the models previously evoked. \cite{Maslov2000} goes further in the modelling of trading mechanisms by taking into account fixed limit orders and market orders that trigger transactions, and really simulating the order book. This model was analytically solved using a mean-field approximation by \cite{Slanina2001}.

Following this trend of modelling, the more or less ``rational'' agents composing models in economics tends to vanish and be replaced by the notion of flows: orders are not submitted any more by an agent following a strategic behaviour, but are viewed as an arriving flow whose properties are to be determined by empirical observations of market mechanisms. Thus, the modelling of order books calls for more ``stylized facts'', i.e. empirical properties that could be observed on a large number of order-driven markets. \cite{Biais1995} is a thorough empirical study of the order flows in the Paris Bourse a few years after its complete computerization. Market orders, limit orders, time of arrivals and placement are studied. \cite{BouchaudPotters2002} and \cite{BouchaudPotters2003} provide statistical features on the order book itself. These empirical studies, that have been reviewed in the first part of this review, are the foundation for ``zero-intelligence'' models, in which ``stylized facts'' are expected to be reproduced by the properties of the order flows and the structure of order book itself, without considering exogenous ``rationality''. \cite{ChalletStinchcombe2001} propose a simple model of order flows: limit orders are deposited in the order book and can be removed if not executed, in a simple deposition-evaporation process. \cite{BouchaudPotters2002} use this type of model with empirical distribution as inputs. As of today, the most complete empirical model is to our knowledge \cite{MikeFarmer2008}, where order placement and cancellation models are proposed and fitted on empirical data. Finally, new challenges arise as scientists try to identify simple mechanisms that allow an agent-based model to reproduce non-trivial behaviours: herding behaviour in\cite{ContBouchaud2000}, dynamic price placement in \cite{Preis2007}, threshold behaviour in \cite{Cont2007}, etc.

In this part we review some of these models. This survey is of course far from exhaustive, and we have just selected models that we feel are representative of a specific trend of modelling.

\subsection{Early order-driven market modelling: Market microstructure and policy issues}

The pioneering works in simulation of financial markets were aimed to study market regulations. The very first one, \cite{Stigler1964}, tries to investigate the effect of regulations of the SEC on American stock markets, using empirical data from the 20's and the 50's. Twenty years later, at the start of the computerization of financial markets, \cite{Hakansson1985} implements a simulator in order to test the feasibility of automated market making. Instead of reviewing the huge microstructure literature, we refer the reader to the well-known books by \cite{Ohara1995} or \cite{Hasbrouck2007}, for example, for a panorama of this branch of finance. However, by presenting a small selection of early models, we here underline the grounding of recent order book modelling.

\subsubsection{A pioneer order book model}
To our knowledge, the first attempt to simulate a financial market was by \cite{Stigler1964}. This paper was a biting and controversial reaction to the Report of the Special Study of the Securities Markets of the SEC (\cite{SEC1963}), whose aim was to ``study the adequacy of rules of the exchange and that the New York stock exchange undertakes to regulate its members in all of their activities'' (\cite{Cohen1963}). According to Stigler, this SEC report lacks rigorous tests when investigating the effects of regulation on financial markets. Stating that ``demand and supply are [...] erratic flows with sequences of bids and asks dependent upon the random circumstances of individual traders'', he proposes a simple simulation model to investigate the evolution of the market. In this model, constrained by simulation capability in 1964, price is constrained within $L=10$ ticks. (Limit) orders are randomly drawn, in trade time, as follows: they can be bid or ask orders with equal probability, and their price level is uniformly distributed on the price grid. Each time an order crosses the opposite best quote, it is a market order. All orders are of size one. Orders not executed $N=25$ time steps after their submission are cancelled. Thus, $N$ is the maximum number of orders available in the order book.

In the original paper, a run of a hundred trades was manually computed using tables of random numbers. Of course, no particular results concerning the ``stylized facts'' of financial time series was expected at that time. However, in his review of some order book models, \cite{Slanina2008} makes simulations of a similar model, with parameters $L=5000$ and $N=5000$, and shows that price returns are not Gaussian: their distribution exhibits power law with exponent $0.3$, far from empirical data. As expected, the limitation $L$ is responsible for a sharp cut-off of the tails of this distribution.

\subsubsection{Microstructure of the double auction}
\cite{Garman1976} provides an early study of the double auction market with a point of view that does not ignore temporal structure, and really defines order flows. Price is discrete and constrained to be within $\{p_1,p_L\}$. Buy and sell orders are assumed to be submitted according to two Poisson processes of intensities $\lambda$ and $\mu$. Each time an order crosses the best opposite quote, it is a market order. All quantities are assumed to be equal to one.
The aim of the author was to provide an empirical study of the market microstructure. The main result of its Poisson model was to support the idea that negative correlation of consecutive price changes is linked the microstructure of the double auction exchange.
This paper is very interesting because it can be seen as precursor that clearly sets the challenges of order book modelling. 
First, the mathematical formulation is promising. With its fixed constrained prices, \cite{Garman1976} can define the state of the order book at a given time as the vector $(n_i)_{i=1,\ldots,L}$ of awaiting orders (negative quantity for bid orders, positive for ask orders). Future analytical models will use similar vector formulations that can be cast it into known mathematical processes in order to extract analytical results -- see e.g. \cite{ContStoikov2009} reviewed below.
Second, the author points out that, although the Poisson model is simple, analytical solution is hard to work out, and he provides Monte Carlo simulation. The need for numerical and empirical developments is a constant in all following models.
Third, the structural question is clearly asked in the conclusion of the paper: ``Does the auction-market model imply the characteristic leptokurtosis seen in empirical security price changes?''. The computerization of markets that was about to take place when this research was published -- Toronto's CATS\footnote{Computer Assisted Trading System} opened a year later in 1977 -- motivated many following papers on the subject. As an example, let us cite here \cite{Hakansson1985}, who built a model to choose the right mechanism for setting clearing prices in a multi-securities market.

\subsubsection{Zero-intelligence}
In the models by \cite{Stigler1964} and \cite{Garman1976}, orders are submitted in a purely random way on the grid of possible prices. Traders do not observe the market here and do not act according to a given strategy. Thus, these two contributions clearly belong to a class of ``zero-intelligence'' models.
To our knowledge, \cite{GodeSunder1993} is the first paper to introduce the expression ``zero-intelligence'' in order to describe non-strategic behaviour of traders. It is applied to traders that submit random orders in a double auction market. The expression has since been widely used in agent-based modelling, sometimes in a slightly different meaning (see more recent models described in this review). 
In \cite{GodeSunder1993}, two types of zero-intelligence traders are studied. The first are unconstrained zero-intelligence traders. These agents can submit random order at random prices, within the allowed price range $\{1,\ldots,L\}$. The second are constrained zero-intelligence traders. These agents submit random orders as well, but with the constraint that they cannot cross their given reference price $p^R_i$: constrained zero-intelligence traders are not allowed to buy or sell at loss.
The aim of the authors was to show that double auction markets exhibit an intrinsic ``allocative efficiency'' (ratio between the total profit earned by the traders divided by the maximum possible profit) even with zero-intelligence traders. An interesting fact is that in this experiment, price series resulting from actions by zero-intelligence traders are much more volatile than the ones obtained with constrained traders. This fact will be confirmed in future models where ``fundamentalists'' traders, having a reference price, are expected to stabilize the market (see \cite{WyartBouchaud2007} or \cite{LuxMarchesi2000} below).
Note that the results have been criticized by \cite{Cliff1997}, who show that the observed convergence of the simulated price towards the theoretical equilibrium price may be an artefact of the model. More precisely, the choice of traders' demand carry a lot of constraints that alone explain the observed results.

Modern works in Econophysics owe a lot to these early models or contributions. Starting in the mid-90's, physicists have proposed simple order book models directly inspired from Physics, where the analogy ``order $\equiv$ particle'' is emphasized. Three main contributions are presented in the next section.

\subsection{Order-driven market modelling in Econophysics}
\label{subsection:EarlyOrderDrivenModels}

\subsubsection{The order book as a reaction-diffusion model}
A very simple model directly taken from Physics was presented in \cite{BPS1997}. The authors consider a market with $N$ noise traders able to exchange one share of stock at a time. Price $p(t)$ at time $t$ is constrained to be an integer (i.e. price is quoted in number of ticks) with an upper bound $\bar p$: $\forall t,\quad p(t)\in\{0,\ldots,\bar p\}$. Simulation is initiated at time $0$ with half of the agents asking for one share of stock (buy orders, bid) with price:
\begin{equation}
p_b^j(0)\in\{0,\bar p/2\},\qquad j=1,\ldots,N/2,
\end{equation}
and the other half offering one share of stock (sell orders, ask) with price:
\begin{equation}
p_s^j(0)\in\{\bar p/2, \bar p\},\qquad j=1,\ldots,N/2 .
\end{equation}

\begin{figure}[b]
\includegraphics[width=\columnwidth]{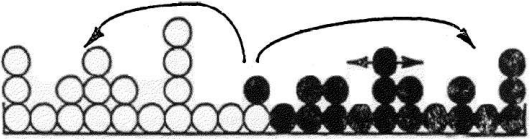}
\caption{\label{figure:BPSModel}Illustration of the Bak, Paczuski and Shubik model: white particles (buy orders, bid) moving from the left, black particles (sell orders, ask) moving from the right. Reproduced from \cite{BPS1997}.}
\end{figure}

At each time step $t$, agents revise their offer by exactly one tick, with equal probability to go up or down. Therefore, at time $t$, each seller (resp. buyer) agent chooses his new price as:
\begin{equation}
p_s^j(t+1) = p_s^j(t) \pm 1 \qquad\textrm{ (resp. } p_b^j(t+1) = p_b^j(t) \pm 1 \textrm{ )}. 
\end{equation}
A transaction occurs when there exists $(i,j)\in\{1,\ldots,N/2\}^2$ such that $p_b^i(t+1)=p_s^j(t+1)$. In such a case the orders are removed and the transaction price is recorded as the new price $p(t)$. Once a transaction has been recorded, two orders are placed at the extreme positions on the grid: $p_b^i(t+1)=0$ and $p_s^j(t+1)=\bar p$. As a consequence, the number of orders in the order book remains constant and equal to the number of agents. In figure \ref{figure:BPSModel}, an illustration of these moving particles is given.

\begin{figure}
\includegraphics[width=\columnwidth]{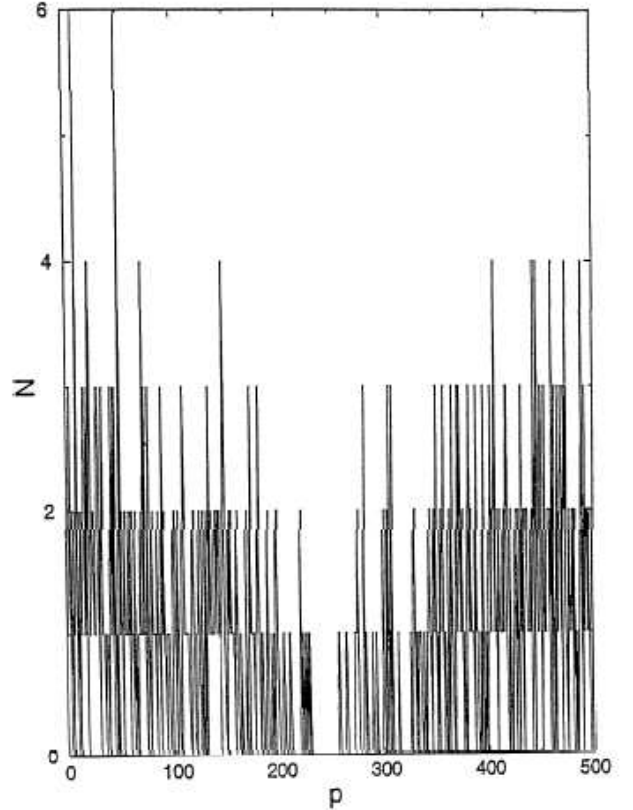}
\caption{\label{figure:BPSSnapshotLOB}Snapshot of the limit order book in the Bak, Paczuski and Shubik model. Reproduced from \cite{BPS1997}.}
\end{figure}

As pointed out by the authors, this process of simulation is similar the reaction-diffusion model $A+B\rightarrow\emptyset$ in Physics. In such a model, two types of particles are inserted at each side of a pipe of length $\bar p$ and move randomly with steps of size 1. Each time two particles collide, they're annihilated and two new particles are inserted. The analogy is summarized in table \ref{table:PhysicsBPSAnalogy}. 
\begin{table}[ht] 
\caption{Analogy between the $A+B\rightarrow\emptyset$ reaction model and the order book in \cite{BPS1997}.}
\centering
	\begin{tabular}{|c|c|} 
	\hline\hline                        
	Physics & \cite{BPS1997} 
	\\
	[0.5ex] 
	\hline\hline
	Particles & Orders
	\\
	Finite Pipe & Order book
	\\
	Collision & Transaction
	\\
	\hline
	\end{tabular} 
\label{table:PhysicsBPSAnalogy}
\end{table} 
Following this analogy, it thus can be showed that the variation $\Delta p(t)$ of the price $p(t)$ verifies :
\begin{equation}
	\Delta p(t) \sim t^{1/4}(\ln(\frac{t}{t_0}))^{1/2}.
\end{equation}
Thus, at long time scales, the series of price increments simulated in this model exhibit a Hurst exponent $H=1/4$. As for the stylized fact $H\approx 0.7$, this sub-diffusive behavior appears to be a step in the wrong direction compared to the random walk $H=1/2$. Moreover, \cite{Slanina2008} points out that no fat tails are observed in the distribution of the returns of the model, but rather fits the empirical distribution with an exponential decay. Other drawbacks of the model could be mentioned. For example, the reintroduction of orders at each end of the pipe leads to unrealistic shape of the order book, as shown on figure~\ref{figure:BPSSnapshotLOB}.
Actually here is the main drawback of the model: ``moving'' orders is highly unrealistic as for modelling an order book, and since it does not reproduce any known financial exchange mechanism, it cannot be the base for any larger model. Therefore, attempts by the authors to build several extensions of this simple framework, in order to reproduce ``stylized facts'' by adding fundamental traders, strategies, trends, etc. are not of interest for us in this review. However, we feel that the basic model as such is very interesting because of its simplicity and its ``particle'' representation of an order-driven market that has opened the way for more realistic models.

\subsubsection{Introducing market orders}
\cite{Maslov2000} keeps the zero-intelligence structure of the \cite{BPS1997} model but adds more realistic features in the order placement and evolution of the market. First, limit orders are submitted and stored in the model, without moving. Second, limit orders are submitted around the best quotes. Third, market orders are submitted to trigger transactions.
More precisely, at each time step, a trader is chosen to perform an action. This trader can either submit a limit order with probability $q_l$ or submit a market order with probability $1-q_l$. Once this choice is made, the order is a buy or sell order with equal probability. All orders have a one unit volume.

As usual, we denote $p(t)$ the current price. In case the submitted order at time step $t+1$ is a limit ask (resp. bid) order, it is placed in the book at price $p(t)+\Delta$ (resp. $p(t)-\Delta$), $\Delta$ being a random variable uniformly distributed in $]0;\Delta^M=4]$. In case the submitted order at time step $t+1$ is a market order, one order at the opposite best quote is removed and the price $p(t+1)$ is recorded.
In order to prevent the number of orders in the order book from large increase, two mechanisms are proposed by the author: either keeping a fixed maximum number of orders (by discarding new limit orders when this maximum is reached), or removing them after a fixed lifetime if they have not been executed.

\begin{figure}
\includegraphics[width=\columnwidth]{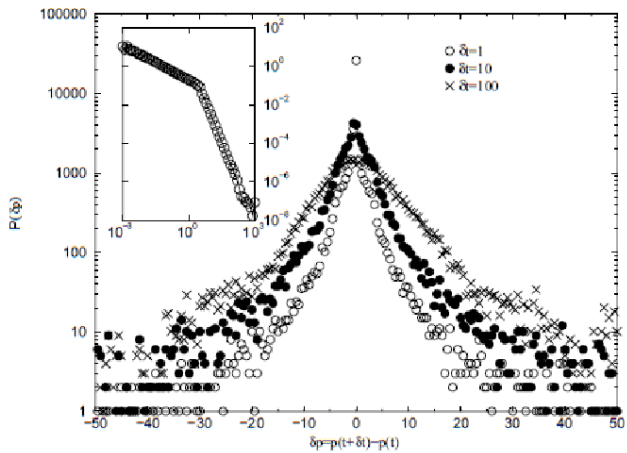}
\caption{\label{figure:MaslovPDFResults}Empirical probability density functions of the price increments in the Maslov model. In inset, log-log plot of the positive increments. Reproduced from \cite{Maslov2000}.}
\end{figure}

Numerical simulations show that this model exhibits non-Gaussian heavy-tailed distributions of returns.  On figure~\ref{figure:MaslovPDFResults}, the empirical probability density of the price increments for several time scales are plotted. 
For a time scale $\delta t=1$, the author fit the tails distribution with a power law with exponent $3.0$, i.e. reasonable compared to empirical value. However, the Hurst exponent of the price series is still $H=1/4$ with this model. It should also be noted that \cite{Slanina2001} proposed an analytical study of the model using a mean-field approximation (See below section \ref{subsection:AnalyticalAttempts}).

This model brings very interesting innovations in order book simulation: order book with (fixed) limit orders, market orders, necessity to cancel orders waiting too long in the order book. These features are of prime importance in any following order book model.

\subsubsection{The order book as a deposition-evaporation process}
\cite{ChalletStinchcombe2001} continue the work of \cite{BPS1997} and \cite{Maslov2000}, and develop the analogy between dynamics of an order book and an infinite one dimensional grid, where particles of two types (ask and bid) are subject to three types of events: \emph{deposition} (limit orders), \emph{annihilation} (market orders) and \emph{evaporation} (cancellation). Note that annihilation occurs when a particle is deposited on a site occupied by a particle of another type. The analogy is summarized in table \ref{table:PhysicsChalletAnalogy}.
\begin{table}[ht] 
\caption{Analogy between the deposition-evaporation process and the order book in \cite{ChalletStinchcombe2001}.}
\centering
	\begin{tabular}{|c|c|} 
	\hline\hline                        
	Physics & \cite{ChalletStinchcombe2001} 
	\\
	[0.5ex] 
	\hline\hline
	Particles & Orders
	\\
	Infinite lattice & Order book
	\\
	Deposition & Limit orders submission
	\\
	Evaporation & Limit orders cancellation
	\\
	Annihilation & Transaction
	\\
	\hline
	\end{tabular} 
\label{table:PhysicsChalletAnalogy}
\end{table}
Hence, the model goes as follows: At each time step, a bid (resp. ask) order is deposited with probability $\lambda$ at a price $n(t)$ drawn according to a Gaussian distribution centred on the best ask $a(t)$ (resp. best bid $b(t)$) and with variance depending linearly on the spread $s(t)=a(t)-b(t)$: $\sigma(t) = Ks(t)+C$. If $n(t)>a(t)$ (resp. $n(t)<b(t)$), then it is a market order: annihilation takes place and the price is recorded. Otherwise, it is a limit order and it is stored in the book. Finally, each limit order stored in the book has a probability $\delta$ to be cancelled (evaporation).

\begin{figure}
\includegraphics[width=\columnwidth]{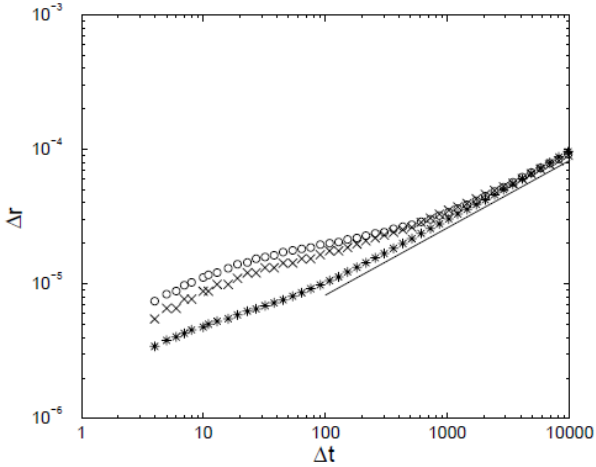}
\caption{\label{figure:ChalletHurstResults}Average return $\langle r_{\Delta t}\rangle$ as a function of $\Delta t$ for different sets of parameters and simultaneous depositions allowed in the Challet and Stinchcombe model. Reproduced from \cite{ChalletStinchcombe2001}.}
\end{figure}

Figure~\ref{figure:ChalletHurstResults} shows the average return as a function of the time scale. It appears that the series of price returns simulated with this model exhibit a Hurst exponent $H=1/4$ for short time scales, and that tends to $H=1/2$ for larger time scales. 
This behaviour might be the consequence of the random evaporation process (which was not modelled in \cite{Maslov2000}, where $H=1/4$ for large time scales). Although some modifications of the process (more than one order per time step) seem to shorten the sub-diffusive region, it is clear that no over-diffusive behaviour is observed. 

\subsection{Empirical zero-intelligence models}

The three models presented in the previous section \ref{subsection:EarlyOrderDrivenModels} have successively isolated essential mechanisms that are to be used when simulating a ``realistic'' market: one order is the smallest entity of the model; the submission of one order is the time dimension (i.e. event time is used, not an exogenous time defined by market clearing and ``tatonnement'' on exogenous supply and demand functions); submission of market orders (as such in \cite{Maslov2000}, as ``crossing limit orders'' in \cite{ChalletStinchcombe2001}) and cancellation of orders are taken into account.
On the one hand, one may try to describe these mechanisms using a small number of parameters, using Poisson process with constant rates for order flows, constant volumes, etc. This might lead to some analytically tractable models, as will be described in section \ref{subsection:AnalyticalAttempts}. On the other hand, one may try to fit more complex empirical distributions to market data without analytical concern. 

This type of modelling is best represented by \cite{MikeFarmer2008}. It is the first model that proposes an advanced calibration on the market data as for order placement and cancellation methods. As for volume and time of arrivals, assumptions of previous models still hold: all orders have the same volume, discrete event time is used for simulation, i.e. one order (limit or market) is submitted per time step.
Following \cite{ChalletStinchcombe2001}, there is no distinction between market and limit orders, i.e. market orders are limit orders that are submitted across the spread $s(t)$. More precisely, at each time step, one trading order is simulated: an ask (resp. bid) trading order is randomly placed at $n(t)=a(t)+\delta a$ (resp. $n(t)=b(t)+\delta b$) according to a Student distribution with scale and degrees of freedom calibrated on market data. If an ask (resp. bid) order satisfies $\delta a<-s(t)=b(t)-a(t)$ (resp. $\delta b>s(t)=a(t)-b(t)$), then it is a buy (resp. sell) market order and a transaction occurs at price $a(t)$ (resp. $b(t)$. 

During a time step, several cancellations of orders may occur. The authors propose an empirical distribution for cancellation based on three components for a given order:
\begin{itemize}
	\item the position in the order book, measured as the ratio $y(t) = {\Delta(t)\over\Delta(0)}$ where $\Delta(t)$ is the distance of the order from the opposite best quote at time $t$,
	\item the order book imbalance, measured by the indicator $N_{imb}(t) = {N_a(t)\over N_a(t)+N_b(t)}$ (resp. $N_{imb}(t) = {N_b(t)\over N_a(t)+N_b(t)}$) for ask (resp. bid) orders, where $N_a(t)$ and $N_b(t)$ are the number of orders at ask and bid in the book at time $t$,
	\item the total number $N(t) = N_a(t) + N_b(t)$ of orders in the book.
\end{itemize}

\begin{figure}
\includegraphics[width=\columnwidth]{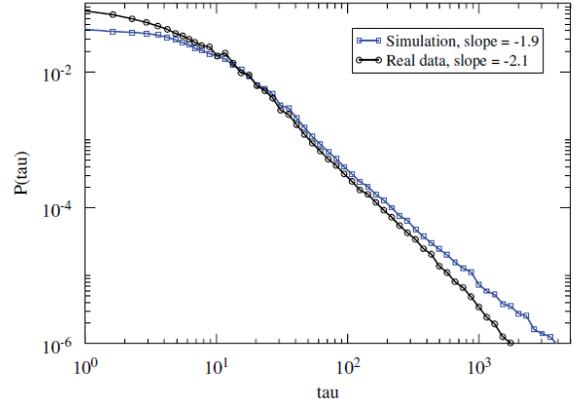}
\caption{\label{figure:MikeFarmerCancellation}Lifetime of orders for simulated data in the Mike and Farmer model, compared to the empirical data used for fitting. Reproduced from \cite{MikeFarmer2008}.}
\end{figure}

Their empirical study leads them to assume that the cancellation probability has an exponential dependance on $y(t)$, a linear one in $N_{imb}$ and finally decreases approximately as $1/N_t(t)$ as for the total number of orders. Thus, the probability $P(C|y(t), N_{imb}(t), N_t(t))$ to cancel an ask order at time $t$ is formally written :
\begin{equation}
	P(C|y(t), N_{imb}(t), N_t(t)) = A(1-e^{-y(t)}) (N_{imb}(t) + B) {1\over N_t(t)},
\end{equation}
where the constants $A$ and $B$ are to be fitted on market data. Figure~\ref{figure:MikeFarmerCancellation} shows that this empirical formula provides a quite good fit on market data.

Finally, the authors mimic the observed long memory of order signs by simulating a fractional Brownian motion. The auto-covariance function $\Gamma(t)$ of the increments of such a process exhibits a slow decay :
\begin{equation}
	\Gamma(k) \sim H(2H-1)t^{2H-2} 
\end{equation}
and it is therefore easy to reproduce exponent $\beta$ of the decay of the empirical autocorrelation function of order signs observed on the market with $H=1-\beta/2$.

\begin{figure}
\includegraphics[width=\columnwidth]{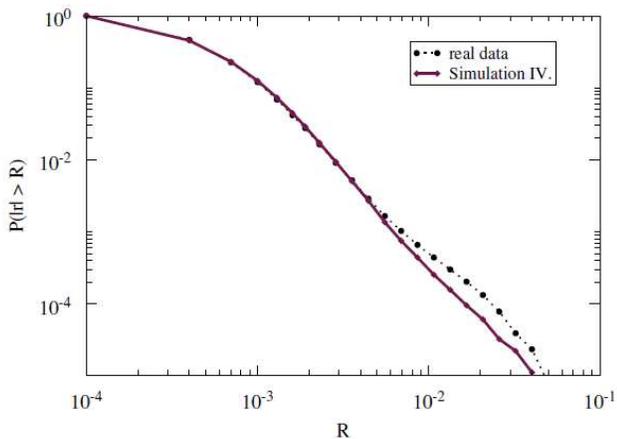}
\caption{\label{figure:MikeFarmerReturnResults}Cumulative distribution of returns in the Mike and Farmer model, compared to the empirical data used for fitting. Reproduced from \cite{MikeFarmer2008}.}
\end{figure}

The results of this empirical model are quite satisfying as for return and spread distribution. The distribution of returns exhibit fat tails which are in agreement with empirical data, as shown on figure~\ref{figure:MikeFarmerReturnResults}.
The spread distribution is also very well reproduced. As their empirical model has been built on the data of only one stock, the authors test their model on 24 other data sets of stocks on the same market and find for half of them a good agreement between empirical and simulated properties. However, the bad results of the other half suggest that such a model is still far from being ``universal''.

Despite these very nice results, some drawbacks have to be pointed out. The first one is the fact that the stability of the simulated order book is far from ensured. Simulations using empirical parameters in the simulations may bring situations where the order book is emptied by large consecutive market orders. Thus, the authors require that there is at least two orders in each side of the book. This exogenous trick might be important, since it is activated precisely in the case of rare events that influence the tails of the distributions.
Also, the original model does not focus on volatility clustering. \cite{GuZhou2009} propose a variant that tackles this feature. 
Another important drawback of the model is the way order signs are simulated. As noted by the authors, using an exogenous fractional Brownian motion leads to correlated price returns, which is in contradiction with empirical stylized facts. We also find that at long time scales it leads to a dramatic increase of volatility. As we have seen in the first part of the review, the correlation of trade signs can be at least partly seen as an artefact of execution strategies. 
Therefore this element is one  of the numerous that should be taken into account when ``programming'' the agents of the model. In order to do so, we have to leave the (quasi) ``zero-intelligence'' world and see how modelling based on heterogeneous agents might help to reproduce non-trivial behaviours. Prior to this development below in \ref{subsection:TowardsNonTrivialBehaviors}, we briefly review some analytical works on the ``zero-intelligence'' models.

\subsection{Analytical treatments of zero-intelligence models}
\label{subsection:AnalyticalAttempts}
In this section we present some analytical results obtained on zero-intelligence models where processes are kept sufficiently simple so that a mean-field approximation may be derived (\cite{Slanina2001}) or probabilities conditionaly to the state of the order book may be computed (\cite{ContStoikov2009}).
The key assumptions here are such that the process describing the order book is stationary. This allows either to write a stable density equation, or to fit the model into a nice mathematical framework such as ergodic Markov chains.

\subsubsection{Mean-field theory}

\cite{Slanina2001} proposes an analytical treatment of the model introduced by \cite{Maslov2000} and reviewed above. Let us briefly described the formalism used. The main hypothesis is the following: on each side of the current price level, the density of limit orders is uniform and constant (and $\rho_+$ on the ask side, $\rho_-$ on the bid side). In that sense, this is a ``mean-field'' approximation since the individual position of a limit order is not taken into account. Assuming we are in a stable state, the arrival of a market order of size $s$ on the ask (resp. bid) side will make the price change by $x_+ = s/\rho_+$ (resp. $x_- = s/\rho_-$). It is then observed that the transformations of the vector $X=\left(x_+,x_-\right)$ occurring at each event (new limit order, new buy market order, new sell market order) are linear transformation that can easily and explicitly be written. Therefore, an equation satisfied by the probability distribution  $P$ of the vector $X$ of price changes can be obtained. Finally, assuming further simplifications (such as $\rho_+=\rho_-$), one can solve this equation for a tail exponent and find that the distribution behaves as $P(x)\approx x^{-2}$ for large $x$.
This analytical result is slightly different from the one obtained by simulation in \cite{Maslov2000}. However, the numerous approximations make the comparison difficult. The main point here is that some sort of mean-field approximation is natural if we assume the existence of a stationary state of the order book, and thus may help handling order book models. 

\cite{SmithFarmer2003} also propose some sort of mean-field approximation for zero-intelligence models. In a similar model (but including a cancellation process), mean field theory and dimensional analysis produces interesting results. For example, it is easy to see that the book depth (i.e. number of orders) $N_e(p)$ at a price $p$ far away from the best quotes is given by $N_e(p)=\lambda/\delta$, where $\lambda$ is the rate of arrival of limit orders per unit of time and per unit of price, and $\delta$ the probability for an order to be cancelled per unit of time. Indeed, far from the best quotes no market orders occurs, so that if a steady-state exists, the number of limit orders par time step $\lambda$ must be balanced by the number of cancellation $\delta N_e(p)$ per unit of time, hence the result.

\subsubsection{Explicit computation of probabilities conditionally on the state of the order book}

\cite{ContStoikov2009} is an original attempt at analytical treatments of limit order books. In their model, the price is contrained to be on a grid $\{1,\ldots,N\}$. The state of the order book can then be described by a vector $X(t)=\left(X_1(t),\ldots,X_N(t)\right)$ where $|X_i(t)|$ is the quantity offered in the order book at price $i$. Conventionaly, $X_i(t), i=1,\ldots,N$ is positive on the ask side and negative on the bid side. As usual, limit orders arrive at level $i$ at a constant rate $\lambda_i$, and market orders arrive at a constant rate $\mu$. Finally, at level $i$, each order can be cancelled at a rate $\theta_i$.
Using this setting, \cite{ContStoikov2009} show that each event (limit order, market order, cancellation) transforms the vector $X$ in a simple linear way. Therefore, it is shown that under reasonable conditions, $X$ is an ergodic Markov chain, and thus admits a stationary state. The original idea is then to use this formalism to compute conditional probabilities on the processes. More precisely, it is shown that using Laplace transform, one may explicitly compute the probability of an increase of the mid price conditionally on the current state of the order book. 

This original contribution could allow explicit evaluation of strategies and open new perspectives in high-frequency trading. However, it is based on a simple model that does not reproduce empirical observations such as volatility clustering. Complex models trying to include market interactions will not fit into these analytical frameworks. We review some of these models in the next section.

\subsection{Towards non-trivial behaviours: modelling market interactions}
\label{subsection:TowardsNonTrivialBehaviors}

In all the models we have reviewed until now, flows of orders are treated as independent processes. Under some (strong) modelling constraints, we can see the order book as a Markov chain and look for analytical results (\cite{ContStoikov2009}). In any case, even if the process is empirically detailed and not trivial (\cite{MikeFarmer2008}), we work with the assumption that orders are independent and identically distributed. This very strong (and false) hypothesis is similar to the ``representative agent'' hypothesis in Economics: orders being successively and independently submitted, we may not expect anything but regular behaviours. Following the work of economists such as \cite{Kirman1992,Kirman1993,Kirman2002}, one has to translate the heterogeneous property of the markets into the agent-based models. Agents are not identical, and not independent.

In this section we present some toy models implementing mechanisms that aim at bringing heterogeneity: herding behaviour on markets in \cite{ContBouchaud2000}, trend following behaviour in \cite{LuxMarchesi2000} or in \cite{Preis2007}, threshold behaviour \cite{Cont2007}.
Most of the models reviewed in this section are not order book models, since a persistent order book is not kept during the simulations. They are rather price models, where the price changes are determined by the aggregation of excess supply and demand. However, they identify essential mechanisms that may clearly explain some empirical data. Incorporating these mechanisms in an order book model is not yet achieved but is certainly a future prospective.

\subsubsection{Herding behaviour}
The model presented in \cite{ContBouchaud2000} considers a market with $N$ agents trading a given stock with price $p(t)$. At each time step, agents choose to buy or sell one unit of stock, i.e. their demand is $\phi_i(t)= \pm 1, i=1,\ldots,N$ with probability $a$ or are idle with probability $1-2a$. The price change is assumed to be linearly linked with the excess demand $D(t)= \sum_{i=1}^N \phi_i(t)$ with a factor $\lambda$ measuring the liquidity of the market :
\begin{equation}
	p(t+1) = p(t) + {1\over\lambda} \sum_{i=1}^N \phi_i(t).
	\label{equation:ContBouchaudPrice}
\end{equation}
$\lambda$ can also be interpreted as a market depth, i.e. the excess demand needed to move the price by one unit. 
In order to evaluate the distribution of stock returns from Eq.(\ref{equation:ContBouchaudPrice}), we need to know the joint distribution of the individual demands $(\phi_{i}(t))_{1\leq i\leq N}$. As pointed out by the authors, if the distribution of the demand $\phi_i$ is independent and identically distributed with finite variance, then the Central Limit Theorem stands and the distribution of the price variation $\Delta p(t) = p(t+1)-p(t)$ will converge to a Gaussian distribution as $N$ goes to infinity.

The idea here is to model the diffusion of the information among traders by randomly linking their demand through clusters. At each time step, agents $i$ and $j$ can be linked with probability $p_{ij}=p={c\over N}$, $c$ being a parameter measuring the degree of clustering among agents. Therefore, an agent is linked to an average number of $(N-1)p$ other traders. Once clusters are determined, the demand are forced to be identical among all members of a given cluster. Denoting $n_c(t)$ the number of cluster at a given time step $t$, $W_k$ the size of the $k$-th cluster, $k=1,\ldots,n_c(t)$ and $\phi_k = \pm 1$ its investement decision, the price variation is then straightforwardly written :
\begin{equation}
	\Delta p(t)={1\over\lambda} \sum_{k=1}^{n_c(t)} W_k \phi_k.
\end{equation}

This modelling is a direct application to the field of finance of the random graph framework as studied in \cite{Erdos1960}. \cite{Kirman1983} previously suggested it in economics. Using these previous theoretical works, and assuming that the size of a cluster $W_k$ and the decision taken by its members $\phi_k(t)$ are independent, the author are able to show that the distribution of the price variation at time $t$ is the sum of $n_c(t)$ independent identically distributed random variables with heavy-tailed distributions :
\begin{equation}
	\Delta p(t)={1\over\lambda} \sum_{k=1}^{n_c(t)} X_k,
\end{equation}
where the density $f(x)$ of $X_k=W_k \phi_k$ is decaying as :
\begin{equation}
	\displaystyle f(x) \sim_{|x|\rightarrow\infty} {A\over |x|^{5/2}} e^{-(c-1)|x|\over W_0}.
\end{equation}
Thus, this simple toy model exhibits fat tails in the distribution of prices variations, with a decay reasonably close to empirical data. Therefore, \cite{ContBouchaud2000} show that taking into account a naive mechanism of communication between agents (herding behaviour) is able to drive the model out of the Gaussian convergence and produce non-trivial shapes of distributions of price returns.

\subsubsection{Fundamentalists and trend followers}
\cite{LuxMarchesi2000} proposed a model very much in line with agent-based models in behavioural finance, but where trading rules are kept simple enough so that they can be identified with a presumably realistic behaviour of agents. This model considers a market with $N$ agents that can be part of two distinct groups of traders: $n_f$ traders are ``fundamentalists'', who share an exogenous idea $p_f$ of the value of the current price $p$; and $n_c$ traders are ``chartists'' (or trend followers), who make assumptions on the price evolution based on the observed trend (mobile average). The total number of agents is constant, so that $n_f+n_c=N$ at any time.
At each time step, the price can be moved up or down with a fixed jump size of $\pm 0.01$ (a tick). The probability to go up or down is directly linked to the excess demand $ED$ through a coefficient $\beta$. The demand of each group of agents is determined as follows :
\begin{itemize}
	\item Each fundamentalist trades a volume $V_f$ proportional, with a coefficient $\gamma$, to the deviation of the current price $p$ from the perceived fundamental value $p_f$: $V_f = \gamma(p_f-p)$. 
	\item Each chartist trades a constant volume $V_c$. Denoting $n_+$ the number of optimistic (buyer) chartists and $n_-$ the number of pessimistic (seller) chartists, the excess demand by the whole group of chartists is written $(n_+-n_-)V_c$. 
\end{itemize}
Therefore, assuming that there exists some noise traders on the market with random demand $\mu$, the global excess demand is written :
\begin{equation}
	ED = (n_+-n_-)V_c + n_f \gamma(p_f-p) + \mu.
\end{equation}
The probability that the price goes up (resp. down) is then defined to be the positive (resp. negative) part of $\beta ED$.

As observed in \cite{WyartBouchaud2007}, fundamentalists are expected to stabilize the market, while chartists should destabilize it. In addition, following \cite{ContBouchaud2000}, the authors expect non-trivial features of the price series to results from herding behaviour and transitions between groups of traders. Referring to Kirman's work as well, a mimicking behaviour among chartists is thus proposed. The $n_c$ chartists can change their view on the market (optimistic, pessimistic), their decision being based on a clustering process modelled by an opinion index $x={n_+-n_-\over n_c}$ representing the weight of the majority. The probabilities $\pi_{+}$ and $\pi_{-}$ to switch from one group to another are formally written :
\begin{equation}
	\pi_{\pm} = v {n_c\over N}e^{\pm U}, \qquad U = \alpha_1 x + \alpha_2 p/v,
\end{equation}
where $v$ is a constant, and $\alpha_1$ and $\alpha_2$ reflect respectively the weight of the majority's opinion and the weight of the observed price in the chartists' decision.
Transitions between fundamentalists and chartists are also allowed, decided by comparison of expected returns (see \cite{LuxMarchesi2000} for details).

The authors show that the distribution of returns generated by their model have excess kurtosis. Using a Hill estimator, they fit a power law to the fat tails of the distribution and observe exponents grossly ranging from 1.9 to 4.6. They also check hints for volatility clustering: absolute returns and squared returns exhibit a slow decay of autocorrelation, while raw returns do not.
It thus appears that such a model can grossly fit some ``stylized facts''. However, the number of parameters involved, as well as the complicated rules of transition between agents, make clear identification of sources of phenomenons and calibration to market data difficult and intractable.

\cite{AlfiPietronero2009, AlfiPietronero2009-1} provide a somewhat simplifying view on the Lux-Marchesi model. They clearly identify the fundamentalist behaviour, the chartist behaviour, the herding effect and the observation of the price by the agents as four essential effects of an agent-based financial model. They show that the number of agents plays a crucial role in a Lux-Marchesi-type model: more precisely, the stylized facts are reproduced only with a finite number of agents, not when the number of agents grows asymptotically, in which case the model stays in a fundamentalist regime. There is a finite-size effect that may prove important for further studies.

\begin{figure}
\includegraphics[width=\columnwidth]{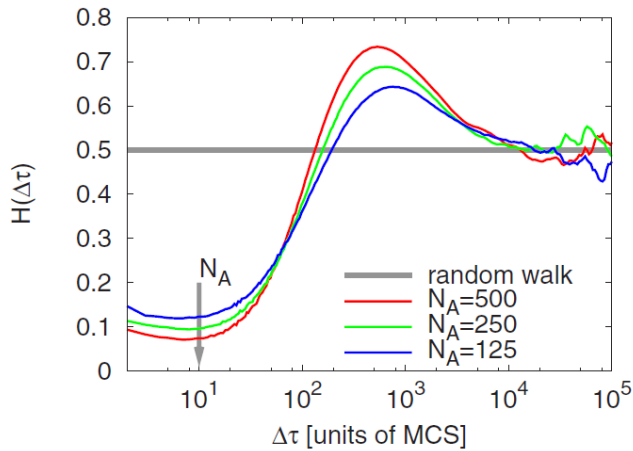}
\caption{\label{figure:PreisHurstResults}Hurst exponent found in the Preis model for different number of agents when including random demand perturbation and dynamic limit order placement depth. Reproduced from \cite{Preis2007}.}
\end{figure}

The role of the trend following mechanism in producing non-trivial features in price time series is also studied in \cite{Preis2007}. The starting point is an order book model similar to \cite{ChalletStinchcombe2001} and \cite{SmithFarmer2003}:  at each time step, liquidity providers submit limit orders at rate $\lambda$ and liquidity takers submit market orders at rate $\mu$. As expected, this zero-intelligence framework does not produce fat tails in the distribution of (log-)returns nor an over-diffusive Hurst exponent.
Then, a stochastic link between order placement and market trend is added: it is assumed that liquidity providers observing a trend in the market will act consequently and submit limit orders at a wider depth in the order book. Although the assumption behind such a mechanism may not be empirically confirmed (a questionable symmetry in order placement is assumed) and should be further discussed, it is interesting enough that it directly provides fat tails in the log-return distributions and an over-diffusive Hurst exponent $H\approx 0.6-0.7$ for medium time-scales, as shown in figure~\ref{figure:PreisHurstResults}.

\subsubsection{Threshold behaviour}

We finally review a model focusing primarily on reproducing the stylized fact of volatility clustering, while most of the previous models we have reviewed were mostly focused on fat tails of log returns. \cite{Cont2007} proposes a model with a rather simple mechanism to create volatility clustering. The idea is that volatility clustering characterizes several regimes of volatility (quite periods vs bursts of activity). Instead of implementing an exogenous change of regime, the author defines the following trading rules.

At each period, an agent $i\in\{1,\ldots,N\}$ can issue a buy or a sell order: $\phi_i(t)=\pm 1$. Information is represented by a series of i.i.d Gaussian random variables. $(\epsilon_t)$. This public information $\epsilon_t$ is a forecast for the value $r_{t+1}$ of the return of the stock. Each agent $i\in\{1,\ldots,N\}$ decides whether to follow this information according to a threshold $\theta_i>0$ representing its sensibility to the public information:
\begin{equation}
	\phi_i(t) = \left\{
		\begin{array}{rcl}
			1 & \textrm{ if } & \epsilon_i(t)>\theta_i(t)
			\\
			0 & \textrm{ if } & |\epsilon_i(t)|<\theta_i(t)
			\\
			-1 & \textrm{ if } & \epsilon_i(t)<-\theta_i(t)
		\end{array}
	\right.
\end{equation}
Then, once every choice is made, the price evolves according to the excess demand $D(t)= \sum_{i=1}^N \phi_i(t)$, in a way similar to \cite{ContBouchaud2000}.
At the end of each time step $t$, threshold are asynchronously updated. Each agent has a probability $s$ to update its threshold $\theta_i(t)$. In such a case, the new threshold $\theta_i(t+1)$ is defined to be the absolute value $|r_t|$of the return just observed. In short:
\begin{equation}
	\theta_i(t+1) = \mbox{\boldmath$1$}_{\{u_i(t)<s\}}|r_t| + \mbox{\boldmath$1$}_{\{u_i(t)>s\}}\theta_i(t).
\end{equation}

The author shows that the time series simulated with such a model do exhibit some realistic facts on volatility. In particular, long range correlations of absolute returns is observed. The strength of this model is that it directly links the state of the market with the decision of the trader. Such a feedback mechanism is essential in order to obtain non trivial characteristics. Of course, the model presented in \cite{Cont2007} is too simple to be fully calibrated on empirical data, but its mechanism could be used in a more elaborate agent-based model in order to reproduce the empirical evidence of volatility clustering.

\subsection{Remarks}
Let us attempt to make some concluding remarks on these developments of agent-based models for order books.
In table \ref{table:LOBSummary}, we summarize some key features of some of the order book models reviewed in this section. Among important elements for future modelling, we may mention the cancellation of orders, which is the less realistic mechanism implemented in existing models ; the order book stability, which is always exogenously enforced (see our review of \cite{MikeFarmer2008} above) ; and the dependence between order flows (see e.g. \cite{MuniToke2010} and reference therein). Empirical estimation of these mechanisms is still challenging.

Emphasis has been put in this section on order book modelling, a field that is at the crossroad of many larger disciplines (market microstructure, behavioural finance and physics). Market microstructure is essential since it defines in many ways the goal of the modelling. We pointed out that it is not a coincidence if the work by \cite{Garman1976} was published when computerization of exchanges was about to make the electronic order book the key of all trading. Regulatory issues that pushed early studies are still very important today. Realistic order book models could be a invaluable tool in testing and evaluating the effects of regulations such as the 2005 Regulation NMS\footnote{National Market System} in the USA, or the 2007 MiFID\footnote{Markets in Financial Instruments Directive} in Europe.

\begingroup
\squeezetable
\begin{table*}[t]
\begin{center}
\begin{tabular}{|p{1cm}|p{2.2cm}|p{2.2cm}|p{2.2cm}|p{2.2cm}|p{2.2cm}|p{2.2cm}|}
\hline
Model & Stigler (1961) & Garman (1976) & Bak, Paczuski and Shubik (1997) & Maslov (2000) & Challet and Stinchcombe (2001) & Mike and Farmer (2008) \\ \hline
 &  &   &  &  &  &  \\ \hline
Price range & Finite grid & Finite grid & Finite grid & Unconstrained & Unconstrained & Unconstrained \\ \hline
Clock & Trade time & Physical Time & Aggregated time & Event time & Aggregated time & Aggregated time \\ \hline
Flows / Agents & One zero-intelligence agent / One flow & One zero-intelligence agent / Two flows (buy/sell) & N agents owning each one limit order & One zero-intelligence flow (limit order with fixed probability, else market order) & One zero-intelligence agent / One flow & One zero-intelligence agent / One flow \\ \hline
Limit orders & Uniform distribution on the price grid & Two Poisson processes for buy and sell orders & Moving at each time step by one tick & Uniformly distributed in a finite interval around last price & Normally distributed around best quote & Student-distributed around best quote \\ \hline
Market orders & Defined as crossing limit orders & Defined as crossing limit orders & Defined as crossing limit orders & Submitted as such & Defined as crossing limit orders & Defined as crossing limit orders \\ \hline
Cancel\-lation orders & Pending orders are cancelled after a fixed number of time steps & None & None (constant number of pending orders) & Pending orders are cancelled after a fixed number of time steps & Pending orders can be cancelled with fixed probability at each time step & Pending orders can be cancelled with 3-parameter conditional probability at each time step \\ \hline
Volume & Unit & Unit & Unit & Unit & Unit & Unit \\ \hline
Order signs & Independent & Independent & Independent & Independent & Independent & Correlated with a fractional Brownian motion \\ \hline
 &  &  &  &  &  &  \\ \hline
Claimed results & Return distribution is power-law ~0.3 / Cut-off because finite grid & Microstructure is responsible for negative correlation of consecutive price changes & No fat tails for returns / Hurst exponent ~ 1/4 for price increments & Fat tails for distributions of returns / Hurst exponent ~ 1/4  & Hurst exponent ~1/4 for short time scales, tending to 1/2 for larger time scales & Fat tails distributions of returns / Realistic spread distribution / Unstable order book \\ \hline
\end{tabular}
\end{center}
\caption{Summary of the characteristics of the reviewed limit order book models.}
\label{table:LOBSummary}
\end{table*}
\endgroup

% =============================================================================
\section{Agent-based modelling for wealth distributions: Kinetic theory models}

The distributions of money, wealth or income, i.e., how such quantities are 
shared among the population of a given country and among different countries, 
is a topic which has been studied by economists for a long time. 
The relevance of the topic to us is twofold:
From the point of view of the science of Complex Systems, wealth distributions 
represent a unique example of a quantitative outcome of a collective behavior
which can be directly compared with the predictions of theoretical models and 
numerical experiments.
Also, there is a basic interest in wealth distributions from the
social point of view, in particular in their degree of (in)equality.
To this aim, the Gini coefficient (or the Gini index, if expressed as a 
percentage), developed by the Italian statistician Corrado Gini, 
represents a concept commonly employed to measure inequality of wealth 
distributions or, more in general, how uneven a given distribution is.
For a cumulative distribution function $F(y)$, that is piecewise differentiable,
has a finite mean $\mu$, and is zero for $y<0$, the Gini coefficient is defined 
as 
\begin{eqnarray}
\label{Gini}
 G & = & 1
     - \frac{1}{\mu} 
       \int^{\infty}_{0} dy \, (1 - F(y))^{2}
      \nonumber \\
  & = & \frac{1}{\mu} 
     \int^{\infty}_{0} dy \, F(y) (1-F(y)) \, .
\end{eqnarray}
It can also be interpreted statistically as half the relative mean difference.  
Thus the Gini coefficient is a number between 0 and 1, where 0 corresponds 
with perfect equality (where everyone has the same income) and 
1 corresponds with perfect inequality (where one person has all the income, 
and everyone else has zero income). 
Some values of $G$ for some countries are listed in Table \ref{tab:gini}.

%______________________________________________________
\begin{table}[ht]
\caption{Gini indices (in percent) of some countries (from {\it Human Development Indicators of
the United Nations Human Development Report} 2004, pp.50-53,
available at {\tt http://hdr.undp.org/en/reports/global/hdr2004}. More recent data are also available from their website.)
}
\centering
{\begin{tabular}{@{}cccc@{}} 
%\toprule
\hline
%\noalign{\smallskip}
Denmark   \hphantom{00} &  $~~~~~~~~~~~$  & \hphantom{0} 24.7 &$~~~~~~~~~~~$ \\
Japan   \hphantom{00} &  $~~~~~~~~~~~$  & \hphantom{0}     24.9 &$~~~~~~~~~~~$ \\
Sweden  \hphantom{00} &  $~~~~~~~~~~~$  & \hphantom{0}     25.0 &$~~~~~~~~~~~$ \\
Norway  \hphantom{00} &  $~~~~~~~~~~~$  & \hphantom{0}     25.8 &$~~~~~~~~~~~$ \\
Germany \hphantom{00} &  $~~~~~~~~~~~$  & \hphantom{0}       28.3 &$~~~~~~~~~~~$ \\
India   \hphantom{00} &  $~~~~~~~~~~~$  & \hphantom{0}       32.5 &$~~~~~~~~~~~$ \\
France  \hphantom{00} &  $~~~~~~~~~~~$  & \hphantom{0}       32.7 &$~~~~~~~~~~~$ \\
Australia \hphantom{00} &  $~~~~~~~~~~~$  & \hphantom{0}     35.2 &$~~~~~~~~~~~$ \\
UK      \hphantom{00} &  $~~~~~~~~~~~$  & \hphantom{0}       36.0 &$~~~~~~~~~~~$ \\
USA     \hphantom{00} &  $~~~~~~~~~~~$  & \hphantom{0}       40.8 &$~~~~~~~~~~~$ \\
Hong Kong  \hphantom{00} &  $~~~~~~~~~~~$  & \hphantom{0}     43.4 &$~~~~~~~~~~~$ \\
China    \hphantom{00} &  $~~~~~~~~~~~$  & \hphantom{0}       44.7 &$~~~~~~~~~~~$ \\
Russia   \hphantom{00} &  $~~~~~~~~~~~$  & \hphantom{0}       45.6 &$~~~~~~~~~~~$ \\
Mexico   \hphantom{00} &  $~~~~~~~~~~~$  & \hphantom{0}       54.6 &$~~~~~~~~~~~$\\
Chile    \hphantom{00} &  $~~~~~~~~~~~$  & \hphantom{0}       57.1 &$~~~~~~~~~~~$\\
Brazil   \hphantom{00} &  $~~~~~~~~~~~$  & \hphantom{0}       59.1 &$~~~~~~~~~~~$\\
South Africa  \hphantom{00} &  $~~~~~~~~~~~$  & \hphantom{0}  59.3 &$~~~~~~~~~~~$\\
Botswana   \hphantom{00} &  $~~~~~~~~~~~$  & \hphantom{0}     63.0 &$~~~~~~~~~~~$\\
Namibia    \hphantom{00} &  $~~~~~~~~~~~$  & \hphantom{0}     70.7 &$~~~~~~~~~~~$\\
%\noalign{\smallskip}
\hline
%\botrule
\end{tabular} \label{tab:gini}}
\end{table}
% ______________________________________________________

Let us start by considering the basic economic quantities: 
money, wealth and income.

% =============================================================================
\subsection{Money, wealth and income}  
A common definition of {\em money} suggests that money is the 
``[c]ommodity accepted by general consent as medium 
of economics exchange''\footnote{In Encyclop{\ae}dia Britannica. Retrieved June 17, 2010, from Encyclop{\ae}dia Britannica Online}.
In fact, money circulates from one economic agent 
(which can represent an individual, firm, country, etc.) 
to another, thus facilitating trade.
It is ``something which all other goods or services are traded for'' 
(for details see \cite{Shostak2000a}).
Throughout history various commodities have been used as money, 
for these cases termed as ``commodity money'',
which include for example rare seashells or beads, and cattle (such as cow in India). 
Recently, ``commodity money'' has been replaced by other forms
referred to as ``fiat money'', which have gradually become the most common ones, 
such as metal coins and paper notes.
Nowadays, other forms of money, such as electronic money, have become the most
frequent form used to carry out transactions.
In any case the most relevant points about money employed are 
its basic functions, which according to standard economic theory are
\begin{itemize}
\item to serve as a medium of exchange, which is universally accepted 
in trade for goods and services;
\item to act as a measure of value, making possible the determination of the 
prices and the calculation of costs, or profit and loss;
\item to serve as a standard of deferred payments, i.e., a tool for the payment
of debt or the unit in which loans are made and future transactions are fixed; 
\item to serve as a means of storing wealth not immediately required for use. 
\end{itemize}
A related feature relevant for the present investigation is that money 
is the medium in which prices or values of all commodities as well as costs, 
profits, and transactions can be determined or expressed.
{\em Wealth} is usually understood as things that have economic utility 
(monetary value or value of exchange), or material goods or property;
it also represents the abundance of objects of value (or riches) 
and the state of having accumulated these objects;
for our purpose, it is important to bear in mind that wealth can be measured 
in terms of money. 
Also {\em income}, defined in \cite{Case2008} as ``the sum of all the wages, salaries, profits, interests payments, rents and other forms of earnings received... in a given period of time'',
is a quantity which can be measured in terms of money (per unit time).

% =============================================================================
\subsection{Modelling wealth distributions}
It was first observed by \cite{Pareto1897a} that in an economy 
the higher end of the distribution of income $f(x)$ follows a power-law,
\begin{equation}
  f(x)\sim x^{-1-\alpha } \, ,
  \label{eq:paretolaw}
\end{equation}
with $\alpha$, now known as the Pareto exponent, estimated by him
to be $\alpha \approx 3/2$.
For the last hundred years the value of $\alpha \sim $ 3/2 seems to have
changed little in time and across the various capitalist economies (see \cite{YakovenkoRosser2009} and references therein). 

\cite{Gibrat1931a} clarified that Pareto's law is valid 
only for the high income range, whereas for the middle income range
he suggested that the income distribution
is described by a log-normal probability density
\begin{equation}
  f(x) 
  \sim 
  \frac{1}{x \sqrt{2\pi \sigma^2}} 
  \exp{\left\{ -\frac{\log^2 (x/x_0)}{2\sigma^2}\right\}} \, ,
  \label{eq:gibratlaw}
\end{equation}
where $\log(x_0) = \langle \log(x) \rangle $ is the mean value 
of the logarithmic variable 
and $\sigma^2 = \langle [\log(x) - \log(x_0)]^2 \rangle $ 
the corresponding variance. 
The factor $\beta=1/\sqrt{2\sigma^2}$, also know an as Gibrat index, 
measures the equality of the distribution.

\begin{figure}
\begin{center}
\includegraphics[width=\columnwidth]{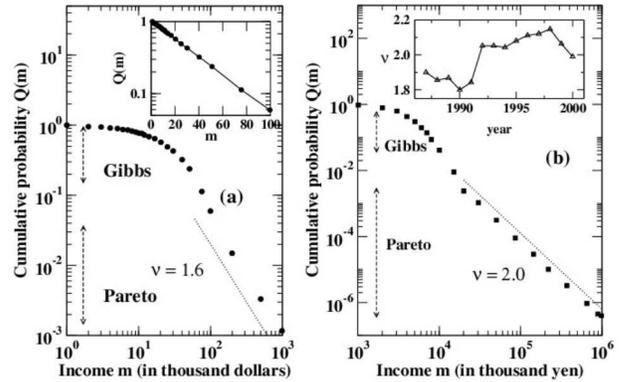}
\end{center}
\caption{Income distributions in the US (left)and Japan (right). 
Reproduced and adapted from \cite{ChatterjeeTokyo2003}, available at {\tt arXiv:cond-mat/0302147}.}
\label{fig:income1}
\end{figure}

More recent empirical studies on income distribution have been carried out
by physicists, e.g. those by \cite{Dragulescu2001a,Dragulescu2001b} for UK and US, 
by \cite{Fujiwara2003a} for Japan, and by \cite{Nirei2007} for US and Japan. For
 an overview see \cite{YakovenkoRosser2009}.
The distributions obtained have been shown to follow either the log-normal (Gamma like) 
or power-law types, depending on the range of wealth, 
as shown in Fig.~\ref{fig:income1}.
%%%%%%%%%%%%%%%
%%%%%%%%%%%%%%%

One of the current challenges is to write down the  ``microscopic equation'' 
which governs the dynamics of the evolution of wealth distributions,
possibly predicting the observed shape of wealth distributions,
including the exponential law at intermediate values of wealth as well as
the century-old Pareto law.  
To this aim, several studies have been made to investigate the characteristics 
of the real income distribution and provide theoretical models or explanations 
(see e.g. reviews by \cite{Lux2005a}, \cite{Chatterjee2007b}, \cite{YakovenkoRosser2009}).

The model of \cite{Gibrat1931a} 
and other models formulated in terms of a Langevin equation 
for a single wealth variable, subjected to multiplicative noise (\cite{Mandelbrot1960a,Levy1996a,Sornette1998a,Burda2003a}),
can lead to equilibrium wealth distributions with a power law tail,
since they converge toward a log-normal distribution.
However, the fit of real wealth distributions does not turn out to be as good as 
that obtained using e.g. a $\mathrm{\Gamma}$- or a $\mathrm{\beta}$-distribution,
in particular due to too large  asymptotic variances (\cite{Angle1986a}).
Other models use a different approach and describe the wealth dynamics as a wealth flow due to exchanges 
between (pairs of) basic units.
In this respect, such models are basically different from the class of models
formulated in terms of a Langevin equation for a single wealth variable.
For example, \cite{Solomon1996a} studied the generalized 
Lotka-Volterra equations in relation to power-law wealth 
distribution.
\cite{Ispolatov1998a} studied random exchange models 
of wealth distributions.
Other models describing wealth exchange have been formulated using 
matrix theory (\cite{Gupta2006a}),
the master equation (\cite{Bouchaud2000b,Dragulescu2000a,Ferrero2004a}),
the Boltzmann equation approach (\cite{Dragulescu2000a,Slanina2004a,Repetowicz2005a,Cordier2005a,Matthes2007a,During2007a,During2008a}),
or Markov chains (\cite{Scalas2006a,Scalas2007a,Garibaldi2007a}). It should be mentioned that one of the earliest modelling efforts were made by \cite{Champernowne1953}. Since then many economists, \cite{GabaixZipf1999} and \cite{Benhabib2009} amongst others, have also studied mechanisms for power laws, and distributions of wealth.

In the two following sections we consider in greater detail a class of models
usually referred to as \emph{kinetic wealth exchange models} (KWEM),
formulated through finite time difference stochastic 
equations (\cite{Angle1986a,Angle2002a,Angle2006a,
Chakraborti2000a,Dragulescu2000a,Chakraborti2002a,
Hayes2002a,
Chatterjee2003a,
Das2003a,
Scafetta2004b,
Iglesias2003a,
Iglesias2004a,
Ausloos2007a}).
From the studies carried out using wealth-exchange models, 
it emerges that it is possible to use them to generate power law distributions.

% =============================================================================
\subsection{Homogeneous kinetic wealth exchange models}

Here and in the next section we consider KWEMs,
which are statistical models of closed economy.
Their goal,
rather then describing the market dynamics in terms of intelligent agents, is to predict the time evolution of the
distribution of some main quantity, such as wealth, by studying the corresponding flow process among individuals. The
underlying idea is that however complicated the detailed rules of wealth exchanges can be, their average behaviour can
be described in a relatively more simple way and will share some universal properties with other transport processes,
due to general conservation constraints and the effect of the fluctuations due to the environment or associated to the
individual behaviour. In this, there is a clear analogy with the general theory of transport phenomena (e.g. of energy).

In these models the states of agents are defined in terms of the 
wealth variables $\{x_n\},~n=1,2,\dots,N$.
The evolution of the system is carried out according to a trading rule 
between agents which, for obtaining the final equilibrium distribution, 
can be interpreted as the actual time evolution of the agent states 
as well as a Monte Carlo optimization.
The algorithm is based on a  simple update rule performed 
at each time step $t$, when two agents $i$ and $j$ are extracted randomly 
and an amount of wealth $\Delta x$ is exchanged, 
\begin{eqnarray}
  x_i' &=& x_i - \Delta x \, ,
  \nonumber \\
  x_j' &=& x_j + \Delta x \, .
  \label{basic0}
\end{eqnarray}
Notice that the quantity $x$ is conserved during single transactions, 
$x_i'+x_j' = x_i + x_j$, where $x_i = x_i(t)$ and $x_j = x_j(t)$ 
are the agent wealth before, 
whereas $x_i' = x_i(t+1)$ and $x_j' = x_j(t+1)$ 
are the final ones after the transaction.
Several rules have been studied for the model defined by Eqs.~(\ref{basic0}).
It is noteworthy, that 
though this theory has been originally derived from the entropy
maximization principle of statistical mechanics, it has recently
been shown that the same could be derived from the utility
maximization principle as well, following a standard exchange-model
with Cobb-Douglas utility function (as explained later), which 
bridge physics and economics together.

\subsubsection{Exchange models without saving}
\label{sec:basic}
\begin{figure} 
  \begin{center}
    \includegraphics[angle=0,width=\columnwidth]{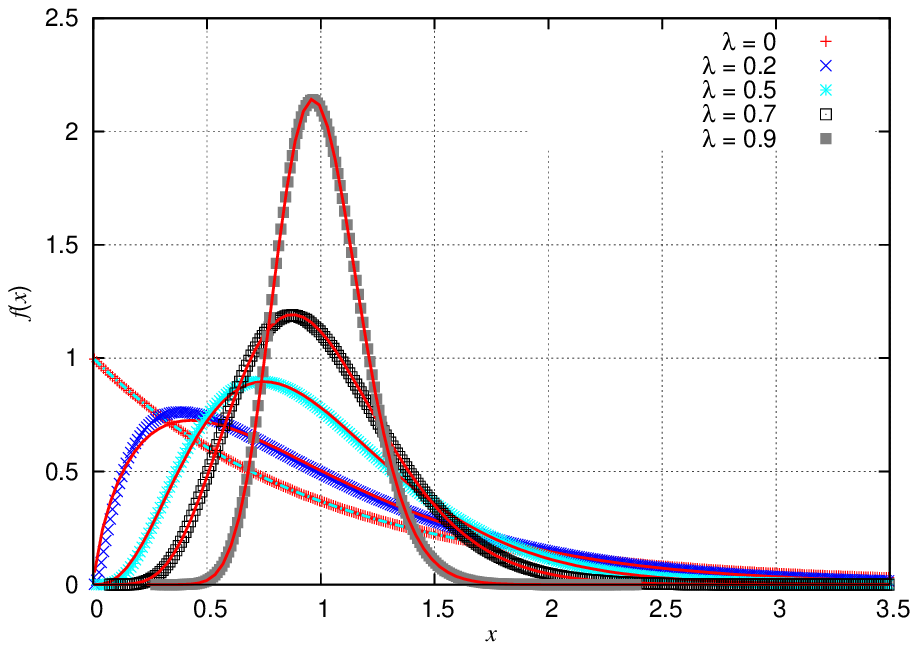}
    \\
    \includegraphics[angle=0,width=\columnwidth]{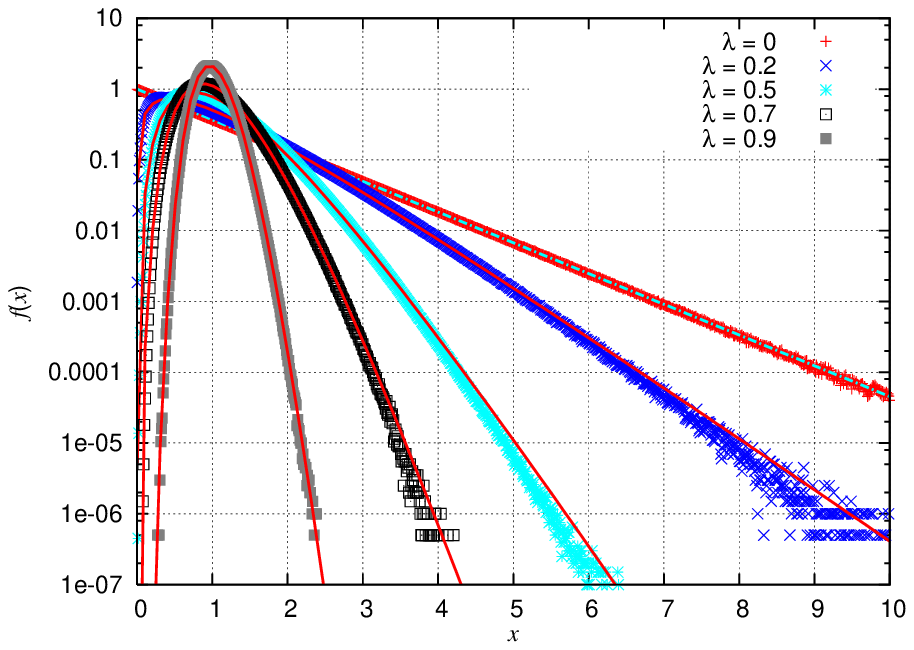}
    \caption{ Probability density for wealth $x$.
    The curve for $\lambda=0$ is the Boltzmann function $f(x)=\langle x \rangle^{-1}\exp(-x/\!\langle x \rangle)$ for the basic model of Sec. \ref{sec:basic}. 
    The other curves correspond to a global saving propensity
    $\lambda>0$, see Sec. \ref{global}.} 
    \label{fig:gamma} 
  \end{center}
\end{figure}

In a simple version of KWEM considered in the works by \cite{Bennati1988a,Bennati1988b,Bennati1993a} 
and also studied by \cite{Dragulescu2000a}
the money difference $\Delta x$ in Eqs.~(\ref{basic0}) 
is assumed to have a constant value, $\Delta x = \Delta x_0$.
Together with the constraint that transactions can take place 
only if $x_i'>0$ and $x_j'>0$, this leads to an equilibrium exponential distribution, 
see the curve for $\lambda=0$ in Fig.~\ref{fig:gamma}.

Various other trading rules were studied 
by \cite{Dragulescu2000a}, choosing
$\Delta x$ as a random fraction of the average money between the two agents,
$\Delta x = \epsilon (x_i + x_j)/2$,
corresponding to a   
$\Delta x =  (1 - \epsilon) x_i - \epsilon x_j$  in (\ref{basic0}),
 or of the average money of the whole system, 
$\Delta x = \epsilon \langle x \rangle$.

The models mentioned, as well as more complicated
ones (\cite{Dragulescu2000a}), lead to an equilibrium wealth 
distribution with an exponential tail
\begin{equation}
  f(x) \sim \beta \exp( -\beta x) \, ,
  \label{BD}
\end{equation}
with the effective temperature $1/\beta$ of the order of the average wealth,
$\beta^{-1} = \langle x \rangle$.
This result is largely independent of the details of the models,
e.g. the multi-agent nature of the interaction, 
the initial conditions, and 
the random or consecutive order of extraction of the interacting agents.
The Boltzmann distribution is 
characterized 
by a majority of poor agents and a few rich agents 
(due to the exponential tail), and has  
a Gini coefficient of $0.5$.

%-------------------------------------------------------------------------------

%
%

%-------------------------------------------------------------------------------

\subsubsection{Exchange models with saving}
\label{global}

As a generalization and more realistic version of the basic exchange models, a saving criterion can be introduced.
\cite{Angle1983a}, motivated by the surplus theory, introduced a unidirectional model of wealth exchange,
in which only a fraction of wealth smaller than one can pass from one agent
to the other, with a $\Delta x = \epsilon x_i$ or $(-\omega x_j)$,
where the direction of the flow is determined by the agent wealth (\cite{Angle1983a,Angle1986a}).
Later Angle introduced the One-Parameter Inequality Process (OPIP) where a constant fraction $1 - \omega$ 
is saved before the transaction (\cite{Angle2002a}) by the agent whose 
wealth decreases, defined by an exchanged wealth amount 
$\Delta x  =  \omega x_i$ or $- \omega x_j$, again with the direction
of the transaction determined by the relative difference between the
agent wealth.

A ``saving parameter'' $0 < \lambda < 1$ representing the fraction of wealth \textit{saved},
was introduced in the model by \cite{Chakraborti2000a}.
In this model (CC) wealth flows simultaneously toward and from each agent
during a single transaction, the dynamics being defined by the equations
\begin{eqnarray}
  x_i' &=& \lambda x_i + \epsilon (1-\lambda) (x_i + x_j) \, ,
  \nonumber \\
  x_j' &=& \lambda x_j + (1 - \epsilon) (1-\lambda) (x_i + x_j) \, ,
  \label{sp1}
\end{eqnarray}
or, equivalently, by a $\Delta x$ in (\ref{basic0}) given by
\begin{equation}
  \Delta x  
  =     (1 - \lambda) [ (1 - \epsilon) x_i - \epsilon x_j ]  \, .
\end{equation}

These models,
apart from the OPIP model of Angle which has the remarkable property
of leading to a power law in a suitable range of $\omega$,
can be well fitted by a $\Gamma$-distribution.
The $\Gamma$-distribution is characterized by a mode $x_m>0$,
in agreement with real data of wealth and income 
distributions (\cite{Dragulescu2001b,Ferrero2004a,Silva2005a,SalaiMartin2002a,SalaiMartin2002b,Aoyama2003a}). 
Furthermore, the limit for small $x$ is zero, 
i.e. $P(x \! \to 0) \! \to 0$, see the example in Fig.~\ref{fig:gamma}.
In the particular case of the model by \cite{Chakraborti2000a}, the
explicit distribution is well fitted by
\begin{eqnarray}
  \label{NGibbs}
  f(x) &=& 
  n \langle x \rangle^{-1} \gamma_n( n x / \langle x \rangle) \nonumber \\&=& 
  \frac{1}{\Gamma(n)} \frac{n}{\langle x \rangle}
  \left( \frac{n x }{\langle x \rangle} \right)^{n-1} 
  \exp\left( - \frac{ n x }{ \langle x \rangle } \right) \ ,
  \\
  \label{n}
  n(\lambda) &\equiv& \frac{D_\lambda}{2} = 1 + \frac{3 \lambda}{1 - \lambda} \ .
\end{eqnarray}
where $\gamma_n(\xi)$ is the standard 
$\Gamma$-distribution.
This particular functional form has been conjectured on the base of the excellent fitting provided
to numerical
data (\cite{Angle1983a,Angle1986a,Patriarca2004b,Patriarca2004a,Patriarca2009a}).
For more information and a comparison of similar fittings for different models see \cite{Patriarca2010b}.
Very recently, \cite{Lallouache2010} have shown using the distributional form of the equation and moment calculations that strictly speaking the Gamma distribution is not the solution of Eq.~(\ref{sp1}), confirming the earlier results of \cite{Repetowicz2005a}. However, the Gamma distribution is a very very good approximation.

%--------------------------------------------
\begin{figure}[tb] 
  \begin{center}
    \includegraphics[angle=0,width=\columnwidth]{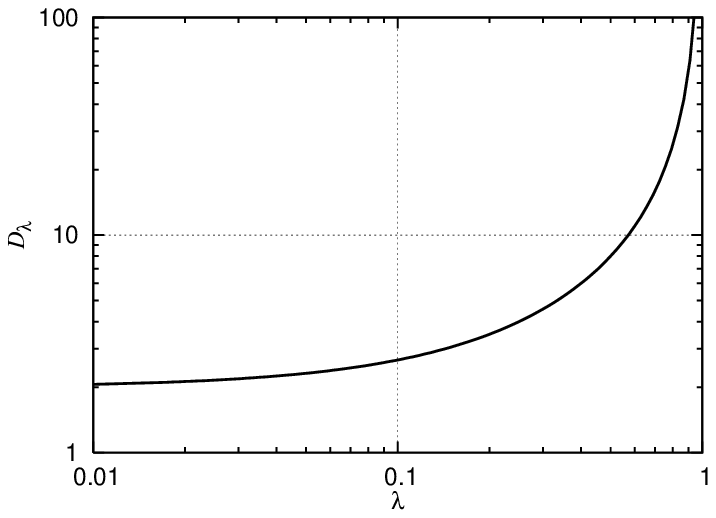}
    \\
    \includegraphics[angle=0,width=\columnwidth]{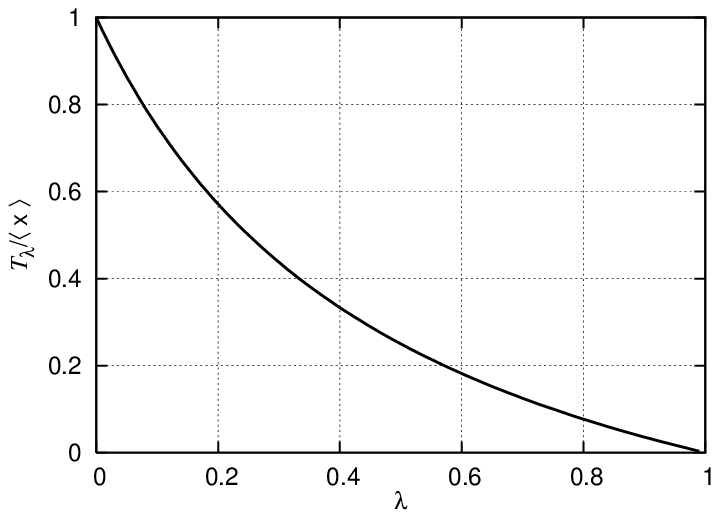}
    \caption{ Effective dimension $D_\lambda$ 
    and temperature $T$ as a function
    of the saving parameter $\lambda$.} 
    \label{fig_Dlambda} 
  \end{center}
\end{figure} 
%--------------------------------------------
The ubiquitous presence of $\Gamma$-functions in the solutions of
kinetic models (see also below heterogeneous models) 
suggests a close analogy with kinetic theory of gases.
In fact, interpreting $D_\lambda = 2n$ as an effective dimension,
the variable $x$ as kinetic energy, 
and introducing the effective temperature 
$\beta^{-1} \equiv T_\lambda = \langle x \rangle / 2 D_\lambda$
according to the equipartition theorem,
Eqs.~(\ref{NGibbs}) and (\ref{n}) define the canonical distribution 
$\beta \gamma_n(\beta x)$ for the kinetic energy of a gas in $D_\lambda = 2n$ 
dimensions, see \cite{Patriarca2004a} for details.
The analogy is illustrated in Table~\ref{tab:analogy}
and the dependences of $D_\lambda = 2n$ and of $\beta^{-1} = T_\lambda$ 
on the saving parameter $\lambda$ are shown in Fig.~\ref{fig_Dlambda}.
%______________________________________________________
\begin{table}
\caption{Analogy between kinetic the theory of gases
and the kinetic exchange model of wealth}
\centering
{\begin{tabular}{lll} 
%\toprule
\hline
%\noalign{\smallskip}
                & Kinetic model       & Economy model  \\
%\noalign{\smallskip}
\hline
%\noalign{\smallskip}
variable        & $K$ (kinetic energy)& $x$ (wealth) \\
units           &  $N$ particles      & $N$ agents\\
interaction     & collisions          & trades\\
dimension       & integer $D$         & real number $D_\lambda$\\
temperature definition
                & $k_\mathrm{B} T = 2\langle K \rangle / D$ & $T_\lambda = 2 \langle x \rangle / D_\lambda$\\
reduced variable& $\xi = K / k_\mathrm{B} T$     & $\xi = x / T_\lambda$\\
equilibrium distribution & $f(\xi) = \gamma_{D/2}(\xi)$ & $f(\xi)=\gamma_{D_\lambda/2}(\xi)$\\
%\noalign{\smallskip}
\hline
%\botrule
\end{tabular} \label{tab:analogy}}
\end{table}
% ______________________________________________________

The exponential distribution is recovered as a special case, for~$n=1$. 
In the limit $\lambda \to 1$, i.e. for $n \to \infty$, 
the distribution $f(x)$ above tends to a Dirac $\delta$-function,
as shown in \cite{Patriarca2004a} and qualitatively illustrated
by the curves in Fig.~\ref{fig:gamma}.
This shows that a large saving criterion leads to a final state in which 
economic agents tend to have similar amounts of money and, 
in the limit of $\lambda \to 1$, exactly the same amount $\langle x \rangle$.

The equivalence between a kinetic wealth-exchange model with saving propensity 
$\lambda \ge 0$ and an $N$-particle system in a space with dimension 
$D_\lambda \ge 2$ is suggested by simple considerations about the kinetics 
of collision processes between two molecules. 
In one dimension, particles undergo head-on collisions in which the whole
amount of kinetic energy can be exchanged.
In a larger number of dimensions the two particles will not travel in general 
exactly along the same line, in opposite verses, and only a fraction of the
energy can be exchanged.
It can be shown that during a binary elastic collision 
in $D$ dimensions only a fraction $1/D$ of the total kinetic energy 
is exchanged on average for kinematic reasons, 
see \cite{Chakraborti2008a} for details. 
The same $1/D$ dependence is in fact obtained inverting Eq.~(\ref{n}), 
which provides for the fraction of exchanged wealth $1-\lambda = 6/(D_\lambda+4)$.

Not all homogeneous models lead to distributions with an exponential
tail.
For instance, in the model studied in \cite{Chakraborti2002a} an
agent $i$ can lose all his wealth, thus becoming unable to trade
again: after a sufficient number of transactions, only one trader
survives in the market and owns the entire wealth.
The equilibrium distribution has a very different shape, as explained below:

In the toy model it is assumed that both the economic agents $i$ and $j$ invest the same amount $ x_{min} $, 
which is taken as the minimum wealth between the two agents, 
$ x_{min} = \mathrm{min}\{x_i,x_j\}$. 
The wealth after the trade are 
$ x_{i}^{\prime }=x_{i} + \Delta x $ and 
$ x_{j}^{\prime }=x_{j} - \Delta x $,
where 
$\Delta x = (2\epsilon -1) x_{min} $.
We note that once an agent has lost all his wealth, he is unable to trade because $ x_{min} $ has become zero. 
Thus, a trader is effectively driven out of the market  once he loses all his wealth. In this way,
after a sufficient number of transactions only one trader survives in the market with the entire
amount of wealth, whereas the rest of the traders have zero wealth. 
In this toy model, only one agent has the entire money of the market and the rest of the traders have zero money, which corresponds to a distribution with Gini coefficient equal to unity. 

Now, a situation is said to be Pareto-optimal ``if by reallocation you cannot make someone better off without making someone else worse off''. 
In Pareto's own words:
\begin{quote}
``We will say that the members of a collectivity enjoy maximum ophelimity in a certain position when it is impossible to find a way of moving from that position very slightly in such a manner that the ophelimity enjoyed by each of the individuals of that collectivity increases or decreases. That is to say, any small displacement in departing from that position necessarily has the effect of increasing the ophelimity which certain individuals enjoy, and decreasing that which others enjoy, of being agreeable to some, and disagreeable to others.''

--- Vilfredo Pareto, Manual of Political Economy (1906), p.261.
\end{quote}
However, as \cite{Sen1971} notes, {\it an economy can be Pareto-optimal, yet still ``perfectly disgusting'' by any ethical standards} . 
It is important to note that Pareto-optimality, is merely a descriptive term, a property of an ``allocation'', and there are no ethical propositions about the desirability of such allocations inherent within that notion. 
Thus, in other words there is nothing inherent in Pareto-optimality that implies the maximization of social welfare. 

This simple toy model thus also produces a Pareto-optimal state (it will be impossible to raise the well-being of anyone except the \emph{winner}, i.e., the agent with all the money, and vice versa ) but the situation
is economically undesirable as far as social welfare is concerned!

Note also, as mentioned above, the OPIP model of \cite{Angle2006a,Angle2002a}, for example, depending on the model
parameters, can also produce a power law tail.
Another general way to produce a power law tail in the equilibrium
distribution seems to diversify the agents, i.e. to consider
heterogeneous models, discussed below.

% =============================================================================
\subsection{Heterogeneous kinetic wealth exchange models}
%
%%%%%%%%%%%%%%%
\begin{figure}
\begin{center}
\includegraphics[width=\columnwidth]{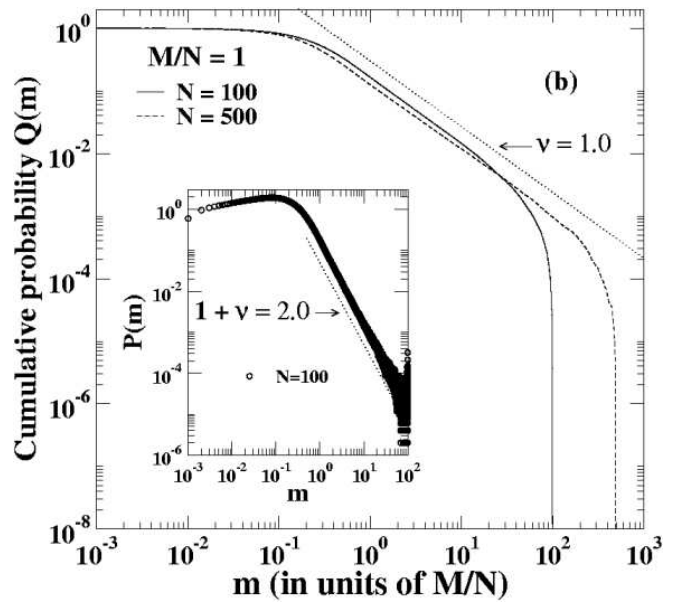}
\end{center}
\caption{Results for randomly assigned saving parameters. Reproduced and adapted from \cite{ChatterjeeTokyo2003}, available at {\tt arXiv:cond-mat/0302147}.}
\label{fig:rsave}
\end{figure}
%%%%%%%%%%%%%%%
% ----------------------------------------------------------
\subsubsection{Random saving propensities}
The models considered above assume the all agents have
the same statistical properties.
The corresponding equilibrium wealth distribution has in most of the cases
an exponential tail, a form which well interpolates real data at small and intermediate 
values of wealth.
However, it is possible to conceive generalized models which lead to
even more realistic equilibrium wealth distributions.
This is the case when agents are diversified by assigning different values of the saving parameter.
For instance, \cite{Angle2002a} studied a model with a trading rule where diversified
parameters $\{\omega_i\}$ occur,
\begin{equation}
  \Delta x  
  =   \omega_i \epsilon x_i 
  \mathrm{~~~~~or~~~~~}  - \omega_j \epsilon x_j \, ,
\end{equation}
with the direction of wealth flow 
determined by the wealth of agents $i$ and $j$.
Diversified saving parameters were independently
introduced by \cite{Chatterjee2003a,Chatterjee2004a}
by generalizing the model introduced in \cite{Chakraborti2000a}:
\begin{eqnarray}
  x_i' &=& 
  \lambda_i x_i + \epsilon [ (1-\lambda_i) x_i + (1-\lambda_j) x_j ] \, ,
  \nonumber \\
  x_j' &=& 
  \lambda x_j + (1 - \epsilon) [(1-\lambda_i) x_i + (1-\lambda_j) x_j ] \, ,
  \label{sp2}
\end{eqnarray}
corresponding to a
\begin{equation}
  \Delta x  
  =  (1 - \epsilon) (1-\lambda_i) x_i - \epsilon (1-\lambda_j) x_j  \, .
\end{equation}
The surprising result is that if the parameters $\{\lambda_i\}$
are suitably diversified, a power law appears in the equilibrium wealth 
distribution, see Fig.~\ref{fig:rsave}.
In particular if the $\lambda_i$ are uniformly distributed in $(0,1)$
the wealth distribution exhibits a robust power-law tail, 
\begin{equation}\label{f-power}
  f(x) \propto x^{-\alpha - 1} \, ,
\end{equation}
with the Pareto exponent $\alpha =1$ largely independent 
of the details of the $\lambda$-distribution.
It may be noted that the exponent value unity is strictly for the
tail end of the distribution and not for small values of the income
or wealth (where the distribution remains exponential). Also, for
finite number $N$ of agents, there is always an exponential (in $N$)
cut off at the tail end of the distribution.
This result is supported by independent theoretical considerations based
on different approaches, such as 
a mean field theory approach (\cite{Mohanty2006a}, see below for further details)
or the Boltzmann equation (\cite{Das2003a,Das2005a,Repetowicz2005a,Chatterjee2005a}). For derivation of the Pareto law from variational principles, using the KWEM context, see \cite{Chakraborti2009PRL}.

% -------------------------------------------------------------------------
\subsubsection{Power-law distribution as an overlap of Gamma distributions}
\begin{figure}[tb]
  \begin{center}
    \includegraphics[angle=0,width=0.45\columnwidth]{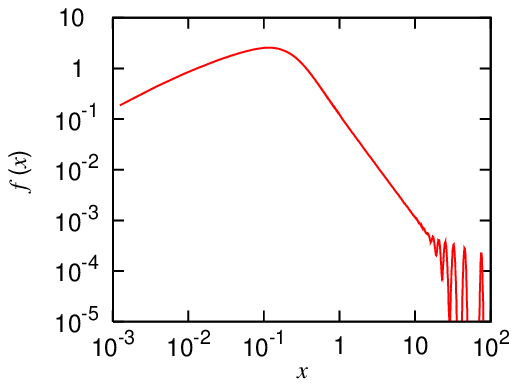}
    \includegraphics[angle=0,width=0.45\columnwidth]{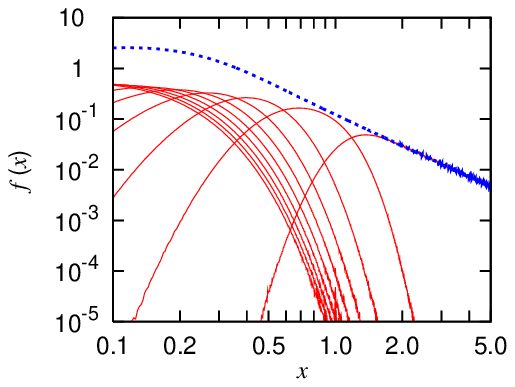}\\
    \includegraphics[angle=0,width=0.45\columnwidth]{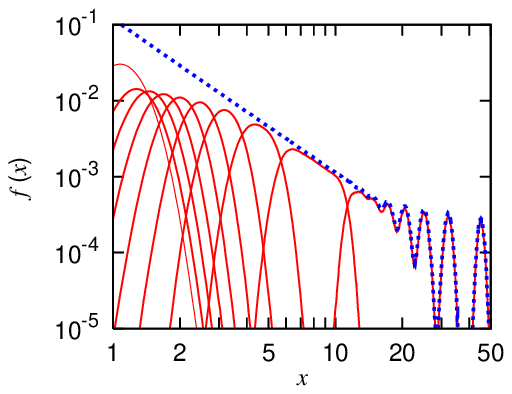}
    \includegraphics[angle=0,width=0.45\columnwidth]{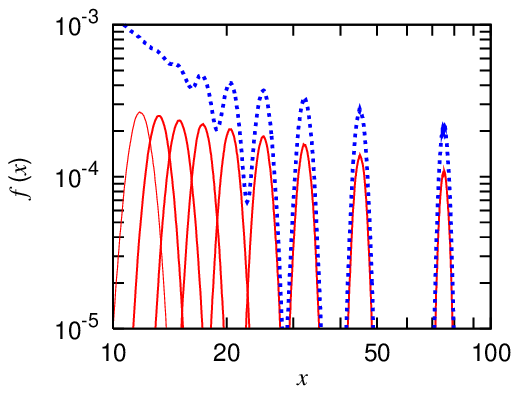}
   \caption{
   Wealth distribution in a system of 1000 agents with saving propensities 
   uniformly distributed in the interval $0 < \lambda < 1$. 
   Top left: marginal distribution. 
   Top right: marginal distribution (dotted line) and  
   distributions of wealth of agents with 
   $\lambda \in (j\Delta\lambda, (j+1)\Delta\lambda)$, 
   $\Delta\lambda=0.1$, $j=0,\dots,9$  (continuous lines). 
   Bottom-left: the distribution of wealth of agents with 
   $\lambda \in (0.9,1)$ has been further resolved into contributions 
   from subintervals  
   $\lambda \in (0.9+j\Delta\lambda, 0.9+(j+1)\Delta\lambda)$,
   $\Delta\lambda=0.01$. 
   Bottom-right: the partial distribution of wealth of agents with 
   $\lambda \in (0.99,1)$ has been further resolved into those 
   from subintervals 
   $\lambda \in (0.99+j\Delta\lambda,0.99+(j+1)\Delta\lambda)$, 
   $\Delta\lambda=0.001$.
   }
    \label{partials_a}
  \end{center}
\end{figure}

A remarkable feature of the equilibrium wealth distribution obtained
from heterogeneous models, noticed in \cite{Chatterjee2004a}, 
is that the individual wealth distribution $f_i(x)$ of the generic $i$-th agent 
with saving parameter $\lambda_i$ has a well defined mode and exponential tail,
in spite of the resulting power-law tail of the marginal distribution
$f(x) = \sum_i f_i(x)$.
In fact, \cite{Patriarca2005a} found by numerical simulation that the marginal distribution 
$f(x)$ can be resolved as an overlap of individual Gamma distributions 
with $\lambda$-dependent parameters;
furthermore, the mode and the average value of the distributions $f_i(x)$ 
both diverge for $\lambda \to 1$ as  
$\left\langle x(\lambda) \right\rangle \sim 1/(1-\lambda)$
(\cite{Chatterjee2004a,Patriarca2005a}).
This fact was justified theoretically by \cite{Mohanty2006a}.
Consider the evolution equations (\ref{sp2}).
In the mean field approximation one can consider that each agents $i$ has 
an (average) wealth $\langle x_i \rangle = y_i$ and 
replace the random number $\epsilon$ with its 
average value $\langle \epsilon \rangle = 1/2$.
Indicating with $y_{ij}$ the new wealth of agent $i$, due to the interaction 
with agent $j$, from Eqs.~(\ref{sp2}) one obtains
\begin{eqnarray}
  y_{ij} &=& 
  (1/2)(1 + \lambda_i) y_i + (1/2)(1-\lambda_j) y_j \, .
  \label{sp2-b}
\end{eqnarray}
At equilibrium, for consistency, average over all the interaction
must give back $y_i$, 
\begin{eqnarray}
  y_i = \sum_j y_{ij} / N \ .
  \label{sp2-c}
\end{eqnarray}
Then summing Eq.~(\ref{sp2-b}) over $j$ and dividing 
by the number of agents $N$, one has
\begin{eqnarray}
  (1 - \lambda_i) y_i = \langle (1 - \lambda) y \rangle \, ,
  \label{sp2-d}
\end{eqnarray}
where $\langle (1 - \lambda) y \rangle = \sum_j (1 - \lambda_j) y_j / N$.  
Since the right hand side is independent of $i$ and this relation holds for
arbitrary distributions of $\lambda_i$, the solution is
\begin{eqnarray}
  y_i = \frac{C}{1 - \lambda_i} \, ,
  \label{sp2-d}
\end{eqnarray}
where $C$ is a constant.
Besides proving the dependence of $y_i = \langle x_i \rangle $ on 
$\lambda_i$, this relation also demonstrates the existence of a power law tail
in the equilibrium distribution.
If, in the continuous limit, $\lambda$ is distributed in $(0,1)$ 
with a density $\phi(\lambda),~(0 \le \lambda < 1)$, 
then using (\ref{sp2-d}) the (average) wealth distribution is given
\begin{eqnarray}
  f(y) = \phi(\lambda) \frac{d\lambda}{dy} 
       = \phi(1 - C/x) \frac{C}{y^2} \, .
  \label{sp2-e}
\end{eqnarray}
Figure~\ref{partials_a} illustrates the phenomenon for a system of $N=1000$ 
agents with random saving propensities uniformly distributed between $0$ 
and $1$.
% ------------------------------
% --------------------------------
The figure confirms the importance of agents with $\lambda$ close to $1$ 
for producing a power-law probability distribution (\cite{Chatterjee2004a,Patriarca2009a}).

\begin{figure}[tb]
  \begin{center}
    \includegraphics[angle=0,width=\columnwidth]{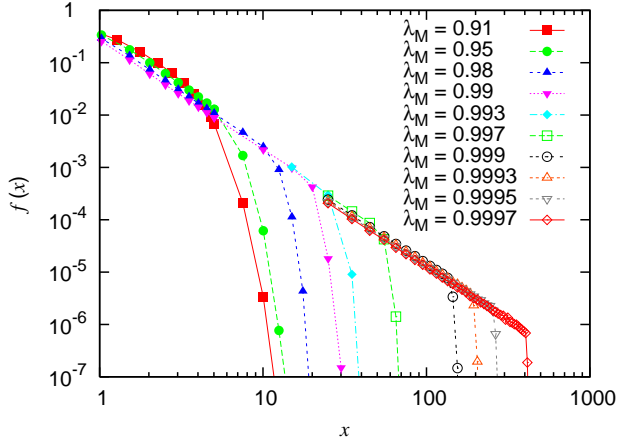}
    \caption{
    Wealth distribution obtained for the uniform saving propensity distributions 
    of $10^5$ agents in the interval $(0,\lambda_\mathrm{M})$.
    }
    \label{cutoff}
  \end{center}
\end{figure}

However, when considering values of $\lambda$ close enough to $1$,
the power law can break down at least for two reasons.
The first one, illustrated in Fig.~\ref{partials_a}-bottom right, is that
the power-law can be resolved into almost disjoint contributions
representing the wealth distributions of single agents.
This follows from the finite number of agents used
and the fact that the distance between the average values of the distributions
corresponding to two consecutive values of $\lambda$ grows faster than 
the corresponding widths (\cite{Patriarca2005a,Bhattacharya2005a}).
The second reason is due to the finite cutoff $\lambda_\mathrm{M}$,
always present in a numerical simulation. However,
to study this effect, one has to consider a system with a number
of agents large enough that it is not possible to resolve the wealth
distributions of single agents for the sub-intervals of $\lambda$ considered.
This was done in by \cite{Patriarca2006c} using a  system with $N = 10^5$ 
agents with saving parameters distributed uniformly between $0$ and 
$\lambda_\mathrm{M}$.
Results are shown in Fig.~\ref{cutoff}, in which curves from left to right 
correspond to increasing values of the cutoff $\lambda_\mathrm{M}$ from $0.9$ 
to $0.9997$. 
% -----------------------------------------------
% ------------------------------------------------
The transition from an exponential to a power-law tail takes place continuously
as the cut-off $\lambda_\mathrm{M}$ is increased beyond a critical value 
$\lambda_\mathrm{M} \approx 0.9$ toward $\lambda_\mathrm{M}=1$, 
through the enlargement of the $x$-interval in which the power-law is observed.

% -------------------------------------------------------------------------
\subsubsection{Relaxation process}
Relaxation in systems with constant $\lambda$ had already been studied by \cite{Chakraborti2000a}, where a systematic increase of the relaxation time with $\lambda$, and eventually a divergence for $\lambda \to 1$, was found.
In fact, for $\lambda = 1$ no exchanges occurs and the system is frozen. 
The relaxation time scale of a heterogeneous system had been studied by \cite{Patriarca2007a}.
The system is observed to relax toward the same equilibrium wealth distribution 
from any given arbitrary initial distribution of wealth.
If time is measured by the number of transactions $n_t$, 
the time scale is proportional to the number of agents $N$, i.e.
defining time $t$ as the ratio $t = n_t/N$ between the number of trades
and the total number of agents $N$ (corresponding to one Monte Carlo cycle 
or one sweep in molecular dynamics simulations)
the dynamics and the relaxation process become independent of $N$. The existence of a
natural time scale independent of the system size provides a foundation for
using simulations of systems with finite $N$ in order to infer properties of
systems with continuous saving propensity distributions and $N \to \infty$.

In a system with uniformly distributed $\lambda$,
the wealth distributions of each agent $i$ with saving parameter $\lambda_i$ 
relaxes toward different states with characteristic shapes 
$f_i(x)$ (\cite{Patriarca2005a,Bhattacharya2005a,Patriarca2006c})
with different relaxation times $\tau_i$ (\cite{Patriarca2007a}).
The differences in the relaxation process can be related to the different
relative wealth exchange rates, that by direct inspection of the evolution
equations appear to be proportional to $1 - \lambda_i$. 
Thus, in general, higher saving propensities are expected to be associated 
to slower relaxation processes with a relaxation time $\propto 1/(1-\lambda)$. 

It is also possible to obtain the relaxation time distribution.
If the saving parameters are distributed in $(0,1)$ with a density 
$\phi(\lambda)$, it follows from probability conservation that 
$\tilde{f}(\bar{x})d\bar{x} = \phi(\lambda) d\lambda$, 
where $\bar{x} \equiv \langle x \rangle_\lambda$ and 
$\tilde{f}(\bar{x})$ the corresponding density of average wealth values. 
In the case of uniformly distributed saving propensities,
one obtains
\begin{equation}
\label{f-phi}
\tilde{f}(\bar{x}) = \phi(\lambda) \frac{d\lambda(\bar{x})}{d\bar{x}}
= \phi\left(1-\frac{k}{\bar{x}}\right) \frac{k}{\bar{x}^2} \, ,
\end{equation}
showing that a uniform saving propensity distribution leads to a power law
$\tilde{f}(\bar{x}) \sim 1/\bar{x}^2$ in the (average) wealth distribution.
In a similar way it is possible to obtain the associated
distribution of relaxation times $\psi(\tau)$ for the global relaxation process
from the relation $\tau_i \propto 1 / (1 - \lambda_i)$,
\begin{equation}
\label{psi-phi}
\psi(\tau) = \phi(\lambda) \frac{d\lambda(\tau)}{d\tau} \propto
\phi\left(1 - \frac{\tau'}{\tau}\right) \frac{\tau'}{\tau^2} \, ,
\end{equation}
where $\tau'$ is a proportionality factor. 
Therefore $\psi(\tau)$ and $\tilde{f}(\bar{x})$ are characterized by power law
tails in $\tau$ and $\bar{x}$ respectively \emph{with the same Pareto
exponent}.

In conclusion, the role of the $\lambda$-cut-off is also related to 
the relaxation process.
This means that the slowest convergence rate is determined 
by the cut-off and is $\propto 1 - \lambda_\mathrm{M}$.
In numerical simulations of heterogeneous KWEMs,
as well as in real wealth distributions,
the cut-off is necessarily finite,
so that the convergence is fast (\cite{Gupta2008a}).
On the other hand, if considering a hypothetical wealth distribution 
with a power law extending to infinite values of $x$,
one cannot find a fast relaxation, due to the infinite time scale
of the system, due to the agents with $\lambda = 1$.

\subsection{Microeconomic formulation of Kinetic theory models}
Very recently, \cite{Chakrabarti20094151} have studied
the framework based on microeconomic theory from which
the kinetic theory market models could be addressed.
They derived the moments of the model by \cite{Chakraborti2000a} and 
reproduced the exchange equations used in the model (with fixed savings parameter).
In the framework considered, the utility function deals with the behaviour of the agents in an exchange economy.

They start by considering two exchange economy, where each agent produces a 
single 
perishable commodity. Each of these goods is different and money exists
in the economy to simply facilitate transactions.
 Each of these agents
are endowed with an initial amount of money $ M_1= m_1(t) $ and $ M_2= m_2(t) $.
Let agent 1 produce $Q_1$ amount of commodity 1 only, and agent 2 produce
$Q_2$ amount of
commodity 2 only.
At each time step $t$, two agents meet randomly to carry out transactions according
to their utility maximization principle. 

The utility functions as defined as follows:
For agent $1$, {$U_1(x_1,x_2,m_1)= x_1^{\alpha_1}x_2^{\alpha_2}m_1^{\alpha_m}$}
and for agent $2$,
$U_2(y_1,y_2,m_2)=y_1^{\alpha_1}y_2^{\alpha_2}m_2^{\alpha_m}$ where the
arguments in both of the utility functions are consumption of the first (i.e.
$x_1$ and $y_1$) and second good (i.e. $x_2$ and $y_2$)
and amount of money in their possession respectively.
For simplicity, they assume that the utility functions are of the above Cobb-Douglas
form with the sum of the powers normalized to $1$ i.e. 
$\alpha_1+\alpha_2+\alpha_m=1$.

Let the commodity prices  
to be determined in the market be denoted by $p_1$ and $p_2$.
Now, the budget constraints are as follows:
For agent $1$
the budget constraint is $p_1x_1+p_2x_2+m_1\leq M_1+p_1Q_1$ and
similarly, for agent $2$ the constraint is
$p_1y_1+p_2y_2+m_2 \leq M_2+p_2Q_2$,
which mean
that the amount that agent $1$ can spend for consuming $x_1$ and $x_2$
added to the amount of money that he holds after trading at time $t+1$
(i.e. $m_1$) cannot exceed the amount of money that he has 
at time $t$ (i.e. $M_1$)
added to what he earns by selling the good he produces (i.e. $Q_1$), and the same is true for agent $2$. 

Then the basic idea is that both of the agents try to
maximize their respective utility subject to their respective budget constraints and
the {\it invisible hand} of the market that is the price mechanism 
works to clear the market for both goods (i.e. total demand equals
total supply for both goods at the equilibrium prices), which means
that agent 1's problem is to maximize his utility subject to
his budget constraint i.e. maximize $U_1(x_1,x_2,m_1)$
subject to $p_1.x_1+p_2.x_2+m_1=M_1+p_1.Q_1$.
Similarly for agent 2, the problem is to maximize $U_1(y_1,y_2,m_2)$
subject to $p_1.y_1+p_2.y_2+m_2=M_2+p_2.Q_2$.
Solving those two maximization exercises by Lagrange multiplier and applying the condition
that the market remains in equilibrium, the competitive
price vector ($\hat p_1, \hat p_2$) as 
$\hat p_i=(\alpha_i/\alpha_m)(M_1+M_2)/Q_i$ for $i=1$, $2$ is found (\cite{Chakrabarti20094151}).
 
The outcomes of such a trading process are then:
\begin{enumerate}
\item 
At optimal prices $(\hat p_1, \hat p_2)$, $m_1(t)+m_2(t)=m_1(t+1)+m_2(t+1)$, 
i.e., demand matches supply in all market at
the market-determined price in equilibrium. Since money is also
treated as a commodity in this framework, its demand (i.e. the total
amount of money held by the two persons after trade) must be equal to what
was supplied (i.e. the total amount of money held by them before trade).
\item If a restrictive assumption is made such that $\alpha_1$ in the utility function 
can vary randomly over time with $\alpha_m$ remaining constant. It readily follows that $\alpha_2$ 
also varies randomly over time with the restriction that the sum of $\alpha_1$ and $\alpha_2$ is a
constant (1-$\alpha_m$). Then in the money demand equations derived, if we suppose $\alpha_m$ is $\lambda$ and $\alpha_1/(\alpha_1+\alpha_2)$ 
is $\epsilon$, it is found that money evolution equations become
$$m_1(t+1)=\lambda m_1(t)+\epsilon(1-\lambda)(m_1(t)+m_2(t))$$
$$m_2(t+1)=\lambda m_2(t)+(1-\epsilon)(1-\lambda)(m_1(t)+m_2(t)). \eqno$$
For a fixed value of $\lambda$, if $\alpha_1$ (or $\alpha_2$) 
is a random variable with uniform distribution over the domain $[0,1-\lambda]$,
then 
$\epsilon$ is also uniformly distributed over the domain $[0,1]$. 
%It may be noted that then $\lambda$ (i.e. $\alpha_m$ in the utility function) 
%is the savings propensity used in the CC model. 
This limit corresponds to the \cite{Chakraborti2000a} model, discussed earlier.
\item For the limiting value of $\alpha_m$ in the utility function
(i.e. $\alpha_m\rightarrow0$
which implies $\lambda\rightarrow0$), the money transfer
equation describing the random sharing of money without saving is obtained, which was studied by \cite{Dragulescu2000a} mentioned earlier.

\end{enumerate}

This actually demonstrates the equivalence of the two maximizations principles of entropy (in physics) and utility (in economics), and is certainly noteworthy.

\section{Agent-based modelling based on Games}

\subsection{Minority Game models}
\subsubsection{El Farol Bar Problem}
\cite{Arthur1994a} introduced the `El Farol Bar' problem as a paradigm of complex economic systems. In this problem, a population of agents have to decide whether to go to the bar opposite Santa Fe, every Thursday night. Due to a limited number of seats, the bar cannot entertain more than $X\%$ of the population. If less than $X\%$ of the
population go to the bar, the time spent in the bar is considered to be satisfying
and it is better to attend the bar rather than staying at home. But if more than $X\%$ of the population go to the bar, then it is too crowded and people in the bar have an unsatisfying time. In this second case, staying at home is considered to be better
choice than attending the bar. 
So, in order to optimise its own utility, each agent has to predict what everybody else will do. 

In particular Arthur was also interested  in agents who have bounds on ``rationality'', i.e. agents who:
\begin{itemize}
\item do not have perfect information about their environment, in general
they will only acquire information through interaction with the dynamically
changing environment;
\item do not have a perfect model of their environment;
\item have limited computational power, so they can't work out all the logical
consequences of their knowledge;
\item have other resource limitations (e.g. memory).
\end{itemize}
In order to take these limitations into account, each agent is randomly given a fixed menu of models potentially suitable to predict the number of people who will go the bar given past data (e.g. the same as two weeks ago, the average of the past few weeks, etc.). 
Each week, each agent evaluates these models against the past data. He chooses the one that was the best predictor on this data and then uses it to predict the number of people who will go to the bar this time. 
If this prediction is less than $X$, then the agent decides to go to the bar as well. If its prediction is more than $X$, the agent stays home. Thus, in order to make decisions on whether to attend
the bar, all the individuals are equipped with certain number of ``strategies'', which provide them the predictions of the attendance in the bar next
week, based on the attendance in the past few weeks.
As a result the number who go to the bar oscillates in an apparently random manner around the critical $X\%$ mark. 

This was one of the first models that led a way different from traditional economics.

\subsubsection{Basic Minority game}
The Minority Games (abbreviated MGs) (\cite{ChalletBook2004}) refer to the multi-agent
models of financial markets with the original formulation introduced by
\cite{Challet1997a}, and all other variants (\cite{Coolen2005,Lamper2002a}), most of which 
share the principal features that the models are repeated games and agents are inductive
in nature. The original formulation of the Minority Game
by \cite{Challet1997a} is sometimes referred as the ``Original Minority
Game'' or the ``Basic Minority Game''.

The basic minority game consists of $N$ (odd natural number) agents, who choose
between one of the two decisions at each round of the game, using their own
simple inductive strategies. The two decisions could be, for example, ``buying'' or ``selling'' commodities/assets, denoted by $0$ or $1$, at a given time $t$.
An agent wins the game if it is one of the members of the minority group, and thus at each round, the minority group of agents win the
game and rewards are given to those strategies that predict the winning side.
All the agents have access to finite amount of public information, which is a common bit-string ``memory'' of the $M$ most recent outcomes, composed of the winning sides in the past few rounds. 
Thus the agents with finite memory are said to exhibit ``bounded rationality'' (\cite{Arthur1994a}).

\begin{figure}
\begin{center}
\includegraphics[width=\columnwidth]{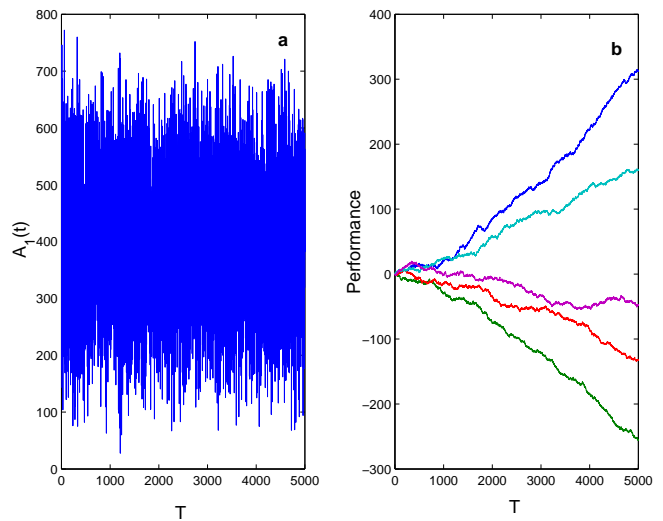}
\end{center}
\caption{Attendance fluctuation and performances of players in Basic Minority Game. Plots of (a) attendance and (b) performance of the players (five curves are: the best, the worst and three randomly chosen) for the
basic minority game with $N=801$; $M=6$; $k=10$ and $T=5000$. Reproduced from \cite{Sysi-Aho2003c}. }
\label{fig:bmg1}
\end{figure}

Consider for example, memory $M=2$; then there are $P=2^{M}=4$ possible ``history'' bit strings: $00$, $01$, $10$ and $11$. 
A ``strategy'' consists of a response, i.e., $0$ or $1$, to each possible history bit strings; therefore, there are $G=2^{P}=2^{2^{M}}=16$ possible strategies which constitute the {}``strategy space''. 
At the beginning of the game, each agent randomly picks $k$ strategies, and after the game, assigns one {}``virtual'' point to a strategy which would have predicted the correct outcome. 
The actual performance $r$ of the player is measured by the number of times the player wins, and the strategy, using which the player wins, gets a {}``real'' point. 
A record of the number of agents who have chosen a particular action, say, {}``selling'' denoted by $1,$ $A_{1}(t)$ as a function of time is kept (see Fig. \ref{fig:bmg1}). 
The fluctuations in the behaviour of $A_{1}(t)$ actually indicate the system's total utility.
For example, we can have a situation where only one player is in the minority and all the other players lose. The other extreme case is when $(N-1)/2$ players are in the minority and $(N+1)/2$ players
lose. 
The total utility of the system is obviously greater for the latter case and from this perspective, the latter situation is more desirable. 
Therefore, the system is more efficient when there are smaller fluctuations around the mean than when the fluctuations are larger.
%%%%%%%%%%%%%
%%%%%%%%%%%%%

As in the El Farol bar problem, unlike most traditional economics models which assume agents are ``deductive'' in nature, here too a ``trial-and-error'' {\it inductive} thinking approach is implicitly implemented in process of decision-making when agents make their choices in the games. 

\subsubsection{Evolutionary minority games}

\begin{figure}
\begin{center}
\includegraphics[width=\columnwidth]{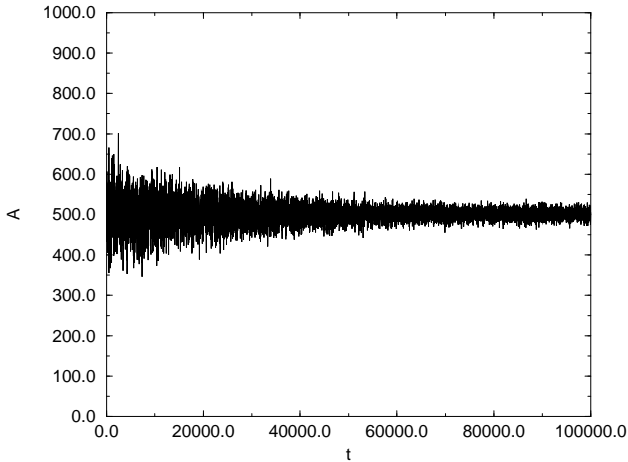}
\end{center}
\caption{Temporal attendance of $A$ for the genetic approach showing a learning process. Reproduced from \cite{Challet1997a}}
\label{fig:challetevol}
\end{figure}

Challet generalized the basic minority game (see \cite{Challet1997a,Challet1998a}) mentioned above to include the Darwinian selection: the worst player is replaced by a new one after some time steps, the new player is a ``clone'' of the best player, i.e. it inherits all the strategies but with corresponding virtual capitals reset to zero (analogous to a new born baby, though having all the predispositions from the parents, does not inherit their knowledge).
To keep a certain diversity they introduced a mutation possibility in cloning. 
They allowed one of the strategies of the best player to be replaced by a new one. 
Since strategies are not just recycled among the players any more, the whole strategy phase space is available for selection. 
They expected this population to be capable of {}``learning'' since bad players are weeded out with time, and fighting is among the so-to-speak the {}``best'' players. 
Indeed in Fig. \ref{fig:challetevol}, they observed that the learning emerged in time. 
Fluctuations are reduced and saturated, this implies the average gain for everybody is improved but never reaches the ideal limit.
%%%%%%%%%%%%%%
%%%%%%%%%%%%%%

\cite{Li2000b,Li2000c} also studied the minority game in the presence of ``evolution''.
In particular, they examined the behaviour in games in which the dimension of the strategy space, $m$, is the same for all agents and fixed for all time. 
They found that for all values of $m$, not too large, evolution results in a substantial improvement in overall system performance.
They also showed that after evolution, results obeyed a scaling relation among games played with different values of $m$ and different numbers of agents, analogous to that found in the non-evolutionary, adaptive games (see remarks on section \ref{subsubsection:Remarks}).
Best system performance still occurred, for a given number of agents, at $m_{c}$, the same value of the dimension of the strategy space as in the non-evolutionary case, but system performance was nearly an order of magnitude better than the non-evolutionary result.
For $m<m_{c}$, the system evolved to states in which average agent wealth was better than in the random choice game.
As $m$ became large, overall systems performance approached that of the random choice game.

\cite{Li2000b,Li2000c} continued the study of evolution in minority games by examining games in which agents with poorly performing strategies can trade in their strategies for new ones from a different strategy space, which meant allowing for strategies that use information from different numbers of time lags, $m$. 
They found, in all the games, that after evolution, wealth per agent is high for agents with strategies drawn from small strategy spaces (small $m$), and low for agents with strategies drawn from large strategy spaces (large $m$). 
In the game played with $N$ agents, wealth per agent as a function of $m$ was very nearly a step function. 
The transition was found to be at $m=m_{t}$, where $m_{t}\simeq m_{c}-1$, and $m_{c}$ is the critical value of $m$ at which $N$ agents playing the game with a fixed strategy space (fixed $m$) have the best emergent coordination and the best utilization of resources. 
They also found that overall system-wide utilization of resources is independent of $N$. 
Furthermore, although overall system-wide utilization of resources after evolution varied somewhat depending on some other aspects of the evolutionary dynamics, in the best cases, utilization of resources was on the order of the best results achieved in evolutionary games with fixed strategy spaces.

\subsubsection{Adaptive minority games}

\begin{figure}
\begin{center}
\includegraphics[width=\columnwidth]{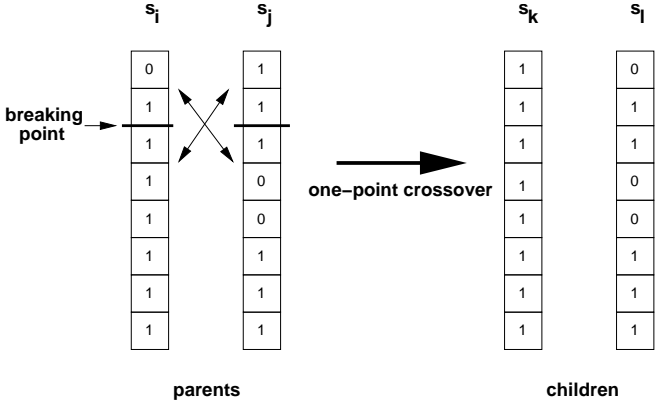}
\end{center}
\caption{Schematic diagram to illustrate the mechanism of one-point genetic crossover for producing new strategies. 
The strategies $s_{i}$ and $s_{j}$ are the parents. 
We choose the breaking point randomly and through this one-point genetic crossover, the children $s_{k}$ and $s_{l}$ are produced and substitute the parents. Reproduced from \cite{Sysi-Aho2003c}. }
\label{fig:crossover}
\end{figure}

\cite{Sysi-Aho2003a,Sysi-Aho2003b,Sysi-Aho2003c,Sysi-Aho2004a} presented a simple modification of the basic minority game where the players modify their strategies periodically after every time interval $\tau $, depending on their performances: if a player finds that he is among the fraction $n$ (where $0<n<1$) who are the worst performing players, he adapts himself and modifies his strategies. 
They proposed that the agents use hybridized one-point genetic crossover mechanism (as shown in Fig. \ref{fig:crossover}), inspired by genetic evolution in biology, to modify the strategies and replace the bad strategies. 
They studied the performances of the agents under different conditions and investigate how they adapt themselves in order to survive or be the best, by finding new strategies using the highly effective mechanism.
%%%%%%%%%%%%%%
%%%%%%%%%%%%%%
They also studied the measure of total utility of the system $U(x_{t})$, which is the number of players in the minority group; the total utility of the system is maximum $U_{max}$ as the highest number of players win is equal to $(N-1)/2$. 
The system is more efficient when the deviations from the maximum total utility $U_{max}$ are smaller, or in other words, the fluctuations in $A_{1}(t)$ around the mean become smaller.

\begin{figure}
\begin{center}
\includegraphics[width=\columnwidth]{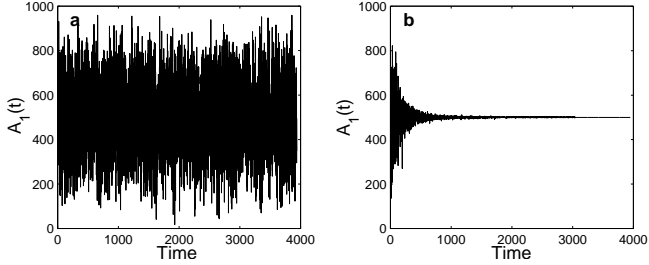}
\end{center}
\caption{Plot to show the time variations of the number of players $A_{1}$ who choose action 1, with the parameters $N=1001$, $m=5$, $s=10$ and $t=4000$ for (a) basic minority game and (b) adaptive game, where
$\tau =25$ and $n=0.6$. Reproduced from \cite{Sysi-Aho2003c}.}
\label{fig:marko1}
\end{figure}

\begin{figure}
\begin{center}
\includegraphics[width=\columnwidth]{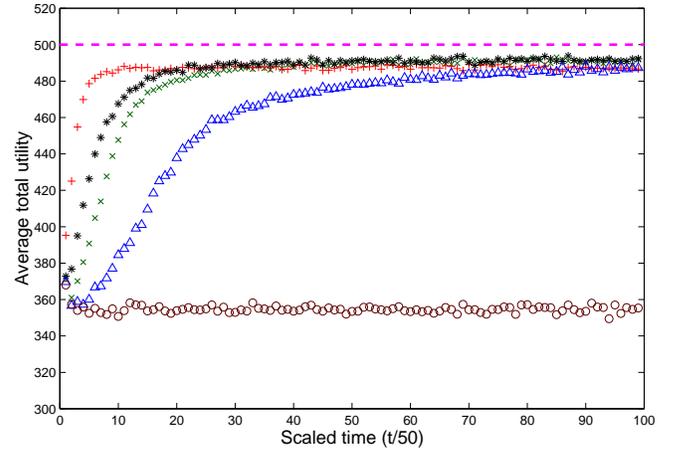}
\end{center}
\caption{Plot to show the variation of total utility of the system with time  for the basic minority game for $N=1001$, $m=5$, $s=10$, $t=5000$, and adaptive game, for the same parameters but different values of
$\tau $ and $n$. 
Each point represents a time average of the total utility for separate bins of size 50 time-steps of the game. 
The maximum total utility ($=(N-1)/2$) is shown as a dashed line. 
The data for the basic minority game is shown in circles. 
The plus signs are for $\tau =10$ and $n=0.6$; the asterisk marks are for $\tau =50$ an $n=0.6$; the cross marks for $\tau =10$ and $n=0.2$ and triangles for $\tau =50$ and $n=0.2$. The ensemble average over 70 different samples was taken in each case. Reproduced from \cite{Sysi-Aho2003c}.}
\label{fig:marko2}
\end{figure}

Interestingly, the fluctuations disappear totally and the system stabilizes to a state where the total utility of the system is at maximum, since at each time step the highest number of players win the game (see Fig. \ref{fig:marko1}). 
As expected, the behaviour depends on the parameter values for the system (see \cite{Sysi-Aho2003c,Sysi-Aho2004a}).
%%%%%%%%%%%%%%
%%%%%%%%%%%%%%
They used the utility function to study the efficiency and dynamics of the game as shown in Fig. \ref{fig:marko2}.
%%%%%%%%%%%%%%
%%%%%%%%%%%%%%
If the parents are chosen randomly from the pool of strategies then the mechanism represents a ``one-point genetic crossover'' and if the parents are the best strategies then the mechanism represents a ``hybridized genetic crossover''.
The children may replace parents or two worst strategies and accordingly four different interesting cases arise:
(a) one-point genetic crossover with parents ``killed'', i.e. parents are replaced by the children,
(b) one-point genetic crossover with parents ``saved'', i.e. the two worst 
strategies are replaced by the children but the parents are retained,
(c) hybridized genetic crossover with parents ``killed''
and (d) hybridized genetic crossover with parents ``saved''.

\begin{figure}
\includegraphics[width=\columnwidth]{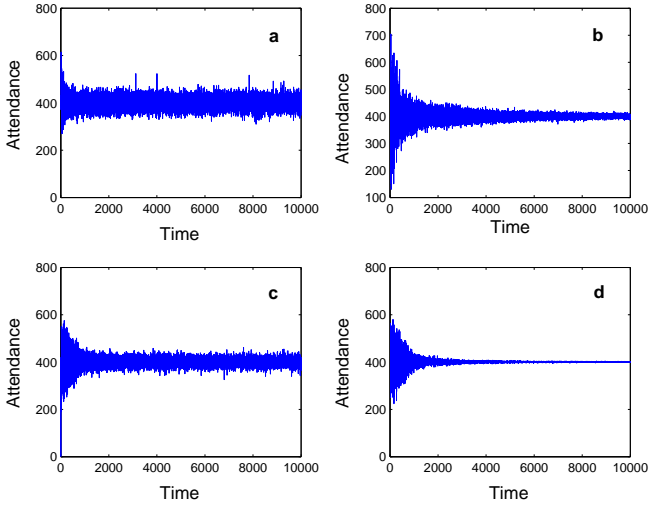}
\caption{ Plots of the attendances by choosing parents randomly (a) and (b), and using the best parents in a player's pool (c) and (d). 
In (a) and (c) case parents are replaced by children and in (b) and (d) case children replace the two worst strategies. 
Simulations have been done with $N=801$, $M=6$, $k=16$, $t=40$, $n=0.4$ and  $T=10000$.} 
\label{all4a}
\end{figure}

\begin{figure}
\includegraphics[width=\columnwidth]{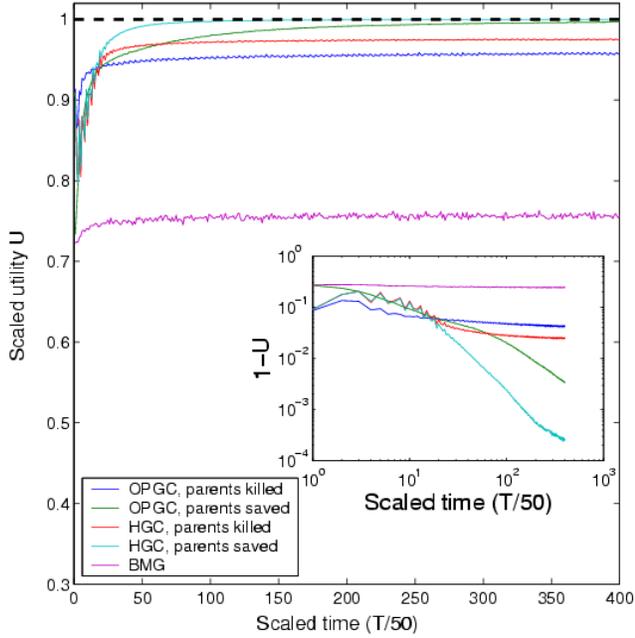}
\caption{ Plots of the scaled utilities of the four different mechanisms in comparison with that of the basic minority game.
Each curve represents an ensemble average over 100 runs and each point in a curve is a time average over a bin of length 50 time-steps. 
In the inset, the quantity ($1-U$) is plotted against  scaled time in the double logarithmic scale. Simulations are done with $N=801$, $M=6$, $k=16$,  $t=40$, $n=0.4$ and  $T=20000$. Reproduced from \cite{Sysi-Aho2003c}.
\label{all4b}}
\end{figure}

In order to determine which mechanism is the most efficient, we have made a comparative study of the 
four cases, mentioned above. We plot the attendance as a function of time for the different mechanisms in Fig. \ref{all4a}. 
%%%%%%%%%%%%%%
%%%%%%%%%%%%%%
In Fig. \ref{all4b} we show the total utility of the system in each of the cases (a)-(d), where we have plotted results of the average over 100 runs and each point in the utility curve represents a time average taken over a bin of length 50  time-steps. 
The simulation time is doubled from those in Fig. \ref{all4a}, in order to expose the asymptotic 
behaviour better. 
On the basis of Figs. \ref{all4a} and \ref{all4b}, we find that the case (d) is the most efficient.
%%%%%%%%%%%%%%
In order to investigate what happens in the level of an individual agent, we created a competitive surrounding-- ``test'' situation where after $T=3120$ time-steps, six players begin to adapt and modify their strategies such that three are using hybridized genetic crossover mechanism and the other three one point genetic crossover, where children replace the parents.
The rest of the players play the basic minority game. 
In this case it turns out that in the end the best players are those who use the hybridized mechanism,
second best are those using the one-point mechanism, and the bad players those who do not adapt at all.
In addition it turns out that the competition amongst the players who adapt using the hybridized genetic
crossover mechanism is severe.

It should be noted that the mechanism of evolution of strategies is considerably different from earlier attempts such as \cite{Challet1997a} or \cite{Li2000b,Li2000c}. 
This is because in this mechanism the strategies are changed by the agents themselves and even though the strategy space evolves continuously, its size and dimensionality remain the same. 

Due to the simplicity of these models (\cite{Sysi-Aho2003a,Sysi-Aho2003b,Sysi-Aho2003c,Sysi-Aho2004a}), a lot of freedom is found in modifying the models to make the situations more realistic and applicable to many real
dynamical systems, and not only financial markets. Many details in the model can be fine-tuned to imitate the
real markets or behaviour of other complex systems.  Many other sophisticated models based on
these games can be setup and implemented, which show a great potential
over the commonly adopted statistical techniques in analyses of financial markets.

\subsubsection{Remarks}
\label{subsubsection:Remarks}
For modelling purposes, the minority game models were meant to serve as a class
of simple models which could produce some macroscopic features observed in the real financial markets, which included the fat-tail price return distribution and volatility clustering (\cite{ChalletBook2004,Coolen2005}). 
Despite the hectic activity (\cite{Challet1998a,Challet2000a}) they have failed to capture or
reproduce most important stylized facts of the real markets. However, in the physicists' community, they have become an
interesting and established class of models where the physics of disordered systems (\cite{Cavagna1999a,Challet2000a}),
lending a large amount of physical insights (\cite{Savit1999a,DeMartino2004}). 
Since in the BMG model a Hamiltonian function could be defined and analytic solutions could be developed in some regimes of the model, the model was viewed with a more physical picture. 
In fact, it is characterized by a clear two-phase structure with very different collective behaviours in the
two phases, as in many known conventional physical systems (\cite{Savit1999a,Cavagna1999a}).

\begin{figure}
\includegraphics[width=\columnwidth]{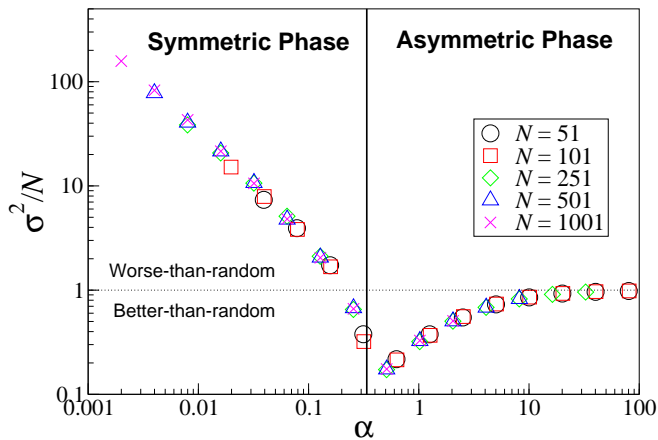}
\caption{
The simulation results of the variance in attendance $\sigma^2/N$ as a function of the control parameter $\alpha=2^M/N$
for games with $k=2$ strategies for each agent, ensemble averaged over 100 sample runs.
Dotted line shows the value of volatility in random choice limit.
Solid line shows the critical value of $\alpha=\alpha_c\approx 0.3374$.
Reproduced from Yeung and Zhang {\tt arxiv:0811.1479}.
}
\label{varVsAlpha}
\end{figure}

\cite{Savit1999a} first found that the macroscopic behaviour of the system
does not depend independently on the parameters $N$ and $M$,
but instead depends on the ratio
\begin{eqnarray}
\label{alpha}
	\alpha \equiv \frac{2^M}{N} = \frac{P}{N}
\end{eqnarray}
which serves as the most important 
control parameter in the game.
The variance in the attendance (see also \cite{Sysi-Aho2003b}) or volatility $\sigma^2/N$, for different values of $N$ and $M$ depend only on the ratio $\alpha$.
 Fig.~\ref{varVsAlpha} shows a plot of $\sigma^2/N$ against the control parameter $\alpha$, where the data collapse of $\sigma^2/N$ for different
values of $N$ and $M$ is clearly evident.
The dotted line in Fig.~\ref{varVsAlpha} corresponds to the ``coin-toss'' limit (random choice or pure chance limit),
in which agents play by simply making random decisions (by coin-tossing) at every rounds of the game.
This value of $\sigma^2/N$  in coin-toss limit can be obtained by simply assuming a binomial distribution of the agents' binary actions,
with probability $0.5$,
such that $\sigma^2/N=0.5(1-0.5)\cdot 4=1$.
When $\alpha$ is small,
the value of $\sigma^2/N$  of the game is larger than the coin-toss limit which implies the collective
behaviours of agents are worse than the random choices.
In the early literature,
it was popularly called as the {\it worse-than-random} regime.
When $\alpha$ increases,
the value of $\sigma^2/N$  decreases and enters a region where agents are performing better than the random choices,
which  was popularly called as the {\it better-than-random} regime.
The  value of $\sigma^2/N$  reaches a minimum value which is substantially smaller than the coin-toss limit.
When $\alpha$ further increases,
the  value of $\sigma^2/N$ increases again and approaches the coin-toss limit.
This allowed one to identify two phases in the Minority Game,
as separated by the minimum  value of $\sigma^2/N$ in the graph.
The value of $\alpha$ where the rescaled volatility attended its minimum 
was denoted by $\alpha_c$,
which represented the phase transition point;
$\alpha_c$ has been shown to have a value of $0.3374\dots$ (for $k=2$) by analytical calculations \cite{Challet2000a}.

Besides
these collective behaviours, physicists became also interested in the dynamics of the
games such as crowd vs anti-crowd movement
of agents, periodic attractors, etc. (\cite{johnson99,johnson99b,hart01}). In this way, the Minority Games
serve as a useful tool and provide a new direction for physicists in viewing and
analysing the underlying dynamics of complex evolving systems such as the financial markets.

\subsection{The Kolkata Paise Restaurant (KPR) problem}
The KPR problem (\cite{anindya,mathematica,proc,asim}) is a repeated game, played between a large number $N$ of agents having no interaction amongst themselves. In KPR problem,  prospective customers (agents) choose from $N$ restaurants each evening simultaneously (in parallel decision mode); $N$ is fixed. Each restaurant has the same price for a meal but a different rank (agreed upon by all  customers) and can serve only one customer any evening. Information regarding the customer distributions for earlier evenings is available to everyone. Each customer's objective is to go to the restaurant with the highest possible rank while avoiding the crowd so as to be able to get dinner there. If more than one customer arrives at any restaurant on any evening, one of them is randomly chosen (each of them are anonymously treated) and is served. The rest do not get dinner that evening.

In Kolkata, there were very cheap and fixed rate ``Paise Restaurants"  that were popular among the daily labourers in the city. During lunch hours, the labourers used to walk (to save the transport costs) to one of these restaurants and would miss lunch if they got to a restaurant where there were too many customers. Walking down to the next restaurant would mean failing to report back to work on time! Paise is the smallest Indian coin and there were indeed some well-known rankings of these restaurants, as some of them would offer tastier items compared to the others. A more general example of such a problem would be when the society provides hospitals (and beds) in every locality but the local patients go to hospitals of better rank (commonly perceived) elsewhere, thereby  competing with the local patients of those hospitals.   Unavailability of treatment in time may be considered as lack of the service for those people and consequently as (social) wastage of service by those unattended hospitals.

A dictator's solution to the KPR problem is the following: the dictator asks everyone to form a queue and then assigns each one a   restaurant  with rank matching the sequence of the person in the queue on the first evening. Then each person is told to go to the next ranked restaurant  in the following evening (for the person in the last ranked restaurant this means going to the first ranked restaurant). This shift proceeds then continuously for successive evenings. This is clearly one of the most efficient solution (with utilization fraction $\bar f$ of the services by the restaurants equal to unity) and the system arrives at this this solution immediately (from the first evening itself). However, in reality this cannot be the true solution of the KPR problem,  where each agent decides on his own (in parallel or democratically) every evening, based on complete information about past events. In this game, the customers try to evolve a learning strategy to eventually get dinners at the best possible ranked restaurant, avoiding the crowd. It is seen, the evolution these strategies take considerable time to converge and even then the eventual utilization fraction $\bar{f}$ is far below unity. 

 Let the symmetric stochastic strategy chosen by each agent be such that at any time $t$, the probability $p_k(t)$ to arrive at the $k$-th ranked restaurant is given by
\begin{eqnarray}
p_k(t) =\frac{1}{z}\left[k^{\alpha}\exp\left(-\frac{n_k(t-1)}{T}\right)\right], \nonumber \\ z=\sum_{k=1}^N\left[k^{\alpha}\exp\left(-\frac{n_k(t-1)}{T}\right)\right],\label{generalstoch}
\end{eqnarray}
where $n_k(t)$ denotes the number of agents arriving at the $k$-th ranked restaurant in period $t$, $T>0$ is a  scaling factor and $\alpha\geq 0$ is an exponent. 

For any natural number $\alpha$ and $T\rightarrow \infty$, an agent goes to the $k$-\rm{th} ranked restaurant with probability $p_k(t)=k^\alpha/\sum k^\alpha$; which means in the limit  $T\rightarrow \infty$ in (\ref{generalstoch}) gives $p_k(t)=k^\alpha/\sum k^\alpha$.  

If an agent selects any restaurant with equal probability $p$ then probability that a single restaurant is chosen by $m$ agents is given by
\begin{eqnarray}
%\label{eq:poisson_mm}
\Delta(m) &=& \left( \begin{array}{c}  N \\ m \end{array}\right)
p^m (1-p)^{N -m}.
\end{eqnarray}

\noindent Therefore, the probability that a restaurant with rank $k$ is not chosen by any of the agents will be given by
    
 \begin{eqnarray}
%\label{eq:poisson_mm1}
\Delta_k(m=0) &=& \left( \begin{array}{c}  N \\ 0 \end{array}\right)
\left(1-p_{k}\right)^{ N }; \ \ p_k=\frac{k^\alpha}{\sum k^\alpha}  \nonumber \\
&\simeq& \exp\left({-k^\alpha N\over \widetilde {N} }\right) \ \ {\rm as} \ \ N \to \infty,
\end{eqnarray}

\noindent where $\widetilde N=\sum_{k=1}^{N} k^\alpha\simeq\int_0^N k^\alpha dk= \frac{N^{\alpha+1}}{(\alpha+1)}.$ Hence
\begin{equation}
 \Delta_k(m=0)=\exp\left(-{k^\alpha \left(\alpha+1\right)\over N^\alpha}\right) .
\end{equation}

\normalsize
\noindent Therefore the average fraction of agents getting dinner in the $k$-{\rm th} ranked restaurant is given by
\begin{equation}
\bar f_k=1- \Delta_k\left(m=0\right).
\end{equation}

Naturally for $\alpha=0$, the problem corresponding to random choice $\bar f_k=1-e^{-1}$, giving  $\bar f=\sum \bar f_k/N \simeq 0.63$ and for  $\alpha=1$, $\bar f_k=1-e^{-2k/N}$ giving  $\bar f=\sum \bar f_k/N \simeq 0.58$.

In summary, in the KPR problem where the decision made by each agent in each evening $t$ is independent and is based on the information about the rank $k$ of the restaurants and their occupancy given by the numbers $n_k(t-1)\ldots n_k(0)$. For several stochastic strategies, only $n_k(t-1)$ is utilized and each agent chooses the $k$-{\rm th} ranked restaurant with probability $p_k(t)$ given by Eq. (\ref{generalstoch}). The utilization fraction $f_k$ of the $k$-{\rm th} ranked  restaurants on every evening is studied and their  average (over $k$) distributions $D(f)$ are studied numerically, as well as analytically,  and one finds (\cite{anindya,mathematica,proc}) their distributions to be Gaussian with the most probable utilization fraction $\bar f\simeq 0.63$, $0.58$ and $0.46$ for the cases with $\alpha=0$, $T\rightarrow\infty$; $\alpha=1$, $T\rightarrow\infty$; and $\alpha=0$, $T\rightarrow0$ respectively. For the stochastic crowd-avoiding strategy discussed in \cite{asim}, 
where $ p_k(t+1)=\frac{1}{n_k(t)} $ for $ k=k_0 $ the restaurant visited by the agent last evening, and $=1/(N-1)$ for all other restaurants ($ k \neq k_0 $),
one gets the best utilization fraction $\bar f\simeq0.8$, and the analytical estimates for $\bar f$ in these limits agree very well with the numerical observations.  
Also, the time required to converge to the above value of $\bar f$ is independent of $ N $.

The KPR problem has similarity with the Minority Game Problem (\cite{Arthur1994a,ChalletBook2004}) as in both the games, herding behaviour is punished and diversity's encouraged. Also, both involves learning of the agents from the past successes etc. Of course, KPR has some simple exact solution limits, a few of which are discussed here. 
The real challenge is, of course, to design algorithms of learning mixed strategies (e.g., from the pool discussed here) by the agents so that the fair social norm emerges eventually (in $ N^0 $ or $ \ln N $ order time) even when every one decides on the basis of their own information independently. As we have seen,      some naive strategies give better values of $\bar f$ compared to most of the ``smarter'' strategies like strict crowd-avoiding strategies, etc. This observation in fact compares well with earlier observation in minority games (see e.g. \cite{Satinover2007}).

It may be noted that all the stochastic strategies, being  parallel in computational mode, have the advantage that they converge to solution at smaller time steps ($\sim N^0 $ or $ \ln N$) while for deterministic strategies the convergence time is typically of order of $N$, which renders such strategies useless in the truly macroscopic ($N\rightarrow\infty$) limits. However, deterministic strategies are useful when $N$ is small and rational agents can design appropriate punishment schemes for the deviators (see \cite{kandori}).
 
The study of the KPR problem shows that while a dictated solution leads to one of the best possible solution to the problem, with each agent getting his dinner at the best ranked restaurant with a period of $N$ evenings,  and  with best possible value of $\bar{f}$ ($=1$) starting from the first evening itself. The parallel decision strategies (employing evolving algorithms by the agents, and past informations,   e.g.,  of $n(t)$), which are necessarily parallel among the agents and stochastic  (as in democracy), are less efficient ($\bar {f}\ll1$; the best one discussed in \cite{asim},  giving $\bar {f}\simeq0.8$ only).  Note here that the time required is not dependent on $ N $. We  also note that most of the  ``smarter'' strategies lead to much lower efficiency.

\section{Conclusions and outlook}
%\label{part:Conclusion}

Agent-based models of order books are a good example of interactions between ideas and methods that are usually linked either to Economics and Finance (microstructure of markets, agent interaction) or to Physics (reaction-diffusion processes, deposition-evaporation process, kinetic theory of gases). As of today, existing models exhibit a trade-off between ``realism'' and calibration in its mechanisms and processes (empirical models such as \cite{MikeFarmer2008}), and explanatory power of simple observed behaviours (\cite{ContBouchaud2000,Cont2007} for example). In the first case, some of the ``stylized facts'' may be reproduced, but using empirical processes that may not be linked to any behaviour observed on the market. In the second case, these are only toy models that cannot be calibrated on data. The mixing of many features, as in \cite{LuxMarchesi2000} and as is usually the case in behavioural finance, leads to poorly tractable models where the sensitivity to one parameter is hardly understandable.
Therefore, no empirical model can tackle properly empirical facts such as volatility clustering. Importing toy model features explaining volatility clustering or market interactions in order book models is yet to be done. Finally, let us also note that to our knowledge, no agent-based model of order books deals with the multidimensional case. Implementing agents trading simultaneously several assets in a way that reproduces empirical observations on correlation and dependence remains an open challenge.

We believe this type of modelling is crucial for future developments in finance. The financial crisis that occurred in 2007-2008 is expected to create a shock in classic modelling in Economics and Finance. Many scientists have expressed their views on this subject (e.g. \cite{Bouchaud2008,Lux2009,Farmer2009}) and we believe as well that agent-based models we have presented here will be at the core of future modelling. As illustrations, let us mention \cite{Iori2006}, which models the interbank market and investigates systemic risk, \cite{Thurner2009}, which investigates the effects of use of leverage and margin calls on the stability of a market and \cite{YakovenkoRosser2009}, which provides a brief overview of the study of wealth distributions and inequalities. No doubt these will be followed by many other contributions.

\section*{Acknowledgements}
The authors would like to thank their collaborators and two anonymous reviewers whose comments greatly helped improving the review. AC is grateful to B.K. Chakrabarti, K. Kaski, J. Kertesz, T. Lux, M. Marsili, D. Stauffer and V. Yakovenko for invaluable suggestions and criticisms.

\bibliographystyle{rQUF}
\bibliography{MarketsPhysicsEconomicsBiblio1,StatisticalPropertiesBiblio1,OrderBookReview1,ConvertedBiblio1,ConvertedBiblio_Zotero1,archive_list1,AddedRevision1,MarketsPhysicsEconomicsBiblio2,StatisticalPropertiesBiblio2,OrderBookReview2,ConvertedBiblio2,ConvertedBiblio_Zotero2,archive_list2,AddedRevision2}

\end{document}